\documentclass[twocolumn,prb,aps,epsf,showpacs,superscriptaddress]{revtex4-1}

\usepackage{graphicx}
\usepackage{dcolumn}
\usepackage{bm}

\newcommand{\D}{\mathrm{d}}
\newcommand{\E}{\mathrm{e}}
\newcommand{\I}{\mathrm{i}}

\newcommand{\ybco}[1]{YBa$_{2}$Cu$_{3}$O$_{#1}$}

\newcommand{\figwidth}{0.85}

\begin{document}

\title{Microwave spectroscopy of vortex dynamics in ortho-II YBa$_{\bm 2}$Cu$_{\bm 3}$O$_{\bm{6.52}}$}

\author{X.-Q.~Zhou}
\altaffiliation{Department of Physics, University of Colorado, Boulder, CO 80309, USA}
\author{C.~J.~S.~Truncik}
\author{W.~A.~Huttema}
\author{N.~C.~Murphy}
\author{P.~J.~Turner}
\author{A.~J.~Koenig}
\affiliation{Department of Physics, Simon Fraser University, Burnaby, BC, V5A~1S6, Canada}
\author{R.-X.~Liang}
\author{D.~A.~Bonn}
\author{W.~N.~Hardy}
\affiliation{Department of Physics and Astronomy, University of British Columbia, Vancouver, BC, V6T~1Z1, Canada}
\affiliation{Canadian Institute for Advanced Research, Toronto, Ontario, MG5 1Z8, Canada}  
\author{D.~M.~Broun}
\affiliation{Department of Physics, Simon Fraser University, Burnaby, BC, V5A~1S6, Canada}

\begin{abstract}
We present measurements of the vortex-state surface impedance, $Z_s = R_s + \I X_s$, of a high quality, ortho-II-ordered single crystal of the cuprate high temperature superconductor \ybco{6.52} \mbox{($T_c = 59$~K)}.   Measurements have been made at four microwave frequencies ($\omega/2 \pi = 2.64$, 4.51, 9.12 and 13.97~GHz), for magnetic fields ranging from 0 to 7~T.  From these data we obtain the field, frequency and temperature dependence of the vortex viscosity, pinning constant, depinning frequency and flux-flow resistivity.  The vortex viscosity, $\eta(\omega,T)$, has a surprisingly strong frequency dependence and bears a striking resemblance to the zero-field quasiparticle conductivity, $\sigma_\mathrm{qp}(\omega,T)$, suggesting that the dominant dissipative mechanism for the flux lines is induced electric fields coupling to bulk, long-lived $d$-wave quasiparticles \emph{outside} the vortex cores. This is in sharp contrast to the conventional Bardeen--Stephen picture, in which dissipation takes place \emph{inside} quasi-normal vortex cores.  The strong frequency dependence of the vortex viscosity in the microwave range requires us to treat it as a complex response function, with an imaginary part that is predicted to contribute to the apparent pinning force on the vortices. Measurements of the frequency dependence of the pinning force confirm that this term is present, and in a form consistent with the requirements of causality.  At low temperatures the flux-flow resistivity, $\rho_\mathrm{ff} \propto 1/\eta$, has the form $\rho_\mathrm{ff}(T) = \rho_0 + \rho_1 \ln(1/T)$, reminiscent of the DC resistivity of cuprates in the pseudogap regime.
\end{abstract}
 
 
\maketitle{} 
  
\section{Introduction}

The frictional force experienced by a quantized flux line moving in a conventional superconductor arises primarily from induced vortex electric fields coupling to charge excitations \emph{within} the vortex core.  This was first captured by Bardeen and Stephen,\cite{Bardeen:1965p151} who treated the vortex core as a cylinder of normal metal embedded in a superconducting background.  Their theory is applicable to conventional superconductors for two reasons:  the vortex cores are large and  support a nearly continuous spectrum of single-particle states; and $s$-wave pairing symmetry results in a low density of extended states surrounding the vortex cores.  In cuprate superconductors the opposite situation holds:  small vortex cores contain at most a few discrete states,\cite{MaggioAprile:1995p3014,Soininen:1994iu} with a continuum of low lying states \emph{outside} the vortex cores due to the nodes in the $d$-wave energy gap.\cite{hardy93,Scalapino:1995p741,Ding:1996p3019}   Bardeen--Stephen theory is therefore unlikely to apply in its original form, but how it should be extended to the cuprates is not at all obvious.

The dissipation associated with a moving flux line is parameterized by a vortex viscosity, $\eta$, giving the linear coefficient of friction per unit length of flux line.  Vortex viscosity, like flux-flow resistivity, is usually thought of as a static property of a type-II superconductor.  However, $\eta$ should more generally be regarded as a frequency-dependent response function,\cite{Choi:1994ga} in a manner similar  to the extension of electrical conductivity to high frequencies.  We will show that $\eta$ has a very strong frequency dependence in ortho-II \ybco{6.52}, and that this frequency dependence carries the fingerprints of the microscopic processes responsible for the vortex dissipation, namely the charge dynamics of $d$-wave quasiparticles in the superconducting state outside the vortex cores. 

As well as being interesting in its own right,\cite{Bardeen:1965p151,Nozieres:1966p667,Larkin:1976p3015} vortex viscosity has additional significance in the cuprates:  low superfluid density \cite{Uemura:1989p962} makes these materials prone to phase disordering by vortex--antivortex fluctuations,\cite{Emery:1995p364,franz01,franz02,herbut02, herbut02a,herbut05,franz06,Tesanovic:2008p2290} with vortex viscosity an important parameter in theoretical models of these effects.\cite{Geshkenbein:1998p3010,Ioffe:2002p717,Lee:2003p7,Melikyan:2005p3011,Nikolic:2006p3012,Bilbro:2011p3009}  There is also the possibility that the viscous response contains dynamical signatures that could identify whether vortex fluctuations are occurring in the pseudogap regime; this would provide information complementary to other experiments that may be probing local pairing and phase-disordered superconductivity.\cite{Corson:1999p716,Xu:2000p609,Wang:2005p2400,Wang:2006p185,Bilbro:2011p2722}

In this paper we report a comprehensive study of the vortex dynamics of high purity, ortho-II-ordered \ybco{6.52}.  
Remarkably, the observed $\eta(\omega,T)$ mimics the behaviour of the zero-field microwave conductivity, $\sigma_\mathrm{qp}(\omega,T)$, in which it has been established that the dominant contribution comes from the charge dynamics of nodal $d$-wave quasiparticles.\cite{Bonn:1992p3021,HIRSCHFELD:1994p570,Hosseini:1999p383,Turner:2003p331,Harris:2006p388}  Our data therefore suggest that bulk $d$-wave quasiparticles \emph{outside} the vortex cores are the primary source of frictional force on vortices in this material, in marked contrast to the situation in conventional $s$-wave materials.  One consequence is that $\eta(T)$, like $\sigma_\mathrm{qp}(T)$, has a characteristic peak at intermediate temperatures, due to a competition between quasiparticle lifetime and quasiparticle density.  This leads to low temperature upturns in the flux-flow resistivity, $\rho_\mathrm{ff}(T) \propto 1/\eta(T)$, which, on closer inspection, are seen to follow a $\log(1/T)$ form, similar to the resistivity observed in the pseudogap regime of the underdoped cuprates.\cite{Ando:1995p148, Boebinger:1996p147}

The paper is organized as follows.  We start with the standard dynamical model of a flux line, and discuss the observability of the vortex Hall effect in microwave measurements.  Vortex pinning is then introduced into the dynamical model, through a redefinition of the vortex viscosity.  Next, we consider a more general situation, in which the vortex viscosity itself is frequency dependent, and show that this leads to a dynamical contribution to the effective pinning constant.  We then present detailed measurements of the surface impedance of a high quality single crystal of ortho-II \ybco{6.52} as a function of field, temperature and frequency.  From these data we obtain  the vortex viscosity, pinning constant, depinning frequency and flux-flow resistivity.  As well as supplying new insights into the origin of the various forces experienced by the vortices, the experiments provide a stringent test of the use of  single-vortex dynamical models in the interpretation of high frequency measurements.

\section{Vortex dynamics}
\label{Sec:vortex_dynamics}

\subsection{Hall effect in a conventional metal}

The vortex Hall effect has useful parallels with that of a normal metal, so we begin by considering the magnetoconductivity of a metal in which the carriers have density $n$, mass $m$ and charge $q$.  If scattering is treated in the relaxation-time approximation, the steady-state force equation is
\begin{equation}
q \left(\mathbf{E} + \mathbf{v} \times \mathbf{B}\right) - \frac{m \mathbf{v}}{\tau} = 0\;,
\label{Eq:normalmetal}
\end{equation}
where $\mathbf{v}$ is the carrier drift velocity and $\tau$ the relaxation time.  We let  the magnetic field $\mathbf{B} = \left(0,0,B_z \right)$.  The resistivity tensor that relates electric field, $\mathbf{E}$, to the current density, $\mathbf{j} = n q \mathbf{v}$, is
\begin{equation}
\bm \rho = \rho_0 \left(\begin{array}{ccc}1 & -\omega_c \tau & 0 \\\omega_c \tau & 1& 0\\ 0 & 0 & 1\end{array}\right)\;,
\end{equation}
where $\omega_c = q B/m$ is the cyclotron frequency and \mbox{$\rho_0 = m/n q^2 \tau$}  the  resistivity.  The Hall angle, $\theta_H$, measures the deflection of the charge currents by the magnetic field: $\tan(\theta_H) = \omega_c \tau$.  The magnetoconductivity tensor is
\begin{equation}
\bm \sigma = \bm \rho^{-1} =  \frac{\sigma_0}{1\!+\! \left(\omega_c \tau\right)^2} \left(\begin{array}{ccc}1 & \omega_c  \tau & 0  \\-\omega_c \tau & 1 & 0\\ 0 & 0 & 1\! +\! \left(\omega_c \tau\right)^2\end{array}\right)\;,
\label{Eq:magnetoconductivity}
\end{equation}
where $\sigma_0 = 1/\rho_0$. Note that $\sigma_{xx}$ and $\sigma_{yy}$ depend on $\omega_c$, while the diagonal components of $\bm \rho$ do not.  Under conditions of constant current bias $\mathbf{j}$, the power dissipation per unit volume is 
\begin{equation}
\mathbf{j}^\top\bm{\rho}\,\mathbf{j} = \rho_0 j^2\;,
\end{equation}
independent of $\omega_c$.  On the other hand, if a constant electric field $\mathbf{E}$ is applied transverse to $\mathbf{B}$, the power dissipation per unit volume is
\begin{equation}
\mathbf{E}^\top\bm{\sigma}\,\mathbf{E} = \frac{\sigma_0}{1 + \left(\omega_c \tau\right)^2} E^2\;,
\end{equation}
which \emph{is} a function of $\omega_c$.  Whether a longitudinal transport experiment is sensitive to the Hall effect therefore depends crucially on  whether constant current or constant electric field is applied.

\subsection{Vortex Hall effect}

The starting point for much work on vortex dynamics is the vortex equation of motion\cite{Nozieres:1966p667,Vinen:1967,JIGittleman:1968p172,KOPNIN:1976ua,KOPNIN:1991va,Hsu:1993ih,Vinokur:1993to,Choi:1994ga,BLATTER:1994p494,Parks:1995p189,GOLOSOVSKY:1996p1}
\begin{equation}
 \eta \mathbf{v}_v + \alpha_H \mathbf{v}_v \times \mathbf{\hat z} = \Phi_0 \mathbf{j} \times \mathbf{\hat z}\;,
\label{Eq:vortex}
\end{equation}
where $\Phi_0 = h/2 e$ is the superconducting flux quantum, $\mathbf{j}$  the applied transport current density, $\mathbf{v}_v$  the vortex velocity, $\eta$  the vortex viscosity, $\alpha_H$  the Hall coefficient, and we assume that magnetic field is applied along the direction $\mathbf{\hat z}$.  (Pinning effects are not included at this point: we show in Sec.~\ref{Sec:complex_viscosity_model} how pinning can be folded into a complex generalization of the vortex viscosity.  Vortex inertia is also ignored, as it is negligible in the microwave frequency range.  Similarly, thermal flux creep is not included in models of microwave-frequency dynamics, as the creep rate is expected to be much lower than the measurement frequency.\cite{GOLOSOVSKY:1996p1})

By requiring that the vortex dynamics be consistent with the magnetoconductivity of the electron fluid, Blatter \emph{et al.}\ argue that the viscosity and Hall coefficient must have the form:\cite{BLATTER:1994p494}
\begin{eqnarray}
\eta & = & \eta_0 \frac{1}{1 + \left(\omega_c \tau\right)^2}\label{Eq:viscosity}\\
\alpha_H & = & \eta_0 \frac{\omega_c \tau}{1 + \left(\omega_c \tau\right)^2}\;,\label{Eq:HallCoefficient}
\end{eqnarray}
where $\eta_0$ is the bare viscosity.  The vortex Hall angle, $\theta_H$, is given by $\tan(\theta_H) = \alpha_H/\eta = \omega_c \tau$.  Here $\omega_c = qB/m$ is the cyclotron frequency corresponding to the relevant magnetic field scale in the vortex core and $\tau$ is the relaxation time of the charge carriers responsible for damping the vortex motion.  Similar results are obtained from microscopic calculations.\cite{KOPNIN:1976vi,KOPNIN:1976ua,KOPNIN:1991va} Note that $\eta$ and $\alpha_H$ have a form similar to that of $\sigma_{xx}$ and $\sigma_{xy}$ in the magnetoconductivity tensor of a normal metal, Eq.~\ref{Eq:magnetoconductivity}.

From the vortex equation of motion, Eq.~\ref{Eq:vortex}, we can read off the vortex viscosity tensor,
\begin{equation}
\bm \eta = \left(\begin{array}{cc}\eta & \alpha_H \\-\alpha_H & \eta\end{array}\right)\;,
\end{equation}
defined so that 
\begin{equation}
{\bm \eta}\, \mathbf{v}_v = \Phi_0 \left(\mathbf{j} \times \mathbf{\hat z}\right)_t\;,
\end{equation}
where $(...)_t$ denotes the component of the vector transverse to the magnetic field and $\mathbf{v}_v$ now refers to the transverse component of vortex velocity.  To obtain the effective resistivity, we solve for the vortex velocity
\begin{equation}
\mathbf{v}_v = \Phi_0 \bm \eta^{-1}\, (\mathbf{j} \times \mathbf{\hat z})_t
\end{equation}
and use the Josephson relation for moving vortices,\cite{Josephson:1965bo}
\begin{equation}
\mathbf{E} = \mathbf{B} \times \mathbf{v}_v\;,
\label{Eq:Josephson}
\end{equation}
 to obtain the average electric field:
\begin{eqnarray}
\mathbf{E} & = & B \mathbf{\hat z} \times \mathbf{v}_v\\
& = & B \Phi_0 \mathbf{\hat z}\times(\bm \eta^{-1}\, \mathbf{j}_t) \times \mathbf{\hat z}\\
& = & B \Phi_0 \bm \eta^{-1}\, \mathbf{j}_t\;.
\end{eqnarray}
(Here it is understood that the cross product of $\mathbf{\hat z}$ with a 2D transverse vector is a $\pi/2$ rotation in the transverse plane.)  The vortex resistivity is then
\begin{equation}
\bm \rho_v  =  B \Phi_0 \bm \eta^{-1} = \frac{B \Phi_0}{\eta^2 + \alpha_H^2} \left(\begin{array}{cc}\eta & -\alpha_H \\\alpha_H & \eta\end{array}\right)\;.
\end{equation}
Under conditions of constant current density $\mathbf{j}$, the power dissipation per unit volume is
\begin{equation}
\mathbf{j}^\top\bm{\rho}_v\,\mathbf{j}  = \Phi_0 B j^2 \frac{\eta}{\eta^2 + \alpha_H^2}\;,
\end{equation}
equivalent to that for an effective vortex viscosity \mbox{$\eta^\ast = \eta + \alpha_H^2/\eta$}.  However, when we substitute for the field dependences of $\eta$ and $\alpha_H$, given by Eqs.~\ref{Eq:viscosity} and \ref{Eq:HallCoefficient}, we obtain a cancellation:
\begin{equation}
\eta^\ast  =  \eta_0 \left( \frac{1}{1 + \left(\omega_c \tau\right)^2}  + \frac{\left(\omega_c \tau\right)^2}{1 + \left(\omega_c \tau\right)^2} \right) = \eta_0\;.
\end{equation}
That is, the relevant viscosity is the \emph{bare viscosity}, independent of the vortex Hall angle. As pointed out by Golosovsky \emph{et al.},\cite{GOLOSOVSKY:1996p1} both the direction of the vortex motion and the magnitude of the viscosity are changed but, if the system is driven by an external source of constant current, the effects cancel and the effective viscosity is the same as if $\theta_H =0$.   The situation we have in the microwave experiments is indeed one of constant $\mathbf{j}$: the superconducting sample has a surface impedance in the m$\Omega$ range, tiny compared to the characteristic impedance of free space.  The sample is placed into the microwave resonator at a magnetic field antinode (electric field node), and the microwave $H$ field imposes a constant surface current density.  

\subsection{Pinning and complex vortex viscosity}
\label{Sec:complex_viscosity_model}

In any real superconductor, local material imperfections lead to pinning, preventing the free flow of flux lines.\cite{Larkin:1979ta} Pinning effects can be particularly strong in the cuprate superconductors, especially at low temperatures, making flux-flow resistivity difficult to measure.  One approach is to use a DC current in excess of the critical current to push the vortices into a state of free flux-flow.\cite{Kim:1965vl,Kunchur:1993ie} However, this is a nonlinear method and must be applied and interpreted carefully.  An alternative approach, which has been used extensively,\cite{JIGittleman:1968p172,Owliaei:1992tt, Pambianchi:1993jv, DCMorgan1993, Morgan:1994p2404, Revenaz:1994fg, GOLOSOVSKY:1994p169, Parks:1995p189, Powell:1996tb, Belk:1997hn, Ghosh:1997p170, Hanaguri:1999fn, Silva:2000ia, Tsuchiya:2001p200, MATSUDA:2002p2718, Silva:2004bh, BMorgan2005, Pompeo:2008cd, Narduzzo:2008io, Ikebe:2009it, Pompeo:2008p2717, Zhou2009} is to probe the \emph{reversible} vortex motion --- the \emph{linear} response of the vortices to a high frequency driving force.  An AC current shakes the flux lines harmonically about their equilibrium positions and, when the measurements are carried out in a manner that is sensitive to both magnitude and phase, allows dissipative and reactive forces to be resolved separately.  The technique permits a clean determination of the viscous and elastic parameters and, since these make contributions of comparable magnitude in the GHz range, is ideally carried out at microwave frequencies.

The usual way to include pinning and elastic forces in the vortex equation of motion is through a pinning force of the form $F_p = - \alpha_p x$, where $\alpha_p$ is the effective pinning constant,\cite{JIGittleman:1968p172} and $x$ is the displacement of the vortex from equilibrium.  This harmonic approximation should work well in the linear-response regime, in which the displacement of the vortex is small compared to the inter-vortex spacing (typically 1~\AA\ vs.\ 100~\AA\ in our experiments).  However, a concern now arises over whether we can continue to describe the dynamics of the system in terms of a single, \emph{average} vortex: in contrast to viscous and electromagnetic forces, which originate from interactions with the electron fluid on a microscopic scale, the elastic forces on a vortex arise from random material imperfections, potentially giving rise to a broad distribution of pinning constants.  Statistical averages over such a distribution do not necessarily correspond to the behaviour of an \emph{average} vortex. 

Nevertheless, there are several situations in which the distribution of local pinning constants should be narrowly defined, and therefore the single-vortex approach valid. At high fields, in the collective-pinning regime, the density of flux lines is much greater than the density of pinning sites: vortices interact predominantly with one another, and only indirectly with the pinning sites, smoothing out point-to-point variations in local pinning constant.  In addition to this, there are two other favourable situations, specific to high frequency experiments.  First, as we will show below, the vortex viscosity can have a substantial imaginary component, which acts as an additional contribution to the effective pinning constant and can even dominate over the elastic component in the microwave frequency range.  This dynamical contribution to pinning arises from interactions with the electron fluid and can be assumed to be the same everywhere in the sample.  Secondly, in a high frequency experiment, the only vortices visible to the microwaves are close to the sample surface.  (Microwave cavity perturbation is a power-absorption technique, so the relevant length scale is \emph{half} the RF penetration depth.)  In clean materials, the interaction with the sample surface can become the dominant elastic force for these near-surface vortices.\cite{Bean:1964wr}  This type of pinning is predominantly electromagnetic, arising from the interaction of the vortex with its image vortex.  In our geometry this provides an intrinsic pinning mechanism that acts along the \emph{entire length} of the flux lines, rather than at particular points.

We therefore proceed with the single-vortex approach, focussing on a one-dimensional model of vortex motion that includes viscous, pinning and Lorentz forces, but ignores the vortex Hall effect, for reasons discussed in the previous section.  In the simplest version of this model we have
\begin{equation}
\eta v + \alpha_p x = \Phi_0 j(t)\;.
\end{equation}
Here the vortex velocity, $v$, is the time derivative of the vortex displacement, $x$.  Using a phasor representation for time-harmonic quantities, in which the transport current density is $j(t) = \mbox{Re}\{J_0 \exp(\I \omega t)\}$, we have 
\begin{equation}
\left(\eta + \frac{\alpha_p}{\I \omega} \right)\tilde v \E^{\I \omega t} = \Phi_0 J_0 \E^{\I \omega t}\;,
\label{Eq:phasorforce}
\end{equation}
where $\tilde v$ is the phasor vortex velocity.  We see that the inclusion of pinning can be incorporated into a redefined, \emph{complex} viscosity, $\tilde \eta = \eta + \alpha_p/\I \omega$.

In addition to the pinning term, there is the possibility that the bare viscosity itself has frequency dependence --- something that indeed occurs in ortho-II \ybco{6.52}.  In this case, $\tilde \eta = \eta(\omega) + \alpha_p/\I \omega$ and, because we are dealing with a physical response function, the bare viscosity must have real and imaginary parts, $\eta(\omega) = \eta^\prime(\omega) - \I \eta^{\prime\prime}(\omega)$.  Causality requires that these be related by Kramers--Kr\"onig relations, \emph{e.g.},
\begin{equation}
\eta^{\prime\prime}(\omega) = \frac{2 \omega}{\pi} \mathcal{P}\int_0^\infty \frac{\eta^\prime(\omega^\prime)}{\omega^2 - {\omega^\prime}^2}\D \omega^\prime\;,
\end{equation}
where $\mathcal{P}$ denotes the principal part of the integral.  The main physical effect is that in systems with a strong frequency dependence of $\eta^\prime$ (\emph{i.e.}, in systems with long-lived charge excitations) the apparent pinning constant will depend on frequency:
\begin{equation}
\alpha_\mathrm{eff}(\omega) = \alpha_p + \omega \eta^{\prime\prime}(\omega)\;.
\end{equation}
As an example, the model we will use below to describe the viscosity of ortho-II \ybco{6.52},
\begin{equation}
\eta^\prime(\omega) = \eta_0 + \eta_1 \frac{1}{1 + \omega^2/\Gamma^2}\;,
\label{Eq:model_viscosity}
\end{equation}
must, by causality, be accompanied by a pinning constant of the form
\begin{equation}
\alpha_\mathrm{eff}(\omega) = \alpha_p + \eta_1 \frac{\omega^2/\Gamma}{1 + \omega^2/\Gamma^2}\;.
\label{Eq:model_pinning}
\end{equation}
We will see that this model provides an excellent description of the data.

Solving Eq.~\ref{Eq:phasorforce} for the phasor vortex velocity, we have
\begin{equation}
\tilde v = \frac{\Phi_0}{\eta(\omega) + \frac{\alpha_p}{\I \omega}} J_0 = \frac{\Phi_0}{\tilde\eta(\omega)} J_0\;.
\end{equation}
By the Josephson relation, Eq.~\ref{Eq:Josephson}, the electric field associated with the vortex motion is 
\begin{equation}
\tilde E = \frac{B \Phi_0}{\tilde\eta(\omega)} J_0\;,
\end{equation}
implying a complex, effective vortex resistivity
\begin{equation}
\tilde \rho_v = \frac{B \Phi_0}{\tilde\eta(\omega)}\;.
\label{Eq:vortex_resistivity}
\end{equation}

\subsection{Vortex electric fields}

The interaction of the vortex with the surrounding electron fluid (including states in the vortex core) arises from the coupling of charge excitations to the electric field induced when the vortex moves.\cite{Bardeen:1965p151}  This electric field can be obtained from the London acceleration equation\cite{London:1935uf}
\begin{equation}
\mathbf{E} = \frac{\partial(\Lambda \mathbf{j}_v)}{\partial t}\;,
\end{equation}
where $\Lambda$ is the London parameter and, in our case, $\mathbf{j}_v$ is the supercurrent density circulating around the vortex.  The time rate of change of $\mathbf{j}_v$ arises solely from the motion of the vortex:
\begin{equation}
\frac{\partial \mathbf{j}_v}{\partial t} = - (\mathbf{v}_v\cdot\mathbf{\nabla}) \mathbf{j}_v\;.
\end{equation}
\begin{figure}[t]
\centering
\includegraphics[width = \columnwidth]{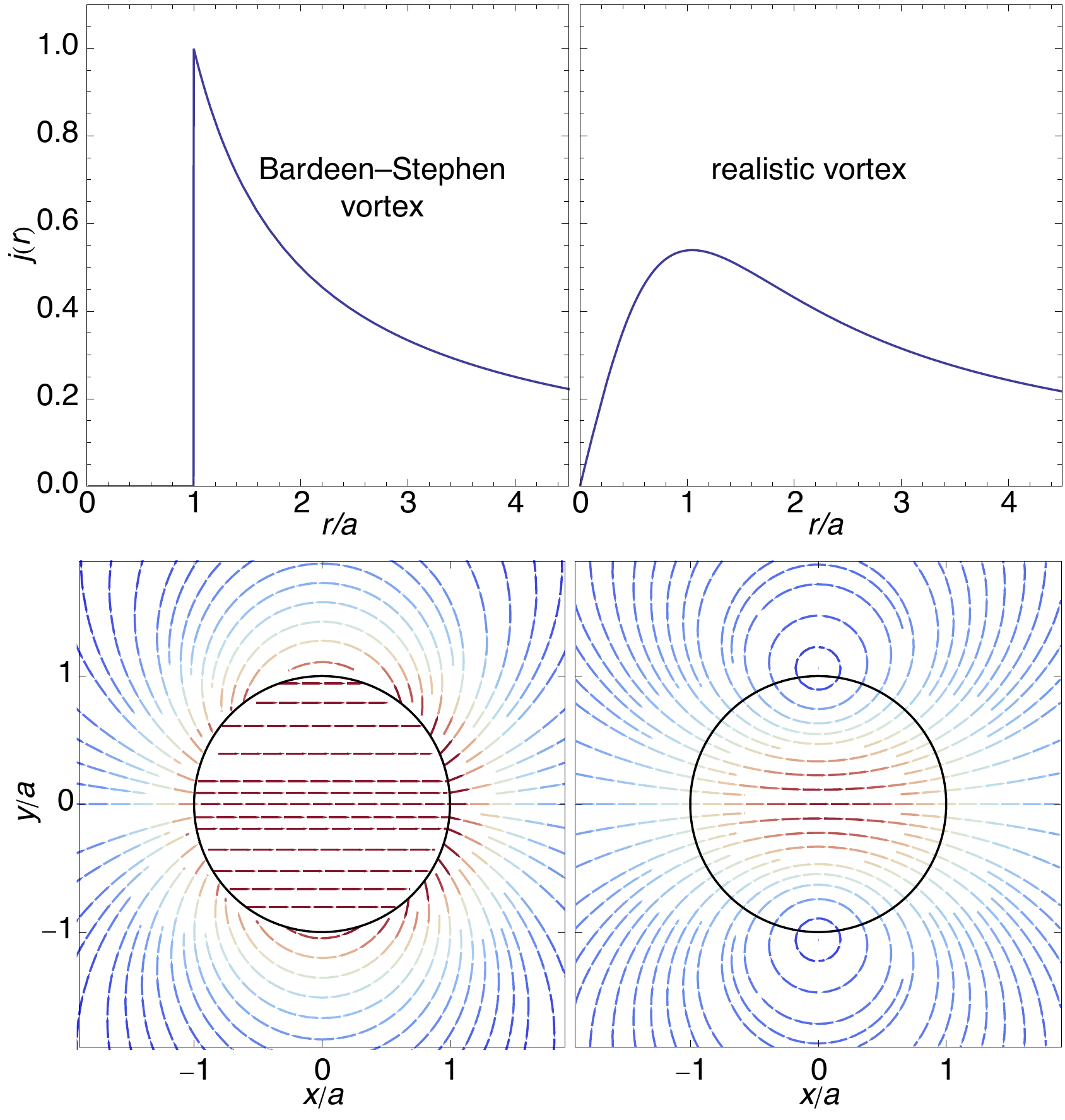}
\caption{(color online).  Supercurrent screening profiles and induced electric fields for vortices with cylindrical symmetry.  The panels on the left show the situation in an idealized Bardeen--Stephen vortex:\cite{Bardeen:1965p151} the supercurrent profile (upper panel) has the form $j(r) \propto 1/r$ outside the core radius $a$, and is zero inside.  For vortex motion in the positive $y$ direction, the resulting electric field (lower panel) is uniform inside the vortex core (solid circle, $r = a$) and has dipolar form outside.  The panels on the right depict a more realistic situation in which the supercurrent profile varies smoothly with radius.\cite{Caroli1964}  The configuration shown (upper panel) uses the approximate form $j(r) \propto \tanh{(r/a)}/\sqrt{r^2 + a^2}$.  As with the Bardeen--Stephen vortex, electric field is also relatively uniform near $r = 0$, with a dipolar form for $r > a$.  The principal differences are a much smoother variation with position, with less intense electric field in the core region $r < a$.} 
\label{fig:electricfields}
\end{figure} 
We will illustrate this in the particular case of a vortex with cylindrical symmetry, moving in the positive $y$ direction at speed $v$.  The screening supercurrent density will be azimuthal, and its magnitude will depend only on distance from the centre of the vortex: \emph{i.e.}, $\mathbf{j}_v = j(r) \mathbf{\hat \theta}$.  In this case we can show that the Cartesian components of electric field are:
\begin{eqnarray}
E_x & = & \frac{v}{\Lambda} \left[\cos^2 \theta \frac{j(r)}{r} + \sin^2 \theta \frac{\partial j}{\partial r} \right]\\
E_y & = & \frac{v}{\Lambda} \sin \theta \cos \theta \left[ \frac{j(r)}{r} -  \frac{\partial j}{\partial r} \right]
\end{eqnarray}
Electric field plots are shown in Fig.~\ref{fig:electricfields} for two cases: a Bardeen--Stephen-like vortex, for which $j(r) \propto 1/r$ outside the core radius $a$ and is zero within; and a more realistic vortex core, in which the supercurrent density varies smoothly through $r = a$, falling linearly to zero as $r \to 0$.  For the latter case, we take the approximate form
\begin{equation}
j(r) \propto \frac{\tanh{(r/a)}}{\sqrt{r^2 + a^2}}\;.
\end{equation}
We see that the qualitative behaviour of the electric fields is similar in both cases: roughly uniform inside the vortex core, with a dipole form outside.  The use of accurately calculated current profiles that break cylindrical symmetry\cite{Ichioka:1999p359} will not substantially change this picture, since the form of the current density ($j(r) \sim 1/r, a < r < \lambda$) is tightly restricted by the vortex topology.  As a result, the electric field profile will always be similar to that shown on the right-hand side of Fig~\ref{fig:electricfields}.  

For the purpose of vortex-dynamics experiments, the important point is that a moving vortex acts as a local concentration of electric field, and that the reaction force experienced by the vortex is the result of the electric field interacting (resistively and reactively) with conducting degrees of freedom in the vicinity of the vortex core.  For the case of ortho-II \ybco{6.52}, a large part of this response appears to be due to bulk $d$-wave quasiparticles \emph{outside} the vortex cores.

\subsection{Extraction of vortex parameters}
\label{Sec:vortex_parameters}

For measurements made at microwave frequencies, in addition to the complex vortex resistivity, $\tilde \rho_v$, given in Eq.~\ref{Eq:vortex_resistivity}, we must also take into account the finite impedance, $\tilde \rho_s$, of the superconducting medium in which the vortices are embedded.  The electrodynamics of this problem have been solved by Coffey and Clem,\cite{COFFEY:1991p156} and Brandt,\cite{BRANDT:1991p149} who find the following simple, additive form to be appropriate in the limit of low temperature and weak field ($B \ll B_{c2}$):
\begin{equation}
\tilde\rho_{\rm{eff}} \approx \tilde\rho_s + \tilde \rho_v = \tilde \rho_s + \frac{B \Phi_0}{\tilde\eta(\omega)}\;.
\label{eq:rff_extract}
\end{equation}
Here the effective complex resistivity, $\tilde \rho_\mathrm{eff}$, is the experimentally accessible quantity in a microwave experiment, being directly related to the complex surface impedance, $Z_s = R_s + \I X_s$, by the local electrodynamic relation, $\tilde \rho = Z_s^2/\I \omega \mu_0$.
\footnote{The local electrodynamic relation, $\rho = Z_s^2/\I \omega \mu_0$, is obtained by solving Maxwell's equations for phasor fields (Amp\`ere and Faraday laws) at the interface between vacuum and a conductor with local electrodynamics, \mbox{$\mathbf{E}(\mathbf{r}) = \rho \mathbf{J}(\mathbf{r})$}.  The surface impedance $Z_s$ is defined as the ratio of the tangential components of electric and magnetic field at the interface}  
To a good approximation, the background contribution from the superconducting medium, $\tilde \rho_s$, can be obtained from a measurement in zero field.  To the extent that nodal quasiparticles in a $d$-wave superconductor give rise to a nonlinear Meissner effect,\cite{YIP:1992p3020} there will be some weak field dependence of $\rho_s$.  However, this effect is known to be much weaker in the YBa$_{2}$Cu$_{3}$O$_{6+y}$ system than theoretically expected,\cite{Bidinosti:1999p2720,Sonier:2007p1185} and is negligible in the current context.  The experimental procedure is then that
\begin{eqnarray}
\tilde \rho_\mathrm{eff} & = & \frac{Z_s^2(B,T)}{\I \omega \mu_0}\;,\\
\tilde \rho_s & \approx & \frac{Z_s^2(B = 0,T)}{\I \omega \mu_0}\;.
\end{eqnarray}
The vortex contribution is isolated by taking the difference, $\tilde \rho_v = \tilde \rho_\mathrm{eff} - \tilde\rho_s$, and from this we obtain the rest of the parameters in the vortex-dynamics model, in the following way: 
\begin{eqnarray}
\tilde \eta & = & \frac{B \Phi_0}{\tilde \rho_v}\;,\\
\eta^\prime & = & B \Phi_0 \mbox{Re}\{\tilde \rho_v^{-1}\}\label{Eq:eta_prime}\;,\\
\alpha_\mathrm{eff} & = & \omega B \Phi_0 \mbox{Im}\{- \tilde \rho_v^{-1}\}\label{Eq:alpha}\;,\\
\rho_{\rm{ff}} & \equiv & \lim_{\omega \to 0} \frac{B \Phi_0}{\eta(\omega)} = \lim_{\omega \to 0} \frac{1}{\mbox{Re}\{\tilde\rho_v^{-1}\}}\;.
\end{eqnarray}
It should be pointed out that an analysis of this sort is only possible if both real and imaginary parts of the surface impedance are measured.  In our experiment these are obtained at the same time, on the same sample.  

\section{Experimental methods}
 
\subsection{YBa$_{\bm 2}$Cu$_{\bm 3}$O$_{\bm{6.52}}$ sample preparation}

Single crystals of high purity YBa$_2$Cu$_3$O$_{6+y}$ were grown using a self-flux method in a chemically inert BaZrO$_3$ crucible.\cite{Liang:2012va}  Oxygen concentration was set to \ybco{6.52}\ by annealing in flowing oxygen at 748~$^\circ$C, followed by a homogenization anneal at 570~$^\circ$C for 10~days in which the crystal was sealed inside a quartz ampoule with a large volume of ceramic at the same oxygen content. The sample was mechanically detwinned at 200~$^\circ$C under uniaxial stress, without changing the oxygen content. Ortho-II ordering, in which the oxygen content of the CuO chains alternates between full and empty, was achieved by annealing at 85~$^\circ$C for one day then 50~$^\circ$C for one week. Note that the highest degree of ortho-II ordering is obtained when the oxygen content is slightly more than needed for filling every other chain, hence the oxygen doping \ybco{6.52}.

\begin{figure}[t]
\centering
\includegraphics[width = \columnwidth]{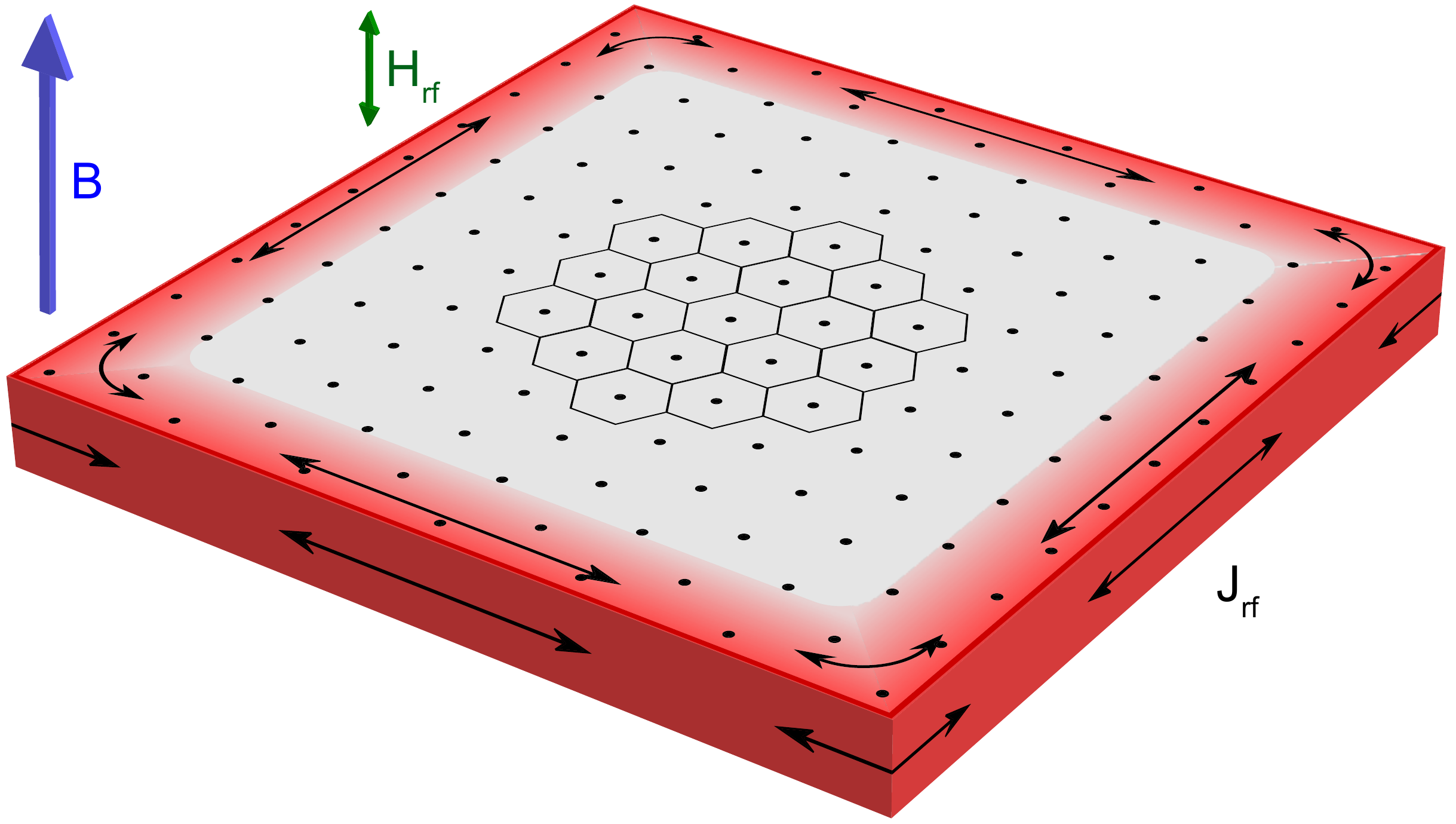}
\caption{(color online).  Sample geometry in the microwave experiment.  The sample is a platelet single crystal, thin in the $c$~direction.  The sample is cooled through $T_c$ in a strong, static magnetic field, $B$, that is applied perpendicular to the CuO$_2$ planes and sets up the vortex lattice.  In this picture, vortex cores are represented schematically by dots on the upper surface of the sample; a number of vortex-lattice unit cells are shown in the centre of the diagram.  A weak microwave field, $H_\mathrm{rf}$, is applied parallel to $B$.  This induces a microwave transport current density, $J_\mathrm{rf}$, which, due to strong demagnetizing effects in this geometry, is concentrated near the edges of the crystal.} 
\label{fig:geometry}
\end{figure} 
 
\subsection{Surface impedance measurements}  

There are several key technical requirements for making accurate microwave measurements of vortex dynamics.  High sensitivity is needed, as the resistive dissipation of sub-mm, high quality single crystals is typically very small.  In addition, the technique must measure both real and imaginary parts of the impedance, to allow an unambiguous separation of viscous and reactive effects.  We have used cavity perturbation of \mbox{high-$Q$} TiO$_2$ (rutile) dielectric resonators, operating in TE$_{0np}$ modes,\cite{Huttema:2006p344} to carry out measurements at four microwave frequencies ($\omega/2 \pi = 2.64$, 4.51, 9.12 and 13.97~GHz).  In contrast to microwave measurements in zero field, for which the resonator is typically a superconducting cavity, the TiO$_2$ dielectric resonator is housed within a normal metal (copper) enclosure, enabling it to be used in an applied magnetic field.  The low loss tangent and high dielectric constant of rutile allow quality factors of $10^6$ to $10^7$ to be achieved.  In addition, the compact size of the resonator gives a much higher filling factor than cavity resonators operating at the same frequencies.  The good mechanical stability of dielectric resonators and the absence of weak superconducting links, combined with a high $Q$ and filling factor, result in a system that has comparable or better surface-impedance resolution than a superconducting cavity system \emph{and} is capable of operating in high magnetic fields.

A single crystal of ortho-II \ybco{6.52}\ with dimensions $a \times b \times c = 0.54$~mm$\times 0.63$~mm$\times 10$~$\mu$m was mounted on the end of a sapphire hot finger\cite{Sridhar:1988p495} and introduced into the resonator through a hole bored along the axis of the rutile cylinder.  The sample was mounted so that the static magnetic field, $B$, was applied along the $c$-direction of the crystal, as shown in Fig.~\ref{fig:geometry}.   The microwave field, $H_\mathrm{rf}$ was also applied along the $c$-direction, in order to induce $a$--$b$ plane screening currents. All measured quantities are $a$--$b$ averages. This geometry has a high demagnetizing factor but offers the advantage of avoiding screening-current loops that close along the $c$-direction, something that is particularly important for electrically anisotropic  materials such as \ybco{6.52}.  The sapphire hot finger allows the temperature of the sample to be regulated separately from that of the resonator, which was kept fixed at 4.2~K.  In our system the hot finger was mounted on a moveable, pumped helium pot, giving a base temperature of 1.1~K.  The introduction of the sample into the resonator causes a change in resonant frequency, $f_0$, and bandwidth, $f_B$.  Subsequent changes of sample temperature, and applied field, lead to further changes in $f_0$ and $f_B$.  The surface impedance of the sample, $Z_s = R_s + \I X_s$, is obtained using the cavity perturbation relation\cite{altshuler1963,Huttema:2006p344}
\begin{equation}
R_s(B,T)+\I\Delta X_s(B, T)\! =\! \Gamma\!\left(\!\!\frac{\Delta f_B(B,T)}{2}- \I\Delta f_0(B,T)\!\!\right)\;,
\label{eq:CP}
\end{equation}
Here $T$ is the sample temperature, $\Gamma$ is an empirically determined scaling factor, $R_s(B,T)$ is the absolute surface resistance, $\Delta X_s(B, T)$ is the shift in surface reactance with respect to zero field and a reference temperature $T_0$, $\Delta f_B(B,T)$ is the shift in resonator bandwidth on introducing the sample into the empty resonator in an applied field $B$, and $\Delta f_0(B,T)$ is the shift in resonant frequency with respect to $B=0$ and $T = T_0$.  Note that the absolute surface reactance cannot be inferred from a measurement of the frequency shift on inserting the sample into the resonator, as the microwave skin depth is much smaller than the effective size of the sample.  Instead, absolute zero-field reactance is set using published penetration depth data.\cite{PeregBarnea:2004p761}  All measurements reported here were made with the sample in a field-cooled state, in order that the sample magnetization be close to its equilibrium value.  Microwave power levels were regulated so that the microwave $H$ field in the resonator was held constant during the course of the experiment, eliminating contributions that might arise from power dependence of the resonator frequency and quality factor.  Field- and temperature-dependent background measurements were made on the empty resonator and sapphire sample holder and used to apply a small correction to the sample signal.

An estimate of the amplitude of the vortex motion can be obtained from the microwave power level and the pinning force constant.  In the following, we use worst-case values for the various quantities (\emph{e.g.}, \mbox{low temperature $Q$}, high temperature $\alpha_p$) to obtain an upper bound on the range of motion.  The input power to the resonator itself is typically $P_\mathrm{in} = 1$~nW or lower.  (This includes a correction for the insertion loss of the cryogenic microwave cables and the fact that the resonator is operated at weak coupling.)  For a quality factor $Q = 10^6$ (characteristic of low temperature operation) and a resonant frequency of 2.64~GHz, the stored energy in the resonator is \mbox{$E = P_\mathrm{in} Q/\omega = 6 \times 10^{-14}$~J.}  For an effective resonator volume of 0.5~cm$^3$ (taking into account the concentrating effect of the dielectric resonator) this corresponds to a peak energy density $U = 1.2 \times 10^{-7}$~J/m$^3$.  From this we obtain the strength of the microwave magnetic field at the centre of the resonator, \mbox{$H_\mathrm{rf} = (U/\mu_0)^{1/2} = 0.3$~A/m.}  We assume that demagnetizing effects enhance this magnetic field by a factor of sample width/sample thickness,\cite{Prozorov:2000tj} giving $H_\mathrm{edge} = 60 H_\mathrm{rf} = 18$~A/m.  For a skin depth $\delta = 0.2$~$\mu$m, this corresponds to a current density $J_\mathrm{rf} = H_\mathrm{edge}/\delta = 10^8$~A/m$^2$.   The force per unit length on the individual vortices is \mbox{$F_\ell = \Phi_0 J_\mathrm{rf} = 2 \times 10^{-7}$~N/m.}  For a pinning constant $\alpha_p = 10^3$~N/m$^2$ (characteristic of high temperatures) we obtain a maximum vortex displacement \mbox{$x_\mathrm{max} = F_\ell/\alpha_p = 2$~\AA.}  We emphasize that this is an upper bound, based on worst-case assumptions, and that we have sufficient signal-to-noise ratio to operate at input powers several orders of magnitude lower.  In every case, data were checked for power dependence in order to avoid nonlinearities. 

 \begin{figure}[t]
\centering
\includegraphics[width= \columnwidth]{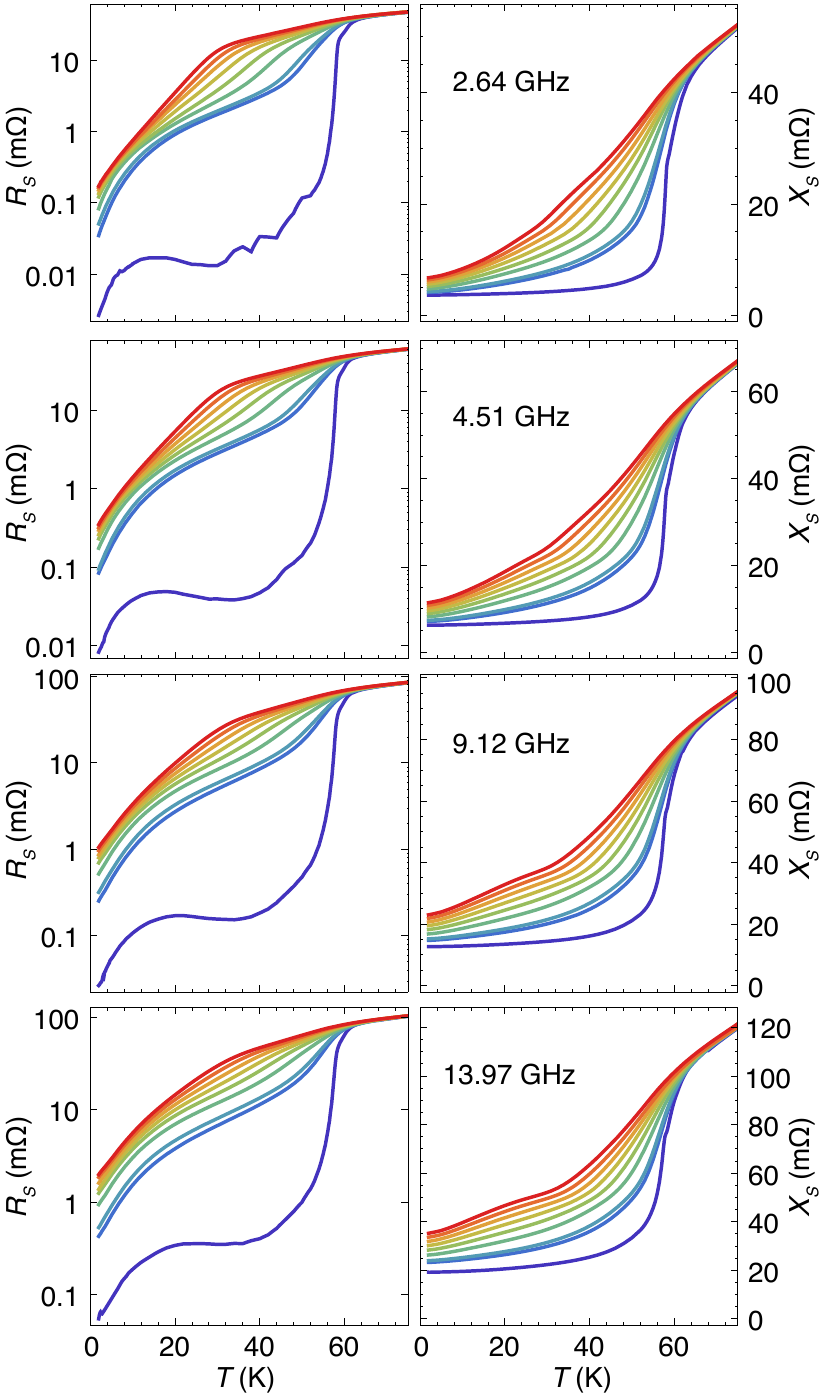}
\caption{(color online).  Surface impedance $Z_s = R_s + \I X_s$ at $\omega/2 \pi = 2.64$, 4.51, 9.12, and 13.97~GHz, for $B = 0$, 0.75, 1, 2, 3, 4, 5, 6 and 7~T (from bottom to top).  Left-hand plots show surface resistance, $R_s$, on a logarithmic scale.  Right-hand plots show surface reactance, $X_s$, on a linear scale.  In each case the field is applied at a temperature $T > T_c$, and held constant during the temperature sweep.} 
\label{fig:Zs}
\end{figure} 

 \begin{figure}[t]
\centering
\includegraphics[width= \columnwidth]{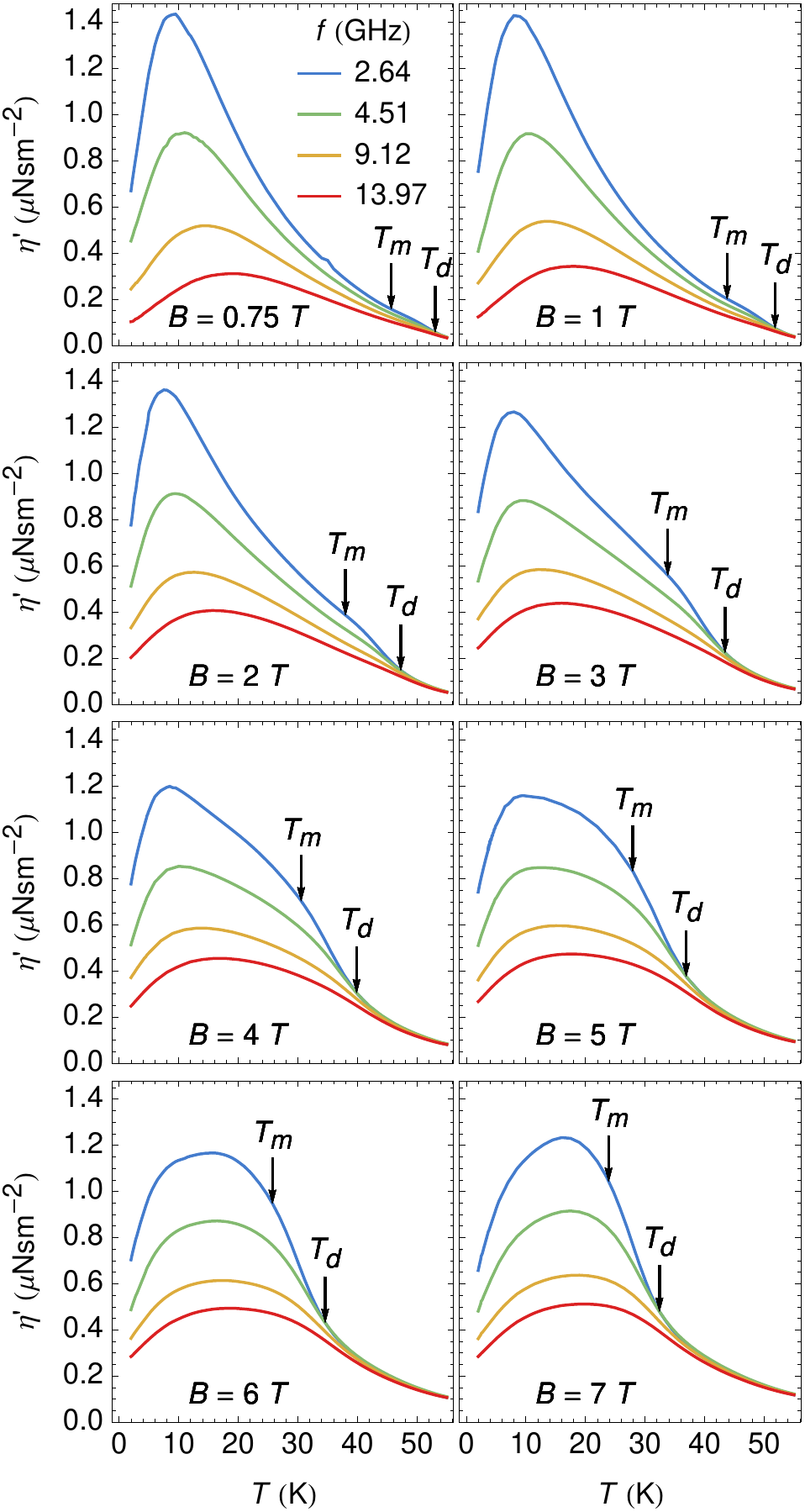}
\caption{(color online).  Real part of the frequency-dependent vortex viscosity, $\eta^\prime(T,B)$, at frequencies $\omega/2 \pi = 2.64$, 4.51, 9.12, and 13.97~GHz, and for magnetic fields from 0.75 to 7~T.  $T_m$ denotes the vortex-lattice melting temperature at each field, obtained from Ref.~\onlinecite{Ramshaw2012}.  $T_d$ denotes the dynamical cross-over temperature, defined to be the point at which the frequency variation of $\eta^\prime$ becomes less than 20\% in our measurement range.} 
\label{fig:viscosity}
\end{figure}

 \begin{figure}[t]
\centering
\includegraphics[width= \figwidth \columnwidth]{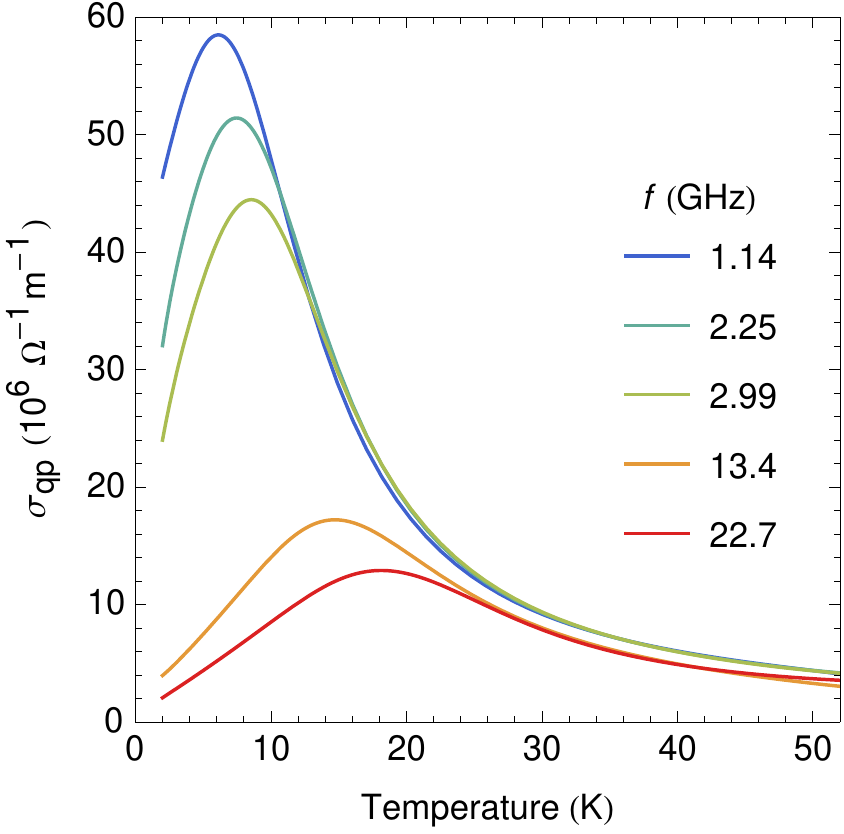}
\caption{(color online).  Zero-field quasiparticle conductivity $\sigma_\mathrm{qp}(\omega,T)$ for ortho-II \ybco{6.52}.  Data are the geometric mean of $a$- and $b$-axis microwave conductivity data from Ref.~\onlinecite{Harris:2006p388}.  The strong rise in $\sigma_\mathrm{qp}(T)$ on cooling is a result of a rapid decrease in inelastic scattering of $d$-wave quasiparticles in the superconducting state.  Low temperature peaks in $\sigma_\mathrm{qp}(T)$ arise from a competition between increasing quasiparticle lifetime and decreasing quasiparticle density on cooling.} 
\label{fig:conductivity}
\end{figure} 

\section{Results and Discussion}

\subsection{Surface impedance}

Surface impedance data, $Z_s = R_s + \I X_s$, are presented in Fig.~\ref{fig:Zs}, at each of the measurement frequencies ($\omega/2 \pi  = 2.64$, 4.51, 9.12 and 13.97~GHz) and for magnetic fields ranging from 0 to 7 T.  In zero field, the superconducting transition is preceded by some rounding due to superconducting fluctuations, but there is a sharp downturn in $Z_s(T)$ at $T_c = 59$~K.  This downturn softens as magnetic field is applied, but remains visible in $X_s(T)$ as a slight kink, even at higher fields.  This is due to the onset of pinning as the vortex lattice freezes. In surface resistance, the system remains strongly dissipative to much lower temperatures, with a substantial decrease in $R_s(T)$ only occurring below the  vortex-lattice melting transition.

 \begin{figure}[t]
\centering
\includegraphics[width= \figwidth \columnwidth]{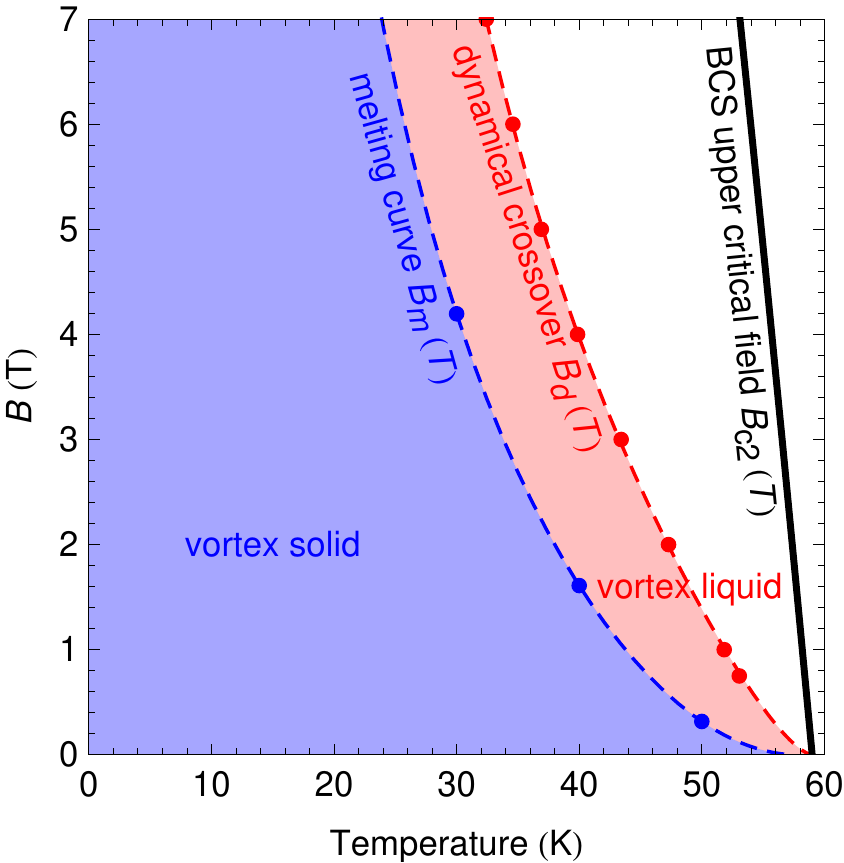}
\caption{(color online).  Field--temperature phase diagram showing the vortex-lattice melting line, $B_m(T)$, from Ref.~\onlinecite{Ramshaw2012}, and the dynamical crossover $B_d(T)$ derived from the frequency dependent vortex viscosity.  $B_{c2}(T)$ is the BCS upper critical field, plotted assuming $B_{c2}(T \to 0) = 40$~T.\cite{Ramshaw2012}} 
\label{fig:BTphasediagram}
\end{figure}

\subsection{Vortex viscosity}

The real part of the frequency-dependent vortex viscosity is obtained from the microwave surface impedance using Eq.~\ref{Eq:eta_prime} and is plotted in Fig.~\ref{fig:viscosity} as a function of temperature, with each panel showing results for a different magnetic field.  The qualitative behaviour of $\eta^\prime(\omega,T)$ is the same in each case: $\eta^\prime(T)$ rises strongly on cooling, with a peak in the 8 to 20~K temperature range.  The peak is highest for the 2.64~GHz data, and decreases in magnitude with increasing frequency up to 13.97~GHz.  The strong frequency variation of $\eta^\prime$ at low temperatures indicates the existence of long-lived charge excitations.\footnote{Note that vortex dissipation is due to the coupling of induced vortex electric fields to charge excitations in the vicinity of the vortex cores: $\eta^\prime(\omega)$ therefore reflects the dynamics of these excitations, rather than the dynamics of the vortices themselves.}  In fact, the frequency and temperature dependence of $\eta^\prime$ is strikingly similar to that of the zero-field microwave conductivity, $\sigma_\mathrm{qp}(\omega,T)$, plotted in Fig.~\ref{fig:conductivity} using data from Ref.~\onlinecite{Harris:2006p388}.  In the case of the zero-field conductivity, the peaks in $\sigma_\mathrm{qp}(T)$ are due to the competing effects of a quasiparticle lifetime that increases rapidly on cooling below $T_c$, and a decreasing normal fraction as quasiparticles condense into the ground state.   The width of the low frequency quasiparticle spectrum,  $\sigma_\mathrm{qp}(\omega)$, provides a measure of the average quasiparticle relaxation rate, with narrow widths in the low GHz range indicating long transport lifetimes and mean free paths of several $\mu$m.\cite{Hosseini:1999p383,Turner:2003p331,Harris:2006p388}  Interestingly, the peak temperatures in $\eta^\prime(T)$ and the peak widths in $\eta^\prime(\omega)$ are very similar to those in $\sigma_\mathrm{qp}(\omega,T)$.  Taken together, these observations strongly suggest that $d$-wave quasiparticles \emph{outside} the vortex cores provide the dominant mechanism for vortex viscosity in this material.  This is completely different from the situation in conventional superconductors, in which the normal-metal cores are  responsible for the vortex viscosity.  

While the qualitative similarities between $\eta^\prime(\omega,T)$ and $\sigma_\mathrm{qp}(\omega,T)$ suggest that $d$-wave quasiparticles are the underlying mechanism, an important consistency check is provided by testing whether the zero-field quasiparticle conductivity is of sufficient magnitude to be responsible for the observed viscosity.   There are several ways this could be done, but a conceptually clear method is to express the viscosity in terms of a length scale, $\ell_\eta$, that represents the area, $\pi\ell_\eta^2$, over which the vortex electric fields would need to couple to the electrical conductivity in order to give rise to the observed viscosity.   In a conventional superconductor the relevant conductivity is the normal-state conductivity and the length scale is the vortex core size, $\xi$.  Bardeen--Stephen theory gives $\eta = \sigma \Phi_0^2/2 \pi \xi^2$, and therefore \mbox{$\ell_\eta = \xi =  \Phi_0 \sqrt{\sigma/2 \pi \eta}$}.  We can now apply a similar analysis to Ortho-II \ybco{6.52}.  We perform the comparison at $T = 8$~K, the temperature of the peak in $\eta^\prime(T)$.  The low field viscosity at this temperature, and at 2.6~GHz, is $1.45 \times 10^{-6}$~Nsm$^{-2}$. Instead of a normal-state conductivity, we use the zero-field quasiparticle conductivity, as our assertion is that the viscous drag arises from quasiparticles \emph{outside} the core. The $a$--$b$ averaged zero-field conductivity\cite{Harris:2006p388} at 8~K and 2.6~GHz is $\approx 4.5 \times 10^7$~$\Omega^{-1}$m$^{-1}$. From this we obtain $\ell_\eta = 48$~\AA.  This is of the same order as the vortex core size in \ybco{6.52},\cite{Sonier:2007p1185, Ramshaw2012} establishing \emph{quantitative} consistency between the observed vortex viscosity and a mechanism based on the electrical conductivity of bulk $d$-wave quasiparticles.

As magnetic field increases, $\eta^\prime(\omega,T)$ undergoes smooth changes in shape, but its low temperature form and overall magnitude are not strongly affected.  The most noticeable change with increasing field is the emergence of a band of temperature, immediately below $T_c$, in which $\eta^\prime(\omega,T)$ has very weak frequency dependence, indicating that the viscous dissipation is being caused by the vortices coupling to charge excitations whose frequency spectrum extends well beyond the microwave range (\emph{i.e.}, excitations that relax more rapidly than on microwave timescales).  To demarcate this regime we define a temperature, $T_d$, above which the frequency variation of $\eta^\prime$ is less than 20\%. This signifies a qualitative change in the relaxation dynamics, which we explore further below, using fits to complex viscosity spectra.   We will see that below $T_d$,  the strongly frequency-dependent part of $\eta^\prime$ rides on top of a broad background.  This indicates that the long-lived excitations coexist with more rapidly relaxing ones.  $T_d$ is marked on Fig.~\ref{fig:viscosity} along with the vortex-lattice melting temperature, $T_m$, from Ramshaw \emph{et al.}\cite{Ramshaw2012}  There is no sharp change in viscosity at the melting transition, just the beginning of a gradual downturn in $\eta^\prime(T)$ that takes place over a roughly 10~K range between $T_m$ and $T_d$.  The melting curve $B_m(T)$ and dynamical crossover $B_d(T)$ are also plotted in a $B$--$T$ phase diagram in Fig.~\ref{fig:BTphasediagram}.

 \begin{figure}[t]
\centering
\includegraphics[width= \figwidth \columnwidth]{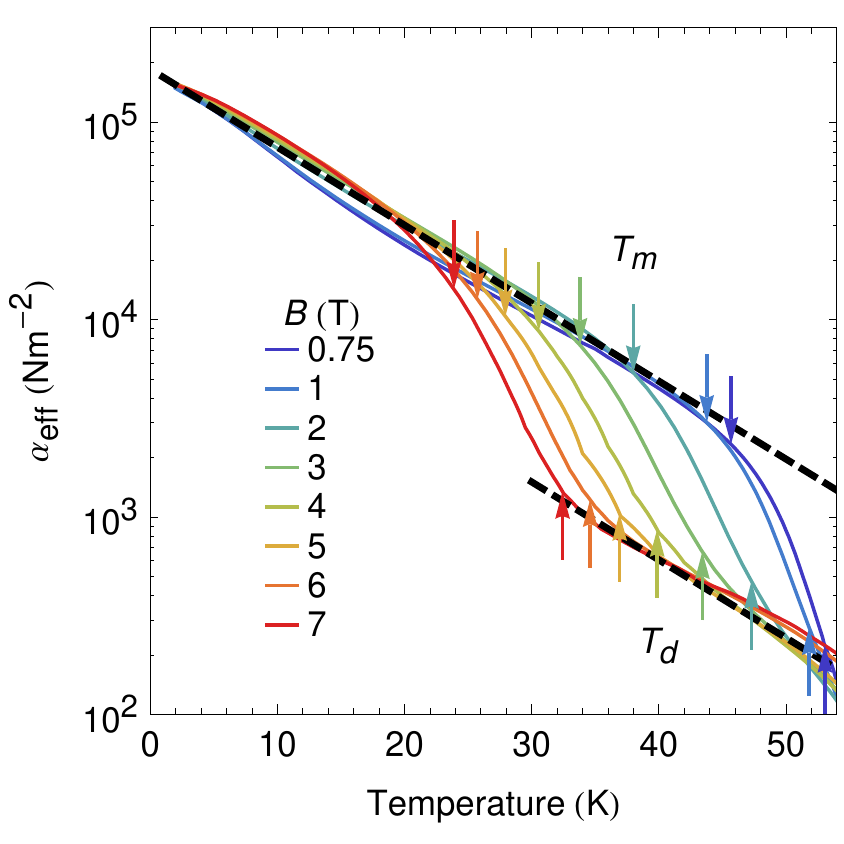}
\caption{(color online).  Temperature dependence of the effective pinning constant, $\alpha_\mathrm{eff}(T)$, at 2.64 GHz, for $B = 0.75$, 1, 2, 3, 4, 5, 6 and 7~T (from right to left).    Dashed lines are guides to the eye and denote $\alpha(T) = \alpha_0 \exp(- T/T_0)$, with $T_0 = 11$~K.  $T_m$ and $T_d$ indicate the melting and dynamical crossover temperatures at each field.} 
\label{fig:alpha264}
\end{figure} 

 \begin{figure}[t]
\centering
\includegraphics[width= \figwidth \columnwidth]{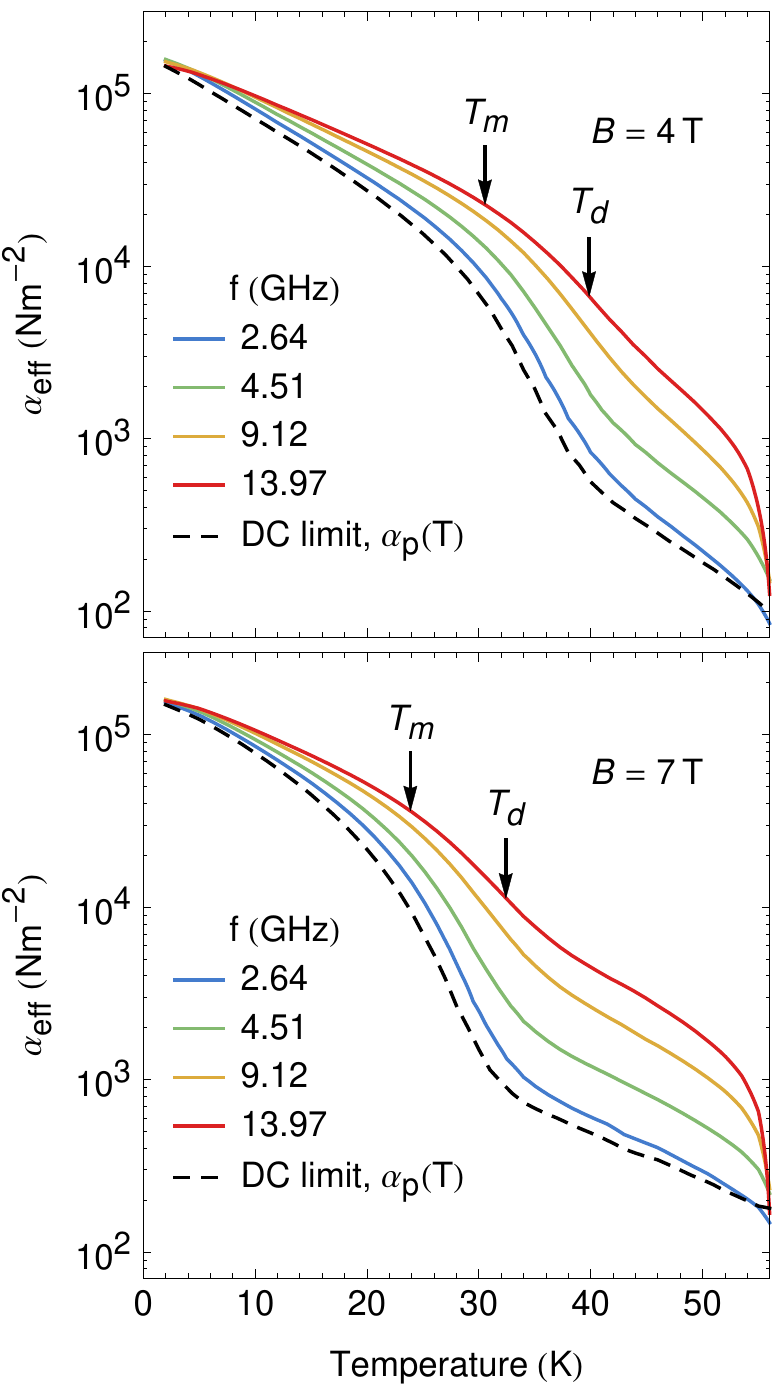}
\caption{(color online).  Frequency-dependent pinning constant, $\alpha_\mathrm{eff}(T)$, for $B =  4$~T (upper panel) and 7~T (lower panel).  Solid curves show data at $\omega/2 \pi = 2.64$, 4.51, 9.12, and 13.97~GHz (from bottom to top).  The dashed curves show the DC limit, $\alpha_p(T)$, obtained from fits to Eqs.~\ref{Eq:model_viscosity} and \ref{Eq:model_pinning}.  $T_m$ and $T_d$ indicate the melting and dynamical crossover temperatures.} 
\label{fig:alphaB}
\end{figure} 

\subsection{Pinning constant}

The effective pinning constant at microwave frequencies, $\alpha_\mathrm{eff}$, is extracted using Eq.~\ref{Eq:alpha}.  The 2.64~GHz data are plotted in Fig.~\ref{fig:alpha264} on a semi-log plot, for each of the magnetic fields.  (We will later see that, over most of the temperature range, the 2.64~GHz traces are close to the static limit of $\alpha_\mathrm{eff}(\omega)$, which, according to the discussion in Sec.~\ref{Sec:complex_viscosity_model}, is the elastic pinning constant $\alpha_p$.)  The pinning constant drops rapidly with increasing temperature, following an approximately exponential temperature dependence, $\alpha_\mathrm{eff}(T) \approx \alpha_0 \exp(- T/T_0)$, with $\alpha_0 = 2 \times 10^5$~N/m$^2$ and $T_0 = 11$~K.  Similar exponential behaviour has been reported in  measurements of pinning constant\cite{GOLOSOVSKY:1994p169,BMorgan2005} and critical current density\cite{Senoussi:1988jf,Shi:1994cn} on optimally doped \ybco{7-\delta}: in that material, a typical value of $\alpha_0$ is 3~to~$4 \times 10^5$~N/m$^2$ (Refs.~\onlinecite{GOLOSOVSKY:1994p169,BMorgan2005}), with $T_0$ in the range 20 to 25~K (Refs.~\onlinecite{GOLOSOVSKY:1994p169,Shi:1994cn,BMorgan2005}).  An elegant theory of this behaviour has been developed by Feigel'man and Vinokur,\cite{Feigelman:1990gp} in which small-amplitude thermal motion of the vortex lattice softens the apparent pinning potential: the specific form of the exponential temperature dependence arises from vortex lattice Debye--Waller factors.  

The vortex-lattice melting transition is clearly visible in the pinning-constant data, with $\alpha_\mathrm{eff}(T)$ dropping by an order of magnitude on passing through $T_m(B)$.  Above the melting transition the pinning constant remains finite, even when we take into account dynamical effects: we will argue below that this is peculiar to surface impedance measurements and is due to surface pinning.  At higher temperatures, above $T_d$, $\alpha_\mathrm{eff}(T)$ reverts back to an exponential trend, with a value of $T_0$ similar to that at low temperature.  

The frequency dependence of $\alpha_\mathrm{eff}(T)$ is shown in Fig.~\ref{fig:alphaB}, for fields of 4 and 7~T.  As we will see in more detail in Sec.~\ref{Sec:complex_viscosity}, $\alpha_\mathrm{eff}$ has substantial frequency dependence at almost all temperatures,  but in Fig.~\ref{fig:alphaB} this is most apparent \emph{above} the melting temperature, due to the low level of static pinning in the vortex-liquid regime.  We will show that the frequency dependence of $\alpha_\mathrm{eff}$ is due to dynamical effects arising from the vortex viscosity. 

 \begin{figure}[t]
\centering
\includegraphics[width= \figwidth \columnwidth]{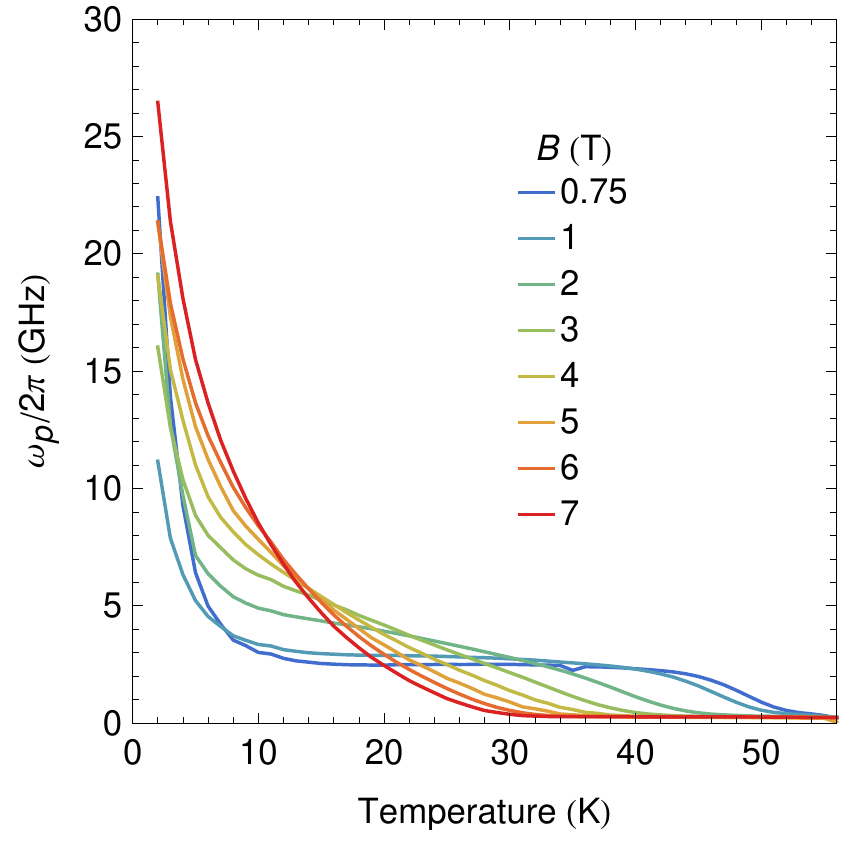}
\caption{(color online).  The depinning frequency, $\omega_p/2 \pi \equiv \alpha_p/\eta_\mathrm{dc}$, for $B = 0.75$, 1, 2, 3, 4, 5, 6 and 7~T.  Here $\alpha_p$ and $\eta_\mathrm{dc} \equiv \eta_0 + \eta_1$ are the DC limits of the pinning and viscous force constants obtained from the fitting procedure.} 
\label{fig:omegap}
\end{figure}

\subsection{Depinning frequency}

In the conventional Gittleman--Rosenblum model of vortex dynamics, viscosity and pinning constant are frequency-independent constants.\cite{JIGittleman:1968p172}  There is then a well-defined depinning frequency, $\omega_p = \alpha_p/\eta$, at which the viscous and elastic forces are equal.  In our measurements, viscosity and pinning constant  are strongly frequency dependent, and a substantial part of the reactive force experienced by the vortex is due to dynamical effects arising from the viscosity, rather than elasticity due to pinning.  The idea of a depinning frequency is therefore not well defined. Nevertheless, we can make an estimate from the DC limits of the pinning and viscous force constants, obtained from the fits to complex vortex viscosity presented in the next section: $\omega_p  \equiv \alpha_p/\eta_\mathrm{dc}$.  Data for $\omega_p$ are plotted in Fig.~\ref{fig:omegap}.  Over most of temperature range, $\omega_p/2 \pi$ is in the low GHz range, indicating that viscous effects are predominant at our measurement frequencies.  At low temperatures, however, $\omega_p/2 \pi$ increases to be 10 to 25~GHz (depending on field) due to the exponential increase in $\alpha_p(T)$ on cooling.  In this regime it is essential that vortex dynamics be probed with a technique that accurately measures both real and imaginary parts of the surface impedance.

\subsection{Complex vortex viscosity}
\label{Sec:complex_viscosity}

One of our initial motivations for carrying out multiple-frequency microwave measurements on ortho-II \ybco{6.52} was as a test of the validity of the single-vortex dynamical model outlined in Sec.~\ref{Sec:vortex_dynamics}.  Single-vortex theories are often treated with skepticism: there is a suspicion that they appear to work only because they have as many adjustable parameters ($\alpha_p$ and $\eta$) as there are degrees of freedom in the data (real and imaginary parts of $\tilde \rho_v)$.  Multiple-frequency data therefore offer the prospect of a stringent test.  A complication arises if the parameters in the vortex model have intrinsic frequency dependence, as appears to be the case in ortho-II \ybco{6.52}.  Testing the vortex model then becomes a more subtle process: as discussed in Sec.~\ref{Sec:complex_viscosity_model}, causality means that the dissipative and reactive parts of the dynamical response must obey Kramers--Kr\"onig relations. 

\begin{figure}[t]
\centering
\includegraphics[width= \columnwidth]{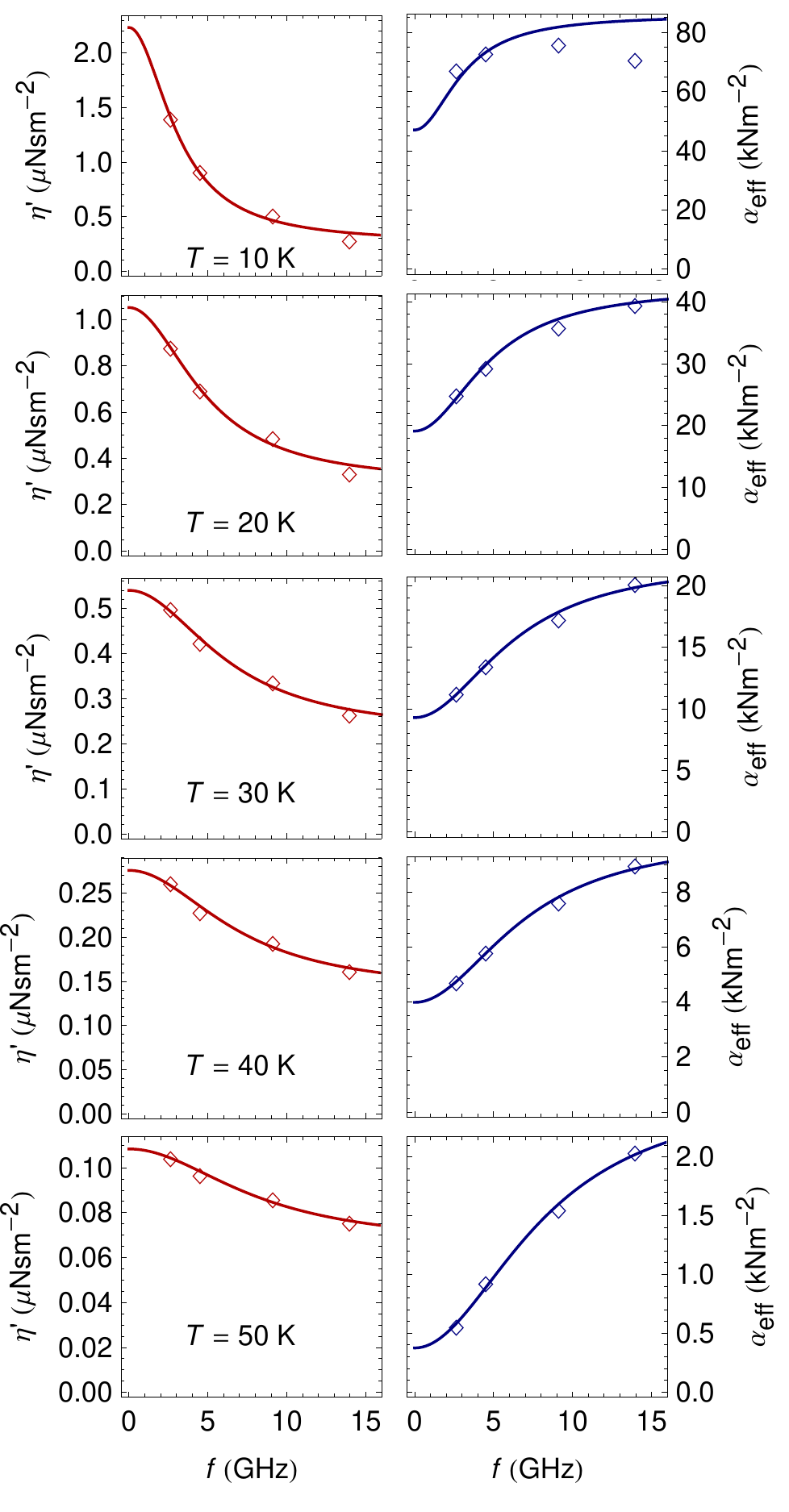}
\caption{(color online).  Simultaneous fits to $\eta^\prime(\omega)$ (left panels) and $\alpha_\mathrm{eff}(\omega)$ (right panels) at $B = 1$~T, using the procedure described in Sec.~\ref{Sec:complex_viscosity}.  } 
\label{fig:visc_fits1}
\end{figure}

\begin{figure}[t]
\centering
\includegraphics[width= \columnwidth]{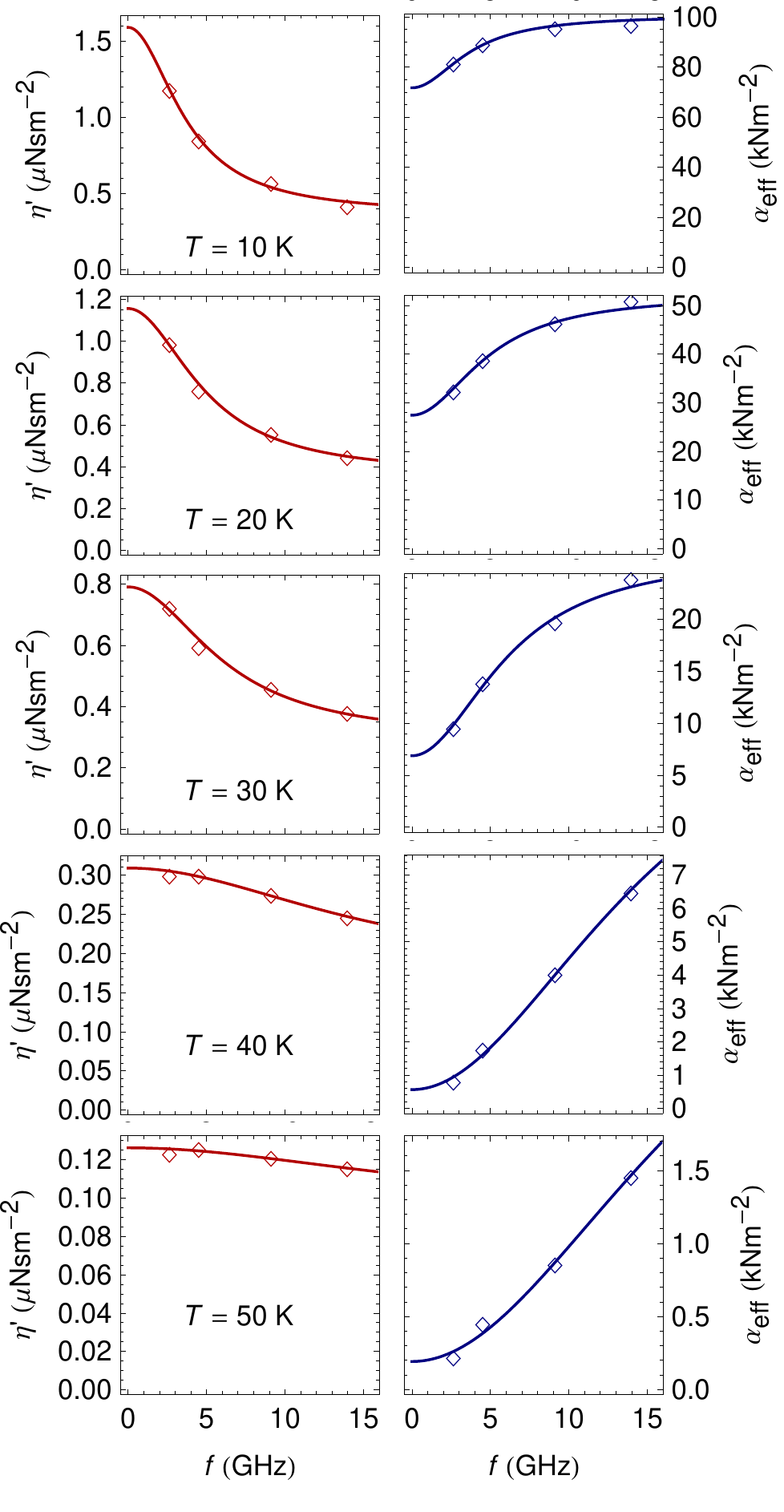}
\caption{(color online).  Simultaneous fits to $\eta^\prime(\omega)$ (left panels) and $\alpha_\mathrm{eff}(\omega)$ (right panels) at $B = 4$~T, using the procedure described in Sec.~\ref{Sec:complex_viscosity}. } 
\label{fig:visc_fits4}
\end{figure}

\begin{figure}[t]
\centering
\includegraphics[width= \columnwidth]{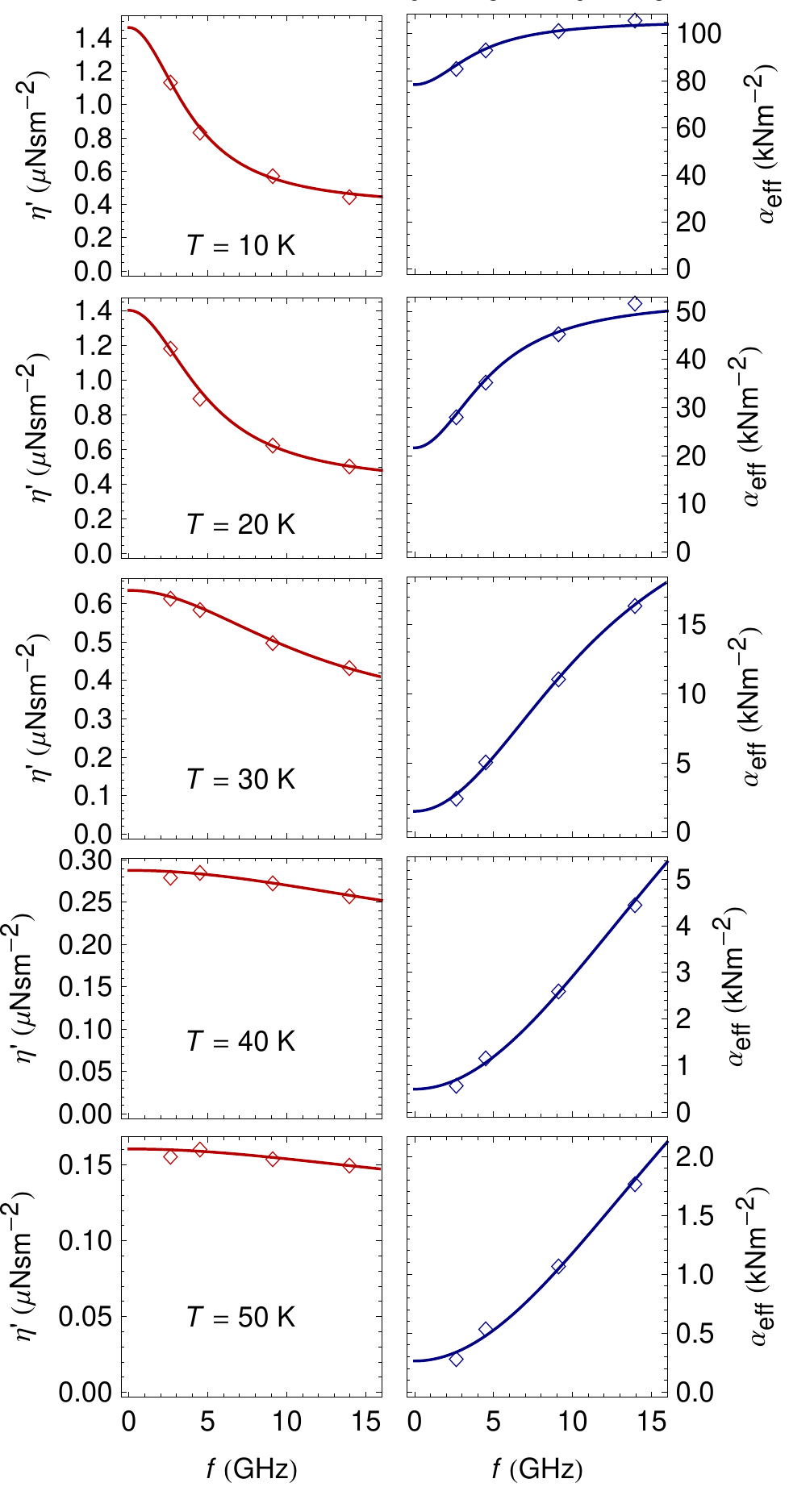}
\caption{(color online).  Simultaneous fits to $\eta^\prime(\omega)$ (left panels) and $\alpha_\mathrm{eff}(\omega)$ (right panels) at $B = 7$~T, using the procedure described in Sec.~\ref{Sec:complex_viscosity}. } 
\label{fig:visc_fits7}
\end{figure}

On purely empirical grounds, an observation of  coexisting fast and slow relaxation processes in the viscosity suggests that we represent $\eta^\prime(\omega)$ by a two-component spectrum.  The simplest possibility is that given by Eq.~\ref{Eq:model_viscosity}, in which a Drude-like Lorenztian spectrum, of magnitude $\eta_1$ and width $\Gamma$, rides on top of a broad background of magnitude $\eta_0$.  By causality, there must be an associated imaginary component, $\eta^{\prime\prime}(\omega)$.  This combines with the elastic contribution, $\alpha_p$, to form the effective pinning constant $\alpha_\mathrm{eff}(\omega)$, given by Eq.~\ref{Eq:model_pinning}.  Using this model, we have carried out simultaneous fits to the measured frequency dependence of $\eta^\prime$ and $\alpha_\mathrm{eff}$, at \emph{all} fields and temperatures.  Representative results,  in 10~K temperature steps up to 50~K, are shown in Figs.~\ref{fig:visc_fits1}, \ref{fig:visc_fits4} and \ref{fig:visc_fits7}, at fields of 1, 4 and 7~T, respectively. 

At all but the lowest temperatures and fields, the ability of the model to capture the observed dynamics is outstanding.  (For the $T = 10$~K, $B = 1$~T data, the discrepancy in $\alpha_\mathrm{eff}(\omega)$ is consistent with proximity to a cyclotron resonance, something that is very interesting in its own right and will be investigated in detail in future measurements.)  Importantly, the multiple frequency data now  overconstrain the model.  And, although these are four-parameter fits, only two of the parameters ($\eta_1$ and $\Gamma$) relate to frequency dependence: $\eta_0$ and $\alpha_p$ are additive offsets.  It is therefore impressive that, at all but the highest temperatures, the frequency dependence of $\eta^\prime$ essentially \emph{predicts} that of $\alpha_\mathrm{eff}$, and \emph{vice versa}. We draw two conclusions from the goodness of the fits: that the measurements themselves are accurate; and that the single-vortex model provides a very good representation of the microwave dynamics.

 \begin{figure}[t]
\centering
\includegraphics[width= \columnwidth]{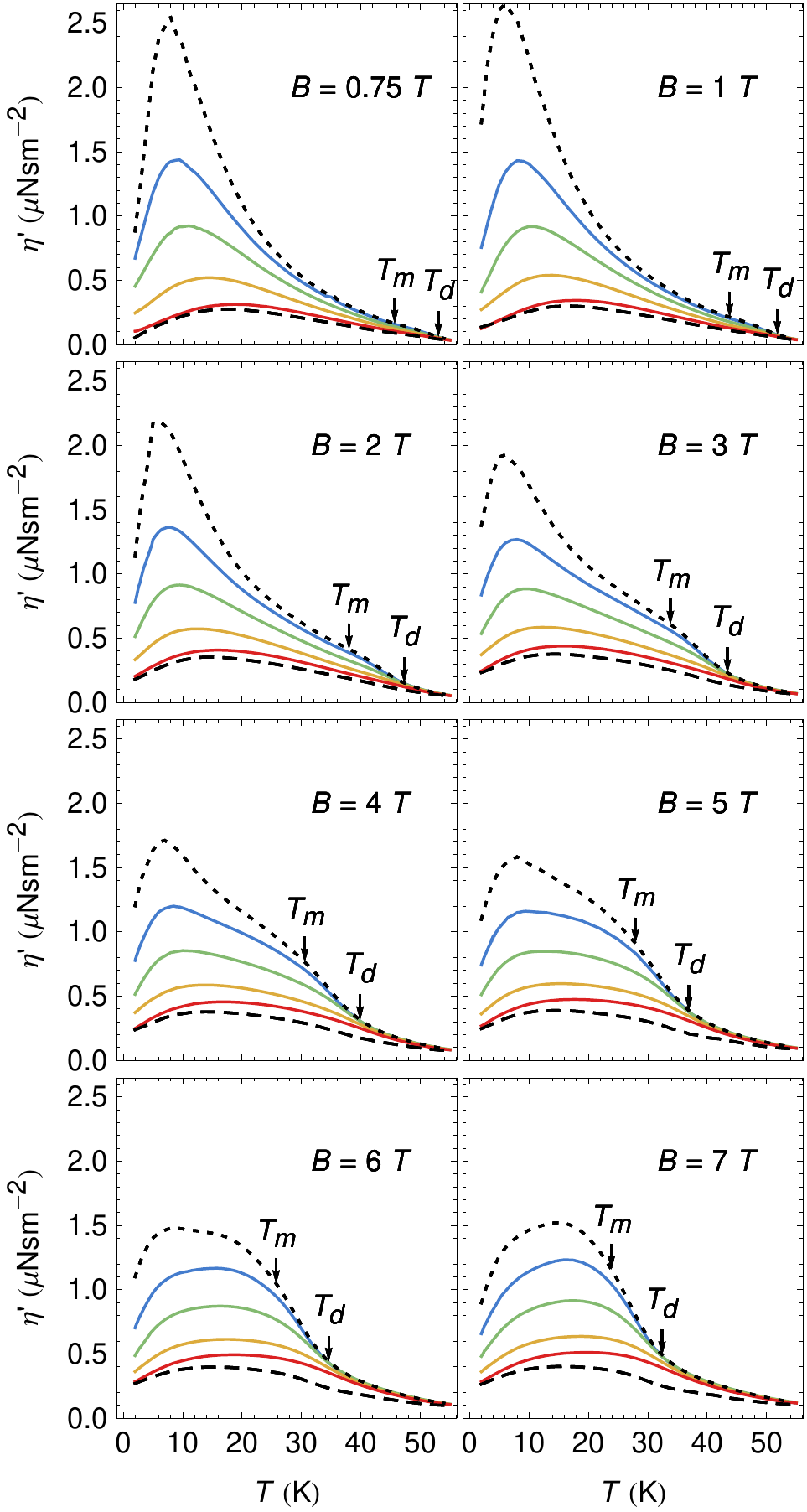}
\caption{(color online).  Real part of the frequency-dependent vortex viscosity, $\eta^\prime(T,B)$, plotted along with the fit parameters $\eta_0$ and $\eta_1$ from Eqs.~\ref{Eq:model_viscosity} and \ref{Eq:model_pinning}.   Long dashes show the fit parameter $\eta_0(T)$; short dashes the DC limit, $\eta_\mathrm{dc}(T) = \eta_0(T) + \eta_1(T)$.   As in Fig.~\ref{fig:viscosity}, $T_m$ denotes the vortex-lattice melting temperature\cite{Ramshaw2012}; $T_d$ gives the dynamical cross-over temperature, above which the frequency dependence of $\eta^\prime$ is very weak.} 
\label{fig:viscosity_params}
\end{figure}

 \begin{figure}[t,h,b]
\centering
\includegraphics[width= \figwidth \columnwidth]{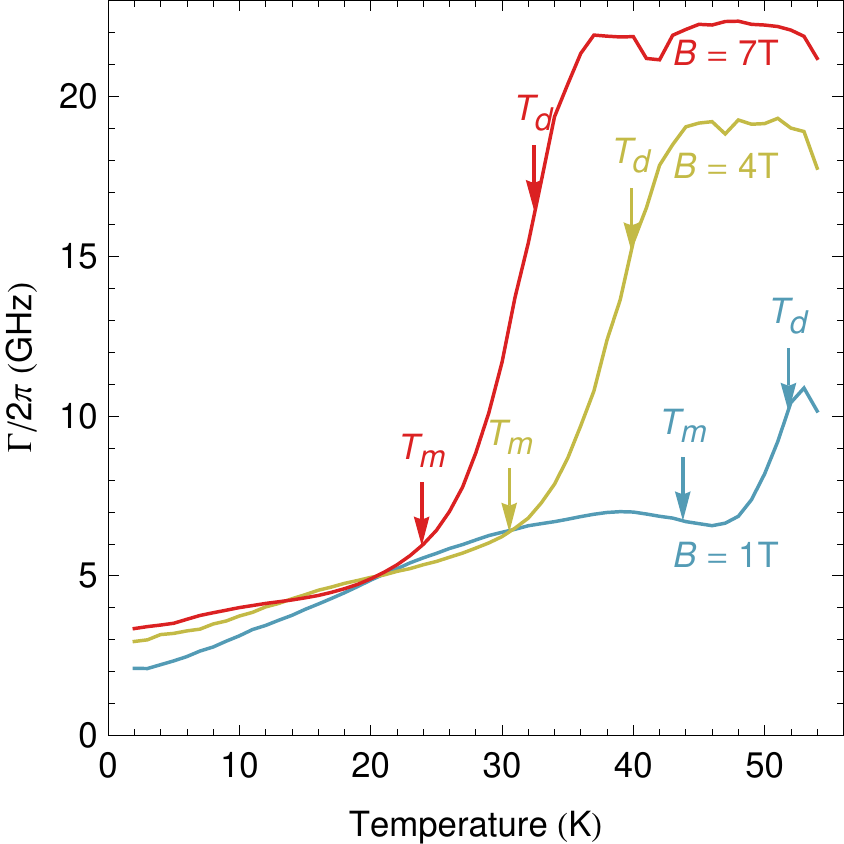}
\caption{(color online). The relaxation rate of the viscous dynamics, $\Gamma(T)$, for $B = 1$, 4 and 7~T, obtained from fitting to the complex viscosity spectra.} 
\label{fig:Gamma}
\end{figure}

The need for a two-component spectrum is a clear indication that fast and slowly relaxing excitations coexist in the vortex state.  At first sight this is not surprising, since quasiparticles in $d$-wave superconductors are known to have strongly energy-dependent relaxation rates as a consequence of the nodal quasiparticle spectrum.\cite{HIRSCHFELD:1994p570}  However, the fast-and-slow dichotomy observed in $\eta^\prime(\omega,T)$, especially at higher fields, is too pronounced to arise from thermally averaging over a distribution of relaxation rates dependent on energy alone.  
An alternative explanation, more in keeping with the disconnected nature of the broad and narrow parts of $\eta^\prime(\omega,T)$, is that the fast and slow processes are occurring in \emph{spatially distinct} regions.  Since the narrow component of $\eta^\prime(\omega,T)$ appears to be characteristic of $d$-wave quasiparticle physics, the slow processes are most naturally associated with regions \emph{outside} the cores. By extension, we would then associate fast relaxation with processes involving the vortex cores.  A picture of spatially separate relaxation mechanisms also helps to make sense of the sudden disappearance of slow relaxation above $T_d$ seen in Fig.~\ref{fig:viscosity}.   As can be seen in the field--temperature phase diagram in Fig.~\ref{fig:BTphasediagram}, $B_d(T)$ lies in the vortex-liquid regime, above the vortex-lattice melting line $B_m(T)$,\cite{Ramshaw2012} showing that, on its own, melting of the vortex-lattice does not eliminate the long-lived excitations.  Instead, $B_d(T)$ is likely a dynamical crossover at which thermally fluctuating vortices begin to move on a timescale similar to that of the microwave measurement frequency. 

The parameters from the fits to Eqs.~\ref{Eq:model_viscosity} and \ref{Eq:model_pinning} are: the elastic pinning constant, $\alpha_p$; the viscosity terms $\eta_0$ and $\eta_1$; and the viscosity relaxation rate, $\Gamma$.  $\alpha_p(T)$ is plotted in Fig.~\ref{fig:alphaB} as the DC limit of $\alpha_\mathrm{eff}(\omega)$: note that $\alpha_p$ fairly closely follows the 2.64~GHz pinning constant data. We see that it remains finite above the melting temperature.  This is at odds with bulk measurements, for which the melting transition marks the resistive onset.\cite{FISHER:1991p681,Charalambous:1992gf,Kwok:1992ug,Safar:1992ep,MACKENZIE:1993p197,Ramshaw2012}  A likely reason for this is that \emph{surface pinning} plays an important role in the microwave measurements: the loss of shear stiffness experienced at the melting transition has a profound effect on the ability of point-like defects to pin the vortex lattice; for planar defects such as the sample surface, shear stiffness is irrelevant.  This observation also serves as a warning that microwave techniques are not necessarily a good probe of bulk pinning.  Nevertheless, by narrowing the distribution of elastic forces experienced by vortices visible to the microwave fields, surface pinning likely works to our advantage, improving the applicability of the single-vortex model.  The dynamical component of $\alpha_\mathrm{eff}$ also helps in this respect, as it arises from interactions with the electron fluid rather than from randomly distributed defects.

The viscosity terms $\eta_0(T)$ and $\eta_1(T)$ are shown in Fig.~\ref{fig:viscosity_params}, with the latter quantity appearing as part of $\eta_\mathrm{dc} \equiv \eta_0 + \eta_1$.  $\eta_0(T)$ closely tracks the 13.97~GHz viscosity, and is very similar in magnitude to the viscosity inferred by Parks \emph{et al.} from terahertz measurements on \ybco{7-\delta} ($T_c = 85$ to 88~K), suggesting that the $\eta_0$ component of the viscosity extends over a very broad frequency range, up to terahertz frequencies.

The final fit parameter from Eqs.~\ref{Eq:model_viscosity} and \ref{Eq:model_pinning}, $\Gamma(T)$,  is plotted separately, in Fig.~\ref{fig:Gamma}.  At the lowest temperatures, $\Gamma/2\pi$ is of the order of several~GHz.  This is comparable to the width of zero-field quasiparticle conductivity spectra in ortho-II \ybco{6.50},\cite{Turner:2003p331, Harris:2006p388} providing one of the pieces of evidence linking vortex dissipation to the $d$-wave quasiparticles.  $\Gamma(T)$ initially grows linearly with temperature, also in accord with the behaviour inferred from zero-field measurements, in which the form of the temperature dependence of the scattering in Ortho-II \ybco{6.52} is taken as being indicative of weak-limit scattering of nodal quasiparticles.\cite{Turner:2003p331, Harris:2006p388}  It is interesting that $\Gamma(T)$ is comparable to the quasiparticle relaxation rate in zero field, as it indicates that the vortex lattice does not contribute strongly to quasiparticle scattering.  Vortices are large on the scale of the Fermi wavelength and would be expected to act as a source of small-angle scattering: such processes are known to be ineffective at relaxing charge currents in $d$-wave superconductors.\cite{Durst:2000p963}    In contrast, small-angle scattering should have a strong effect on the rate of \emph{intra-nodal} transitions, and be very effective at randomizing the quasiparticle group velocity.\cite{Durst:2000p963}  As a result, quasiparticles should become diffusively confined in the vortex lattice, even while  electrical transport measurements, such as the ones presented here, indicate mean free paths much larger than the inter-vortex spacing.  The vortex lattice should instead be responsible for new quasiparticle effects, such as pair-breaking induced by Doppler shifting of quasiparticle energies.\cite{VOLOVIK:1993p201} These should grow in importance with $B$, and may be responsible for the increase in $\eta_0(T)$ at higher fields.  On passing through the melting temperature, $\Gamma(T)$ starts to increase rapidly, then appears to plateau above the dynamical crossover with $\Gamma/2\pi \approx 20$~GHz in the higher-field data sets.  This frequency scale is substantially smaller than that expected for quasiparticle scattering in this temperature range, and is perhaps connected to superconducting fluctuations.

\begin{figure}[t]
\centering
\includegraphics[width= \columnwidth]{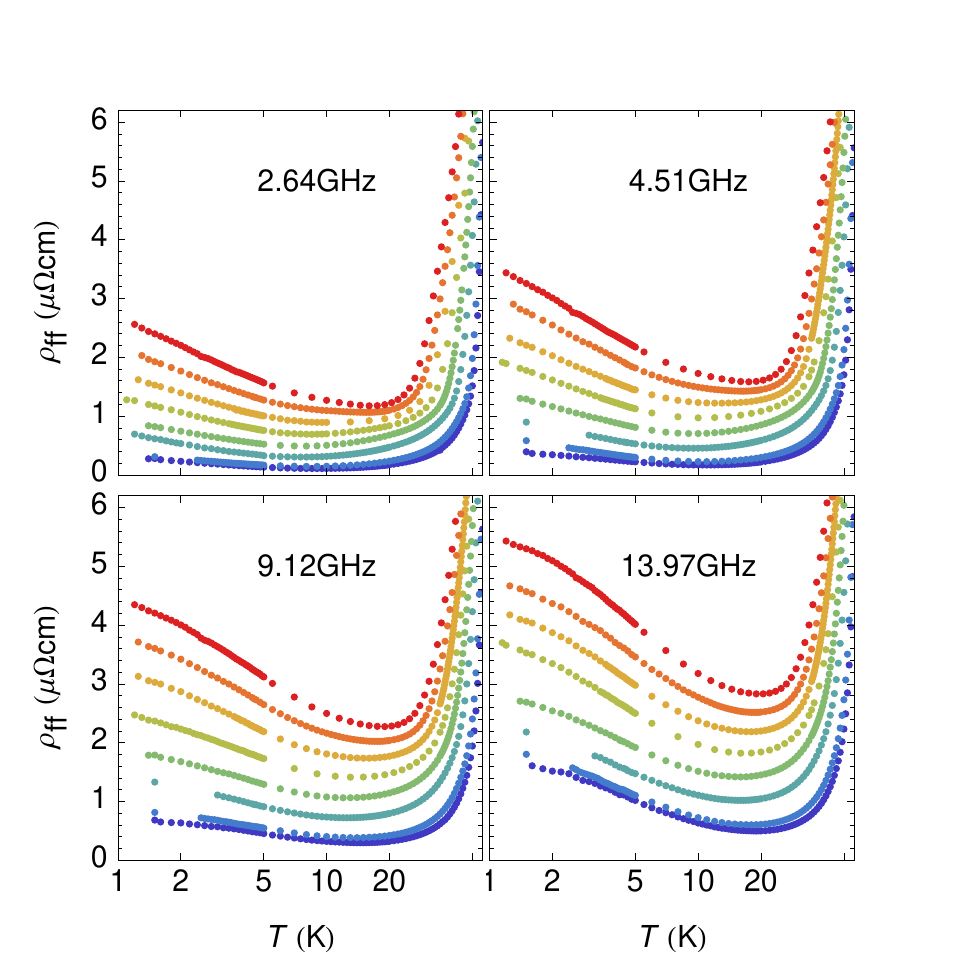}
\caption{(color online).  Frequency-dependent flux-flow resistivity, $\rho_\mathrm{ff}(\omega) \equiv B \Phi_0/\eta^\prime(\omega)$, for $\omega/2 \pi = 2.64$, 4.51, 9.12, and 13.97~GHz, on semi-logarithmic axes, for $B = 0$, 0.75, 1, 2, 3, 4, 5, 6 and 7~T (from bottom to top). At the lower frequencies, the low temperature behaviour follows the form $\rho_\mathrm{ff}(T) = \rho_0 + A \log(1/T)$.   At 9.12 and 13.97~GHz, the low temperature divergence appears to be cut off by finite frequency effects.} 
\label{fig:rho_ff}
\end{figure}

\subsection{Flux-flow resistivity}

Finally, we obtain an estimate of the flux-flow resistivity, $\rho_\mathrm{ff}$, from the vortex-dynamics data.   As mentioned in the introduction, we are faced with a trade-off when measuring $\rho_\mathrm{ff}$ in any real material, due to the presence of pinning: we must either use large currents that depin the vortices (for cuprates, an increasingly difficult prospect as temperature is lowered\cite{Kunchur:1993ie}); or measure the linear response at high frequencies and work back towards the static limit. Having taken the latter approach, we define a frequency-dependent flux-flow resistivity, $\rho_\mathrm{ff}(\omega) \equiv B \Phi_0/\eta^\prime(\omega)$, which is plotted in Fig.~\ref{fig:rho_ff}.  At all fields and frequencies, $\rho_\mathrm{ff}(T)$ shows an initial drop on cooling through $T_c$,  a broad minimum in the 8~to~20~K range, and an upturn at low temperatures --- this is a reflection of the peaked structure in $\eta^\prime(T)$.  The most remarkable aspect of the data is that, at the lower frequencies, the low temperature upturns appear to follow a $\log(1/T)$ form. This is reminiscent of the behaviour of the DC resistivity in the pseudogap regime,\cite{Ando:1995p148, Boebinger:1996p147} raising the interesting possibility that the two effects are connected.

\section{Conclusions}

In summary, this work builds on a number of technical advances in microwave spectroscopy of the vortex state: the use of high $Q$ dielectric resonators, allowing high sensitivity measurements on small single crystals; the ability to simultaneously measure real and imaginary parts of the impedance at each frequency, allowing a clean separation of viscous and elastic effects; measurement over a wide range of microwave frequencies; and measurements as a function of magnetic field.  Applied to a high-purity single crystal of ortho-II \ybco{6.52}, these developments come together to reveal the vortex dynamics in unprecedented detail, uncovering a close connection between the vortex viscosity $\eta^\prime(\omega,T)$ and the zero-field dynamics of the $d$-wave quasiparticles.  The vortex viscosity is revealed to have strong frequency dependence and, when treated as a complex-valued quantity, is consistent with the tight constraints causality places on physical response functions.  This gives us a great deal of confidence in both the experimental technique, and the use of single-vortex models of microwave-frequency vortex dynamics.

\section{Acknowledgements}

The authors thank W.~A.~Atkinson, J.~S.~Dodge, M.~P.~Kennett, B.~Morgan, J.~E. Sonier, Z.~Te\u{s}anovi\'c and J.~R.~Waldram for useful discussions.  Research support for the experiments was provided by the Natural Science and Engineering Research Council of Canada (NSERC) and the Canadian Foundation for Innovation.  Research support for sample preparation was provided by NSERC and the Canadian Institute for Advanced Research.\\


\begin{thebibliography}{98}%
\makeatletter
\providecommand \@ifxundefined [1]{%
 \@ifx{#1\undefined}
}%
\providecommand \@ifnum [1]{%
 \ifnum #1\expandafter \@firstoftwo
 \else \expandafter \@secondoftwo
 \fi
}%
\providecommand \@ifx [1]{%
 \ifx #1\expandafter \@firstoftwo
 \else \expandafter \@secondoftwo
 \fi
}%
\providecommand \natexlab [1]{#1}%
\providecommand \enquote  [1]{``#1''}%
\providecommand \bibnamefont  [1]{#1}%
\providecommand \bibfnamefont [1]{#1}%
\providecommand \citenamefont [1]{#1}%
\providecommand \href@noop [0]{\@secondoftwo}%
\providecommand \href [0]{\begingroup \@sanitize@url \@href}%
\providecommand \@href[1]{\@@startlink{#1}\@@href}%
\providecommand \@@href[1]{\endgroup#1\@@endlink}%
\providecommand \@sanitize@url [0]{\catcode `\\12\catcode `\$12\catcode
  `\&12\catcode `\#12\catcode `\^12\catcode `\_12\catcode `\%12\relax}%
\providecommand \@@startlink[1]{}%
\providecommand \@@endlink[0]{}%
\providecommand \url  [0]{\begingroup\@sanitize@url \@url }%
\providecommand \@url [1]{\endgroup\@href {#1}{\urlprefix }}%
\providecommand \urlprefix  [0]{URL }%
\providecommand \Eprint [0]{\href }%
\providecommand \doibase [0]{http://dx.doi.org/}%
\providecommand \selectlanguage [0]{\@gobble}%
\providecommand \bibinfo  [0]{\@secondoftwo}%
\providecommand \bibfield  [0]{\@secondoftwo}%
\providecommand \translation [1]{[#1]}%
\providecommand \BibitemOpen [0]{}%
\providecommand \bibitemStop [0]{}%
\providecommand \bibitemNoStop [0]{.\EOS\space}%
\providecommand \EOS [0]{\spacefactor3000\relax}%
\providecommand \BibitemShut  [1]{\csname bibitem#1\endcsname}%
\let\auto@bib@innerbib\@empty
\bibitem [{\citenamefont {Bardeen}\ and\ \citenamefont
  {Stephen}(1965)}]{Bardeen:1965p151}%
  \BibitemOpen
  \bibfield  {author} {\bibinfo {author} {\bibfnamefont {J.}~\bibnamefont
  {Bardeen}}\ and\ \bibinfo {author} {\bibfnamefont {M.~J.}\ \bibnamefont
  {Stephen}},\ }\href@noop {} {\bibfield  {journal} {\bibinfo  {journal}
  {Phys.\ Rev.}\ }\textbf {\bibinfo {volume} {140}},\ \bibinfo {pages} {1197}
  (\bibinfo {year} {1965})}\BibitemShut {NoStop}%
\bibitem [{\citenamefont {Maggio-Aprile}\ \emph {et~al.}(1995)\citenamefont
  {Maggio-Aprile}, \citenamefont {Renner}, \citenamefont {Erb}, \citenamefont
  {Walker},\ and\ \citenamefont {Fischer}}]{MaggioAprile:1995p3014}%
  \BibitemOpen
  \bibfield  {author} {\bibinfo {author} {\bibfnamefont {I.}~\bibnamefont
  {Maggio-Aprile}}, \bibinfo {author} {\bibfnamefont {C.}~\bibnamefont
  {Renner}}, \bibinfo {author} {\bibfnamefont {A.}~\bibnamefont {Erb}},
  \bibinfo {author} {\bibfnamefont {E.}~\bibnamefont {Walker}}, \ and\ \bibinfo
  {author} {\bibfnamefont {O.}~\bibnamefont {Fischer}},\ }\href@noop {}
  {\bibfield  {journal} {\bibinfo  {journal} {Phys.\ Rev.\ Lett.}\ }\textbf
  {\bibinfo {volume} {75}},\ \bibinfo {pages} {2754} (\bibinfo {year}
  {1995})}\BibitemShut {NoStop}%
\bibitem [{\citenamefont {Soininen}\ \emph {et~al.}(1994)\citenamefont
  {Soininen}, \citenamefont {Kallin},\ and\ \citenamefont
  {Berlinsky}}]{Soininen:1994iu}%
  \BibitemOpen
  \bibfield  {author} {\bibinfo {author} {\bibfnamefont {P.~I.}\ \bibnamefont
  {Soininen}}, \bibinfo {author} {\bibfnamefont {C.}~\bibnamefont {Kallin}}, \
  and\ \bibinfo {author} {\bibfnamefont {A.~J.}\ \bibnamefont {Berlinsky}},\
  }\href@noop {} {\bibfield  {journal} {\bibinfo  {journal} {Phys.\ Rev.\ B}\
  }\textbf {\bibinfo {volume} {50}},\ \bibinfo {pages} {R13883} (\bibinfo
  {year} {1994})}\BibitemShut {NoStop}%
\bibitem [{\citenamefont {Hardy}\ \emph {et~al.}(1993)\citenamefont {Hardy},
  \citenamefont {Bonn}, \citenamefont {Morgan}, \citenamefont {Liang},\ and\
  \citenamefont {Zhang}}]{hardy93}%
  \BibitemOpen
  \bibfield  {author} {\bibinfo {author} {\bibfnamefont {W.~N.}\ \bibnamefont
  {Hardy}}, \bibinfo {author} {\bibfnamefont {D.~A.}\ \bibnamefont {Bonn}},
  \bibinfo {author} {\bibfnamefont {D.~C.}\ \bibnamefont {Morgan}}, \bibinfo
  {author} {\bibfnamefont {R.}~\bibnamefont {Liang}}, \ and\ \bibinfo {author}
  {\bibfnamefont {K.}~\bibnamefont {Zhang}},\ }\href@noop {} {\bibfield
  {journal} {\bibinfo  {journal} {Phys.\ Rev.\ Lett.}\ }\textbf {\bibinfo
  {volume} {70}},\ \bibinfo {pages} {3999} (\bibinfo {year}
  {1993})}\BibitemShut {NoStop}%
\bibitem [{\citenamefont {Scalapino}(1995)}]{Scalapino:1995p741}%
  \BibitemOpen
  \bibfield  {author} {\bibinfo {author} {\bibfnamefont {D.~J.}\ \bibnamefont
  {Scalapino}},\ }\href@noop {} {\bibfield  {journal} {\bibinfo  {journal}
  {Phys.\ Rep.}\ }\textbf {\bibinfo {volume} {250}},\ \bibinfo {pages} {330}
  (\bibinfo {year} {1995})}\BibitemShut {NoStop}%
\bibitem [{\citenamefont {Ding}\ \emph {et~al.}(1996)\citenamefont {Ding},
  \citenamefont {Norman}, \citenamefont {Campuzano}, \citenamefont {Randeria},
  \citenamefont {Bellman}, \citenamefont {Yokoya}, \citenamefont {Takahashi},
  \citenamefont {Mochiku},\ and\ \citenamefont {Kadowaki}}]{Ding:1996p3019}%
  \BibitemOpen
  \bibfield  {author} {\bibinfo {author} {\bibfnamefont {H.}~\bibnamefont
  {Ding}}, \bibinfo {author} {\bibfnamefont {M.~R.}\ \bibnamefont {Norman}},
  \bibinfo {author} {\bibfnamefont {J.~C.}\ \bibnamefont {Campuzano}}, \bibinfo
  {author} {\bibfnamefont {M.}~\bibnamefont {Randeria}}, \bibinfo {author}
  {\bibfnamefont {A.~F.}\ \bibnamefont {Bellman}}, \bibinfo {author}
  {\bibfnamefont {T.}~\bibnamefont {Yokoya}}, \bibinfo {author} {\bibfnamefont
  {T.}~\bibnamefont {Takahashi}}, \bibinfo {author} {\bibfnamefont
  {T.}~\bibnamefont {Mochiku}}, \ and\ \bibinfo {author} {\bibfnamefont
  {K.}~\bibnamefont {Kadowaki}},\ }\href@noop {} {\bibfield  {journal}
  {\bibinfo  {journal} {Phys.\ Rev.\ B}\ }\textbf {\bibinfo {volume} {54}},\
  \bibinfo {pages} {R9678} (\bibinfo {year} {1996})}\BibitemShut {NoStop}%
\bibitem [{\citenamefont {Choi}\ \emph {et~al.}(1994)\citenamefont {Choi},
  \citenamefont {Lihn}, \citenamefont {Drew},\ and\ \citenamefont
  {Hsu}}]{Choi:1994ga}%
  \BibitemOpen
  \bibfield  {author} {\bibinfo {author} {\bibfnamefont {E.-J.}\ \bibnamefont
  {Choi}}, \bibinfo {author} {\bibfnamefont {H.-T.~S.}\ \bibnamefont {Lihn}},
  \bibinfo {author} {\bibfnamefont {H.~D.}\ \bibnamefont {Drew}}, \ and\
  \bibinfo {author} {\bibfnamefont {T.~C.}\ \bibnamefont {Hsu}},\ }\href@noop
  {} {\bibfield  {journal} {\bibinfo  {journal} {Phys.\ Rev.\ B}\ }\textbf
  {\bibinfo {volume} {49}},\ \bibinfo {pages} {13271} (\bibinfo {year}
  {1994})}\BibitemShut {NoStop}%
\bibitem [{\citenamefont {Nozi{\`e}res}\ and\ \citenamefont
  {Vinen}(1966)}]{Nozieres:1966p667}%
  \BibitemOpen
  \bibfield  {author} {\bibinfo {author} {\bibfnamefont {P.}~\bibnamefont
  {Nozi{\`e}res}}\ and\ \bibinfo {author} {\bibfnamefont {W.~F.}\ \bibnamefont
  {Vinen}},\ }\href@noop {} {\bibfield  {journal} {\bibinfo  {journal} {Phil.\
  Mag.}\ }\textbf {\bibinfo {volume} {14}},\ \bibinfo {pages} {667} (\bibinfo
  {year} {1966})}\BibitemShut {NoStop}%
\bibitem [{\citenamefont {Larkin}\ and\ \citenamefont
  {Ovchinnikov}(1976)}]{Larkin:1976p3015}%
  \BibitemOpen
  \bibfield  {author} {\bibinfo {author} {\bibfnamefont {A.~I.}\ \bibnamefont
  {Larkin}}\ and\ \bibinfo {author} {\bibfnamefont {Y.~N.}\ \bibnamefont
  {Ovchinnikov}},\ }\href@noop {} {\bibfield  {journal} {\bibinfo  {journal}
  {JETP Lett.}\ }\textbf {\bibinfo {volume} {23}},\ \bibinfo {pages} {187}
  (\bibinfo {year} {1976})}\BibitemShut {NoStop}%
\bibitem [{\citenamefont {Uemura}\ \emph {et~al.}(1989)\citenamefont {Uemura},
  \citenamefont {Luke}, \citenamefont {Sternlieb}, \citenamefont {Brewer},
  \citenamefont {Carolan}, \citenamefont {Hardy}, \citenamefont {Kadono},
  \citenamefont {Kempton}, \citenamefont {Kiefl}, \citenamefont {Kreitzman},
  \citenamefont {Mulhern}, \citenamefont {Riseman}, \citenamefont {Williams},
  \citenamefont {Yang}, \citenamefont {Uchida}, \citenamefont {Takagi},
  \citenamefont {Gopalakrishnan}, \citenamefont {Sleight}, \citenamefont
  {Subramanian}, \citenamefont {Chien}, \citenamefont {Cielplak}, \citenamefont
  {Xiao}, \citenamefont {Lee}, \citenamefont {Statt}, \citenamefont {Stronach},
  \citenamefont {Kossler},\ and\ \citenamefont {Yu}}]{Uemura:1989p962}%
  \BibitemOpen
  \bibfield  {author} {\bibinfo {author} {\bibfnamefont {Y.~J.}\ \bibnamefont
  {Uemura}}, \bibinfo {author} {\bibfnamefont {G.~M.}\ \bibnamefont {Luke}},
  \bibinfo {author} {\bibfnamefont {B.~J.}\ \bibnamefont {Sternlieb}}, \bibinfo
  {author} {\bibfnamefont {J.~H.}\ \bibnamefont {Brewer}}, \bibinfo {author}
  {\bibfnamefont {J.~F.}\ \bibnamefont {Carolan}}, \bibinfo {author}
  {\bibfnamefont {W.~N.}\ \bibnamefont {Hardy}}, \bibinfo {author}
  {\bibfnamefont {R.}~\bibnamefont {Kadono}}, \bibinfo {author} {\bibfnamefont
  {J.~R.}\ \bibnamefont {Kempton}}, \bibinfo {author} {\bibfnamefont {R.~F.}\
  \bibnamefont {Kiefl}}, \bibinfo {author} {\bibfnamefont {S.~R.}\ \bibnamefont
  {Kreitzman}}, \bibinfo {author} {\bibfnamefont {P.}~\bibnamefont {Mulhern}},
  \bibinfo {author} {\bibfnamefont {T.~M.}\ \bibnamefont {Riseman}}, \bibinfo
  {author} {\bibfnamefont {D.~L.}\ \bibnamefont {Williams}}, \bibinfo {author}
  {\bibfnamefont {B.~X.}\ \bibnamefont {Yang}}, \bibinfo {author}
  {\bibfnamefont {S.}~\bibnamefont {Uchida}}, \bibinfo {author} {\bibfnamefont
  {H.}~\bibnamefont {Takagi}}, \bibinfo {author} {\bibfnamefont
  {J.}~\bibnamefont {Gopalakrishnan}}, \bibinfo {author} {\bibfnamefont
  {A.~W.}\ \bibnamefont {Sleight}}, \bibinfo {author} {\bibfnamefont {M.~A.}\
  \bibnamefont {Subramanian}}, \bibinfo {author} {\bibfnamefont {C.~L.}\
  \bibnamefont {Chien}}, \bibinfo {author} {\bibfnamefont {M.~Z.}\ \bibnamefont
  {Cielplak}}, \bibinfo {author} {\bibfnamefont {G.}~\bibnamefont {Xiao}},
  \bibinfo {author} {\bibfnamefont {V.~Y.}\ \bibnamefont {Lee}}, \bibinfo
  {author} {\bibfnamefont {B.~W.}\ \bibnamefont {Statt}}, \bibinfo {author}
  {\bibfnamefont {C.~E.}\ \bibnamefont {Stronach}}, \bibinfo {author}
  {\bibfnamefont {W.~J.}\ \bibnamefont {Kossler}}, \ and\ \bibinfo {author}
  {\bibfnamefont {X.~H.}\ \bibnamefont {Yu}},\ }\href@noop {} {\bibfield
  {journal} {\bibinfo  {journal} {Phys.\ Rev.\ Lett.}\ }\textbf {\bibinfo
  {volume} {62}},\ \bibinfo {pages} {2317} (\bibinfo {year}
  {1989})}\BibitemShut {NoStop}%
\bibitem [{\citenamefont {Emery}\ and\ \citenamefont
  {Kivelson}(1995)}]{Emery:1995p364}%
  \BibitemOpen
  \bibfield  {author} {\bibinfo {author} {\bibfnamefont {V.~J.}\ \bibnamefont
  {Emery}}\ and\ \bibinfo {author} {\bibfnamefont {S.~A.}\ \bibnamefont
  {Kivelson}},\ }\href@noop {} {\bibfield  {journal} {\bibinfo  {journal}
  {Nature}\ }\textbf {\bibinfo {volume} {374}},\ \bibinfo {pages} {434}
  (\bibinfo {year} {1995})}\BibitemShut {NoStop}%
\bibitem [{\citenamefont {Franz}\ and\ \citenamefont
  {Te\u{s}anovi\'c}(2001)}]{franz01}%
  \BibitemOpen
  \bibfield  {author} {\bibinfo {author} {\bibfnamefont {M.}~\bibnamefont
  {Franz}}\ and\ \bibinfo {author} {\bibfnamefont {Z.}~\bibnamefont
  {Te\u{s}anovi\'c}},\ }\href@noop {} {\bibfield  {journal} {\bibinfo
  {journal} {Phys.\ Rev.\ Lett.}\ }\textbf {\bibinfo {volume} {87}},\ \bibinfo
  {pages} {257003} (\bibinfo {year} {2001})}\BibitemShut {NoStop}%
\bibitem [{\citenamefont {Franz}\ \emph {et~al.}(2002)\citenamefont {Franz},
  \citenamefont {Te\u{s}anovi\'c},\ and\ \citenamefont {Vafek}}]{franz02}%
  \BibitemOpen
  \bibfield  {author} {\bibinfo {author} {\bibfnamefont {M.}~\bibnamefont
  {Franz}}, \bibinfo {author} {\bibfnamefont {Z.}~\bibnamefont
  {Te\u{s}anovi\'c}}, \ and\ \bibinfo {author} {\bibfnamefont {O.}~\bibnamefont
  {Vafek}},\ }\href@noop {} {\bibfield  {journal} {\bibinfo  {journal} {Phys.\
  Rev.\ B}\ }\textbf {\bibinfo {volume} {66}},\ \bibinfo {pages} {054535}
  (\bibinfo {year} {2002})}\BibitemShut {NoStop}%
\bibitem [{\citenamefont {Herbut}(2002{\natexlab{a}})}]{herbut02}%
  \BibitemOpen
  \bibfield  {author} {\bibinfo {author} {\bibfnamefont {I.~F.}\ \bibnamefont
  {Herbut}},\ }\href@noop {} {\bibfield  {journal} {\bibinfo  {journal} {Phys.\
  Rev.\ Lett.}\ }\textbf {\bibinfo {volume} {88}},\ \bibinfo {pages} {047006}
  (\bibinfo {year} {2002}{\natexlab{a}})}\BibitemShut {NoStop}%
\bibitem [{\citenamefont {Herbut}(2002{\natexlab{b}})}]{herbut02a}%
  \BibitemOpen
  \bibfield  {author} {\bibinfo {author} {\bibfnamefont {I.~F.}\ \bibnamefont
  {Herbut}},\ }\href@noop {} {\bibfield  {journal} {\bibinfo  {journal} {Phys.\
  Rev.\ B}\ }\textbf {\bibinfo {volume} {66}},\ \bibinfo {pages} {094504}
  (\bibinfo {year} {2002}{\natexlab{b}})}\BibitemShut {NoStop}%
\bibitem [{\citenamefont {Herbut}(2005)}]{herbut05}%
  \BibitemOpen
  \bibfield  {author} {\bibinfo {author} {\bibfnamefont {I.~F.}\ \bibnamefont
  {Herbut}},\ }\href@noop {} {\bibfield  {journal} {\bibinfo  {journal} {Phys.\
  Rev.\ Lett.}\ }\textbf {\bibinfo {volume} {94}},\ \bibinfo {pages} {237001}
  (\bibinfo {year} {2005})}\BibitemShut {NoStop}%
\bibitem [{\citenamefont {Franz}\ and\ \citenamefont
  {Iyengar}(2006)}]{franz06}%
  \BibitemOpen
  \bibfield  {author} {\bibinfo {author} {\bibfnamefont {M.}~\bibnamefont
  {Franz}}\ and\ \bibinfo {author} {\bibfnamefont {A.~P.}\ \bibnamefont
  {Iyengar}},\ }\href@noop {} {\bibfield  {journal} {\bibinfo  {journal}
  {Phys.\ Rev.\ Lett.}\ }\textbf {\bibinfo {volume} {96}},\ \bibinfo {pages}
  {047007} (\bibinfo {year} {2006})}\BibitemShut {NoStop}%
\bibitem [{\citenamefont {Te\u{s}anovi\'c}(2008)}]{Tesanovic:2008p2290}%
  \BibitemOpen
  \bibfield  {author} {\bibinfo {author} {\bibfnamefont {Z.}~\bibnamefont
  {Te\u{s}anovi\'c}},\ }\href@noop {} {\bibfield  {journal} {\bibinfo
  {journal} {Nat.\ Phys.}\ }\textbf {\bibinfo {volume} {4}},\ \bibinfo {pages}
  {408} (\bibinfo {year} {2008})}\BibitemShut {NoStop}%
\bibitem [{\citenamefont {Geshkenbein}\ \emph {et~al.}(1998)\citenamefont
  {Geshkenbein}, \citenamefont {Ioffe},\ and\ \citenamefont
  {Millis}}]{Geshkenbein:1998p3010}%
  \BibitemOpen
  \bibfield  {author} {\bibinfo {author} {\bibfnamefont {V.~B.}\ \bibnamefont
  {Geshkenbein}}, \bibinfo {author} {\bibfnamefont {L.~B.}\ \bibnamefont
  {Ioffe}}, \ and\ \bibinfo {author} {\bibfnamefont {A.~J.}\ \bibnamefont
  {Millis}},\ }\href@noop {} {\bibfield  {journal} {\bibinfo  {journal} {Phys.\
  Rev.\ Lett}\ }\textbf {\bibinfo {volume} {80}},\ \bibinfo {pages} {5778}
  (\bibinfo {year} {1998})}\BibitemShut {NoStop}%
\bibitem [{\citenamefont {Ioffe}\ and\ \citenamefont
  {Millis}(2002)}]{Ioffe:2002p717}%
  \BibitemOpen
  \bibfield  {author} {\bibinfo {author} {\bibfnamefont {L.~B.}\ \bibnamefont
  {Ioffe}}\ and\ \bibinfo {author} {\bibfnamefont {A.~J.}\ \bibnamefont
  {Millis}},\ }\href@noop {} {\bibfield  {journal} {\bibinfo  {journal} {Phys.\
  Rev.\ B}\ }\textbf {\bibinfo {volume} {66}},\ \bibinfo {pages} {094513}
  (\bibinfo {year} {2002})}\BibitemShut {NoStop}%
\bibitem [{\citenamefont {Lee}(2003)}]{Lee:2003p7}%
  \BibitemOpen
  \bibfield  {author} {\bibinfo {author} {\bibfnamefont {P.~A.}\ \bibnamefont
  {Lee}},\ }\href@noop {} {\bibfield  {journal} {\bibinfo  {journal} {Physica
  C}\ }\textbf {\bibinfo {volume} {388}},\ \bibinfo {pages} {7} (\bibinfo
  {year} {2003})}\BibitemShut {NoStop}%
\bibitem [{\citenamefont {Melikyan}\ and\ \citenamefont
  {Te\u{s}anovi\'c}(2005)}]{Melikyan:2005p3011}%
  \BibitemOpen
  \bibfield  {author} {\bibinfo {author} {\bibfnamefont {A.}~\bibnamefont
  {Melikyan}}\ and\ \bibinfo {author} {\bibfnamefont {Z.}~\bibnamefont
  {Te\u{s}anovi\'c}},\ }\href@noop {} {\bibfield  {journal} {\bibinfo
  {journal} {Phys.\ Rev.\ B}\ }\textbf {\bibinfo {volume} {71}},\ \bibinfo
  {pages} {214511} (\bibinfo {year} {2005})}\BibitemShut {NoStop}%
\bibitem [{\citenamefont {Nikoli{\'c}}\ and\ \citenamefont
  {Sachdev}(2006)}]{Nikolic:2006p3012}%
  \BibitemOpen
  \bibfield  {author} {\bibinfo {author} {\bibfnamefont {P.}~\bibnamefont
  {Nikoli{\'c}}}\ and\ \bibinfo {author} {\bibfnamefont {S.}~\bibnamefont
  {Sachdev}},\ }\href@noop {} {\bibfield  {journal} {\bibinfo  {journal}
  {Phys.\ Rev.\ B}\ }\textbf {\bibinfo {volume} {73}},\ \bibinfo {pages}
  {134511} (\bibinfo {year} {2006})}\BibitemShut {NoStop}%
\bibitem [{\citenamefont {Bilbro}\ \emph
  {et~al.}(2011{\natexlab{a}})\citenamefont {Bilbro}, \citenamefont
  {ValdesAguilar}, \citenamefont {Logvenov}, \citenamefont {Bozovic},\ and\
  \citenamefont {Armitage}}]{Bilbro:2011p3009}%
  \BibitemOpen
  \bibfield  {author} {\bibinfo {author} {\bibfnamefont {L.~S.}\ \bibnamefont
  {Bilbro}}, \bibinfo {author} {\bibfnamefont {R.}~\bibnamefont
  {ValdesAguilar}}, \bibinfo {author} {\bibfnamefont {G.}~\bibnamefont
  {Logvenov}}, \bibinfo {author} {\bibfnamefont {I.}~\bibnamefont {Bozovic}}, \
  and\ \bibinfo {author} {\bibfnamefont {N.~P.}\ \bibnamefont {Armitage}},\
  }\href@noop {} {\bibfield  {journal} {\bibinfo  {journal} {Phys.\ Rev.\ B}\
  }\textbf {\bibinfo {volume} {84}},\ \bibinfo {pages} {100511} (\bibinfo
  {year} {2011}{\natexlab{a}})}\BibitemShut {NoStop}%
\bibitem [{\citenamefont {Corson}\ \emph {et~al.}(1999)\citenamefont {Corson},
  \citenamefont {Mallozzi}, \citenamefont {Orenstein}, \citenamefont
  {Eckstein},\ and\ \citenamefont {Bozovic}}]{Corson:1999p716}%
  \BibitemOpen
  \bibfield  {author} {\bibinfo {author} {\bibfnamefont {J.}~\bibnamefont
  {Corson}}, \bibinfo {author} {\bibfnamefont {R.}~\bibnamefont {Mallozzi}},
  \bibinfo {author} {\bibfnamefont {J.}~\bibnamefont {Orenstein}}, \bibinfo
  {author} {\bibfnamefont {J.~N.}\ \bibnamefont {Eckstein}}, \ and\ \bibinfo
  {author} {\bibfnamefont {I.}~\bibnamefont {Bozovic}},\ }\href@noop {}
  {\bibfield  {journal} {\bibinfo  {journal} {Nature}\ }\textbf {\bibinfo
  {volume} {398}},\ \bibinfo {pages} {221} (\bibinfo {year}
  {1999})}\BibitemShut {NoStop}%
\bibitem [{\citenamefont {Xu}\ \emph {et~al.}(2000)\citenamefont {Xu},
  \citenamefont {Ong}, \citenamefont {Wang}, \citenamefont {Kakeshita},\ and\
  \citenamefont {Uchida}}]{Xu:2000p609}%
  \BibitemOpen
  \bibfield  {author} {\bibinfo {author} {\bibfnamefont {Z.~A.}\ \bibnamefont
  {Xu}}, \bibinfo {author} {\bibfnamefont {N.~P.}\ \bibnamefont {Ong}},
  \bibinfo {author} {\bibfnamefont {Y.}~\bibnamefont {Wang}}, \bibinfo {author}
  {\bibfnamefont {T.}~\bibnamefont {Kakeshita}}, \ and\ \bibinfo {author}
  {\bibfnamefont {S.}~\bibnamefont {Uchida}},\ }\href@noop {} {\bibfield
  {journal} {\bibinfo  {journal} {Nature}\ }\textbf {\bibinfo {volume} {406}},\
  \bibinfo {pages} {486} (\bibinfo {year} {2000})}\BibitemShut {NoStop}%
\bibitem [{\citenamefont {Wang}\ \emph {et~al.}(2005)\citenamefont {Wang},
  \citenamefont {Li}, \citenamefont {Naughton}, \citenamefont {Gu},
  \citenamefont {Uchida},\ and\ \citenamefont {Ong}}]{Wang:2005p2400}%
  \BibitemOpen
  \bibfield  {author} {\bibinfo {author} {\bibfnamefont {Y.}~\bibnamefont
  {Wang}}, \bibinfo {author} {\bibfnamefont {L.}~\bibnamefont {Li}}, \bibinfo
  {author} {\bibfnamefont {M.~J.}\ \bibnamefont {Naughton}}, \bibinfo {author}
  {\bibfnamefont {G.~D.}\ \bibnamefont {Gu}}, \bibinfo {author} {\bibfnamefont
  {S.}~\bibnamefont {Uchida}}, \ and\ \bibinfo {author} {\bibfnamefont {N.~P.}\
  \bibnamefont {Ong}},\ }\href@noop {} {\bibfield  {journal} {\bibinfo
  {journal} {Phys.\ Rev.\ Lett.}\ }\textbf {\bibinfo {volume} {95}},\ \bibinfo
  {pages} {247002} (\bibinfo {year} {2005})}\BibitemShut {NoStop}%
\bibitem [{\citenamefont {Wang}\ \emph {et~al.}(2006)\citenamefont {Wang},
  \citenamefont {Li},\ and\ \citenamefont {Ong}}]{Wang:2006p185}%
  \BibitemOpen
  \bibfield  {author} {\bibinfo {author} {\bibfnamefont {Y.}~\bibnamefont
  {Wang}}, \bibinfo {author} {\bibfnamefont {L.}~\bibnamefont {Li}}, \ and\
  \bibinfo {author} {\bibfnamefont {N.~P.}\ \bibnamefont {Ong}},\ }\href@noop
  {} {\bibfield  {journal} {\bibinfo  {journal} {Phys.\ Rev.\ B}\ }\textbf
  {\bibinfo {volume} {73}},\ \bibinfo {pages} {024510} (\bibinfo {year}
  {2006})}\BibitemShut {NoStop}%
\bibitem [{\citenamefont {Bilbro}\ \emph
  {et~al.}(2011{\natexlab{b}})\citenamefont {Bilbro}, \citenamefont {Aguilar},
  \citenamefont {Logvenov}, \citenamefont {Pelleg}, \citenamefont
  {Boovi{\'c}},\ and\ \citenamefont {Armitage}}]{Bilbro:2011p2722}%
  \BibitemOpen
  \bibfield  {author} {\bibinfo {author} {\bibfnamefont {L.~S.}\ \bibnamefont
  {Bilbro}}, \bibinfo {author} {\bibfnamefont {R.~V.}\ \bibnamefont {Aguilar}},
  \bibinfo {author} {\bibfnamefont {G.}~\bibnamefont {Logvenov}}, \bibinfo
  {author} {\bibfnamefont {O.}~\bibnamefont {Pelleg}}, \bibinfo {author}
  {\bibfnamefont {I.}~\bibnamefont {Boovi{\'c}}}, \ and\ \bibinfo {author}
  {\bibfnamefont {N.~P.}\ \bibnamefont {Armitage}},\ }\href@noop {} {\bibfield
  {journal} {\bibinfo  {journal} {Nature Physics}\ }\textbf {\bibinfo {volume}
  {7}},\ \bibinfo {pages} {298} (\bibinfo {year}
  {2011}{\natexlab{b}})}\BibitemShut {NoStop}%
\bibitem [{\citenamefont {Bonn}\ \emph {et~al.}(1992)\citenamefont {Bonn},
  \citenamefont {Dosanjh}, \citenamefont {Liang},\ and\ \citenamefont
  {Hardy}}]{Bonn:1992p3021}%
  \BibitemOpen
  \bibfield  {author} {\bibinfo {author} {\bibfnamefont {D.~A.}\ \bibnamefont
  {Bonn}}, \bibinfo {author} {\bibfnamefont {P.}~\bibnamefont {Dosanjh}},
  \bibinfo {author} {\bibfnamefont {R.}~\bibnamefont {Liang}}, \ and\ \bibinfo
  {author} {\bibfnamefont {W.~N.}\ \bibnamefont {Hardy}},\ }\href@noop {}
  {\bibfield  {journal} {\bibinfo  {journal} {Phys.\ Rev.\ Lett.}\ }\textbf
  {\bibinfo {volume} {68}},\ \bibinfo {pages} {2390} (\bibinfo {year}
  {1992})}\BibitemShut {NoStop}%
\bibitem [{\citenamefont {Hirschfeld}\ \emph {et~al.}(1994)\citenamefont
  {Hirschfeld}, \citenamefont {Putikka},\ and\ \citenamefont
  {Scalapino}}]{HIRSCHFELD:1994p570}%
  \BibitemOpen
  \bibfield  {author} {\bibinfo {author} {\bibfnamefont {P.~J.}\ \bibnamefont
  {Hirschfeld}}, \bibinfo {author} {\bibfnamefont {W.~O.}\ \bibnamefont
  {Putikka}}, \ and\ \bibinfo {author} {\bibfnamefont {D.~J.}\ \bibnamefont
  {Scalapino}},\ }\href@noop {} {\bibfield  {journal} {\bibinfo  {journal}
  {Phys. Rev. B}\ }\textbf {\bibinfo {volume} {50}},\ \bibinfo {pages} {10250}
  (\bibinfo {year} {1994})}\BibitemShut {NoStop}%
\bibitem [{\citenamefont {Hosseini}\ \emph {et~al.}(1999)\citenamefont
  {Hosseini}, \citenamefont {Harris}, \citenamefont {Kamal}, \citenamefont
  {Dosanjh}, \citenamefont {Preston}, \citenamefont {Liang}, \citenamefont
  {Hardy},\ and\ \citenamefont {Bonn}}]{Hosseini:1999p383}%
  \BibitemOpen
  \bibfield  {author} {\bibinfo {author} {\bibfnamefont {A.}~\bibnamefont
  {Hosseini}}, \bibinfo {author} {\bibfnamefont {R.}~\bibnamefont {Harris}},
  \bibinfo {author} {\bibfnamefont {S.}~\bibnamefont {Kamal}}, \bibinfo
  {author} {\bibfnamefont {P.}~\bibnamefont {Dosanjh}}, \bibinfo {author}
  {\bibfnamefont {J.}~\bibnamefont {Preston}}, \bibinfo {author} {\bibfnamefont
  {R.}~\bibnamefont {Liang}}, \bibinfo {author} {\bibfnamefont {W.~N.}\
  \bibnamefont {Hardy}}, \ and\ \bibinfo {author} {\bibfnamefont {D.~A.}\
  \bibnamefont {Bonn}},\ }\href@noop {} {\bibfield  {journal} {\bibinfo
  {journal} {Phys.\ Rev.\ B}\ }\textbf {\bibinfo {volume} {60}},\ \bibinfo
  {pages} {1349} (\bibinfo {year} {1999})}\BibitemShut {NoStop}%
\bibitem [{\citenamefont {Turner}\ \emph {et~al.}(2003)\citenamefont {Turner},
  \citenamefont {Harris}, \citenamefont {Kamal}, \citenamefont {Hayden},
  \citenamefont {Broun}, \citenamefont {Morgan}, \citenamefont {Hosseini},
  \citenamefont {Dosanjh}, \citenamefont {Mullins}, \citenamefont {Preston},
  \citenamefont {Liang}, \citenamefont {Bonn},\ and\ \citenamefont
  {Hardy}}]{Turner:2003p331}%
  \BibitemOpen
  \bibfield  {author} {\bibinfo {author} {\bibfnamefont {P.~J.}\ \bibnamefont
  {Turner}}, \bibinfo {author} {\bibfnamefont {R.}~\bibnamefont {Harris}},
  \bibinfo {author} {\bibfnamefont {S.}~\bibnamefont {Kamal}}, \bibinfo
  {author} {\bibfnamefont {M.~E.}\ \bibnamefont {Hayden}}, \bibinfo {author}
  {\bibfnamefont {D.~M.}\ \bibnamefont {Broun}}, \bibinfo {author}
  {\bibfnamefont {D.~C.}\ \bibnamefont {Morgan}}, \bibinfo {author}
  {\bibfnamefont {A.}~\bibnamefont {Hosseini}}, \bibinfo {author}
  {\bibfnamefont {P.}~\bibnamefont {Dosanjh}}, \bibinfo {author} {\bibfnamefont
  {G.~K.}\ \bibnamefont {Mullins}}, \bibinfo {author} {\bibfnamefont {J.~S.}\
  \bibnamefont {Preston}}, \bibinfo {author} {\bibfnamefont {R.}~\bibnamefont
  {Liang}}, \bibinfo {author} {\bibfnamefont {D.~A.}\ \bibnamefont {Bonn}}, \
  and\ \bibinfo {author} {\bibfnamefont {W.~N.}\ \bibnamefont {Hardy}},\
  }\href@noop {} {\bibfield  {journal} {\bibinfo  {journal} {Phys.\ Rev.\
  Lett.}\ }\textbf {\bibinfo {volume} {90}},\ \bibinfo {pages} {237005}
  (\bibinfo {year} {2003})}\BibitemShut {NoStop}%
\bibitem [{\citenamefont {Harris}\ \emph {et~al.}(2006)\citenamefont {Harris},
  \citenamefont {Turner}, \citenamefont {Kamal}, \citenamefont {Hosseini},
  \citenamefont {Dosanjh}, \citenamefont {Mullins}, \citenamefont {Bobowski},
  \citenamefont {Bidinosti}, \citenamefont {Broun}, \citenamefont {Liang},
  \citenamefont {Hardy},\ and\ \citenamefont {Bonn}}]{Harris:2006p388}%
  \BibitemOpen
  \bibfield  {author} {\bibinfo {author} {\bibfnamefont {R.}~\bibnamefont
  {Harris}}, \bibinfo {author} {\bibfnamefont {P.~J.}\ \bibnamefont {Turner}},
  \bibinfo {author} {\bibfnamefont {S.}~\bibnamefont {Kamal}}, \bibinfo
  {author} {\bibfnamefont {A.~R.}\ \bibnamefont {Hosseini}}, \bibinfo {author}
  {\bibfnamefont {P.}~\bibnamefont {Dosanjh}}, \bibinfo {author} {\bibfnamefont
  {G.~K.}\ \bibnamefont {Mullins}}, \bibinfo {author} {\bibfnamefont {J.~S.}\
  \bibnamefont {Bobowski}}, \bibinfo {author} {\bibfnamefont {C.~P.}\
  \bibnamefont {Bidinosti}}, \bibinfo {author} {\bibfnamefont {D.~M.}\
  \bibnamefont {Broun}}, \bibinfo {author} {\bibfnamefont {R.}~\bibnamefont
  {Liang}}, \bibinfo {author} {\bibfnamefont {W.~N.}\ \bibnamefont {Hardy}}, \
  and\ \bibinfo {author} {\bibfnamefont {D.~A.}\ \bibnamefont {Bonn}},\
  }\href@noop {} {\bibfield  {journal} {\bibinfo  {journal} {Phys.\ Rev.\ B}\
  }\textbf {\bibinfo {volume} {74}},\ \bibinfo {pages} {104508} (\bibinfo
  {year} {2006})}\BibitemShut {NoStop}%
\bibitem [{\citenamefont {Ando}\ \emph {et~al.}(1995)\citenamefont {Ando},
  \citenamefont {Boebinger}, \citenamefont {Passner}, \citenamefont {Kimura},\
  and\ \citenamefont {Kishio}}]{Ando:1995p148}%
  \BibitemOpen
  \bibfield  {author} {\bibinfo {author} {\bibfnamefont {Y.}~\bibnamefont
  {Ando}}, \bibinfo {author} {\bibfnamefont {G.~S.}\ \bibnamefont {Boebinger}},
  \bibinfo {author} {\bibfnamefont {A.}~\bibnamefont {Passner}}, \bibinfo
  {author} {\bibfnamefont {T.}~\bibnamefont {Kimura}}, \ and\ \bibinfo {author}
  {\bibfnamefont {K.}~\bibnamefont {Kishio}},\ }\href@noop {} {\bibfield
  {journal} {\bibinfo  {journal} {Phys.\ Rev.\ Lett.}\ }\textbf {\bibinfo
  {volume} {75}},\ \bibinfo {pages} {4662} (\bibinfo {year}
  {1995})}\BibitemShut {NoStop}%
\bibitem [{\citenamefont {Boebinger}\ \emph {et~al.}(1996)\citenamefont
  {Boebinger}, \citenamefont {Ando}, \citenamefont {Passner}, \citenamefont
  {Kimura}, \citenamefont {Okuya}, \citenamefont {Shimoyama}, \citenamefont
  {Kishio}, \citenamefont {Tamasaku}, \citenamefont {Ichikawa},\ and\
  \citenamefont {Uchida}}]{Boebinger:1996p147}%
  \BibitemOpen
  \bibfield  {author} {\bibinfo {author} {\bibfnamefont {G.~S.}\ \bibnamefont
  {Boebinger}}, \bibinfo {author} {\bibfnamefont {Y.}~\bibnamefont {Ando}},
  \bibinfo {author} {\bibfnamefont {A.}~\bibnamefont {Passner}}, \bibinfo
  {author} {\bibfnamefont {T.}~\bibnamefont {Kimura}}, \bibinfo {author}
  {\bibfnamefont {M.}~\bibnamefont {Okuya}}, \bibinfo {author} {\bibfnamefont
  {J.}~\bibnamefont {Shimoyama}}, \bibinfo {author} {\bibfnamefont
  {K.}~\bibnamefont {Kishio}}, \bibinfo {author} {\bibfnamefont
  {K.}~\bibnamefont {Tamasaku}}, \bibinfo {author} {\bibfnamefont
  {N.}~\bibnamefont {Ichikawa}}, \ and\ \bibinfo {author} {\bibfnamefont
  {S.}~\bibnamefont {Uchida}},\ }\href@noop {} {\bibfield  {journal} {\bibinfo
  {journal} {Phys.\ Rev.\ Lett.}\ }\textbf {\bibinfo {volume} {77}},\ \bibinfo
  {pages} {5417} (\bibinfo {year} {1996})}\BibitemShut {NoStop}%
\bibitem [{\citenamefont {Vinen}\ and\ \citenamefont
  {Warren}(1967)}]{Vinen:1967}%
  \BibitemOpen
  \bibfield  {author} {\bibinfo {author} {\bibfnamefont {W.~F.}\ \bibnamefont
  {Vinen}}\ and\ \bibinfo {author} {\bibfnamefont {A.~C.}\ \bibnamefont
  {Warren}},\ }\href@noop {} {\bibfield  {journal} {\bibinfo  {journal} {Proc.\
  Phys.\ Soc.}\ }\textbf {\bibinfo {volume} {91}},\ \bibinfo {pages} {409}
  (\bibinfo {year} {1967})}\BibitemShut {NoStop}%
\bibitem [{\citenamefont {Gittleman}\ and\ \citenamefont
  {Rosenblum}(1968)}]{JIGittleman:1968p172}%
  \BibitemOpen
  \bibfield  {author} {\bibinfo {author} {\bibfnamefont {J.~I.}\ \bibnamefont
  {Gittleman}}\ and\ \bibinfo {author} {\bibfnamefont {B.}~\bibnamefont
  {Rosenblum}},\ }\href@noop {} {\bibfield  {journal} {\bibinfo  {journal} {J.\
  Appl.\ Phys.}\ }\textbf {\bibinfo {volume} {39}},\ \bibinfo {pages} {2617}
  (\bibinfo {year} {1968})}\BibitemShut {NoStop}%
\bibitem [{\citenamefont {Kopnin}\ and\ \citenamefont
  {Kravtsov}(1976{\natexlab{a}})}]{KOPNIN:1976ua}%
  \BibitemOpen
  \bibfield  {author} {\bibinfo {author} {\bibfnamefont {N.~B.}\ \bibnamefont
  {Kopnin}}\ and\ \bibinfo {author} {\bibfnamefont {V.~E.}\ \bibnamefont
  {Kravtsov}},\ }\href@noop {} {\bibfield  {journal} {\bibinfo  {journal}
  {Sov.\ Phys.\ JETP}\ }\textbf {\bibinfo {volume} {44}},\ \bibinfo {pages}
  {861} (\bibinfo {year} {1976}{\natexlab{a}})}\BibitemShut {NoStop}%
\bibitem [{\citenamefont {Kopnin}\ and\ \citenamefont
  {Salomaa}(1991)}]{KOPNIN:1991va}%
  \BibitemOpen
  \bibfield  {author} {\bibinfo {author} {\bibfnamefont {N.~B.}\ \bibnamefont
  {Kopnin}}\ and\ \bibinfo {author} {\bibfnamefont {M.~M.}\ \bibnamefont
  {Salomaa}},\ }\href@noop {} {\bibfield  {journal} {\bibinfo  {journal}
  {Phys.\ Rev.\ B}\ }\textbf {\bibinfo {volume} {44}},\ \bibinfo {pages} {9667}
  (\bibinfo {year} {1991})}\BibitemShut {NoStop}%
\bibitem [{\citenamefont {Hsu}(1993)}]{Hsu:1993ih}%
  \BibitemOpen
  \bibfield  {author} {\bibinfo {author} {\bibfnamefont {T.~C.}\ \bibnamefont
  {Hsu}},\ }\href@noop {} {\bibfield  {journal} {\bibinfo  {journal} {Physica
  C}\ }\textbf {\bibinfo {volume} {213}},\ \bibinfo {pages} {305} (\bibinfo
  {year} {1993})}\BibitemShut {NoStop}%
\bibitem [{\citenamefont {Vinokur}\ \emph {et~al.}(1993)\citenamefont
  {Vinokur}, \citenamefont {Geshkenbein}, \citenamefont {Feigel'man},\ and\
  \citenamefont {Blatter}}]{Vinokur:1993to}%
  \BibitemOpen
  \bibfield  {author} {\bibinfo {author} {\bibfnamefont {V.~M.}\ \bibnamefont
  {Vinokur}}, \bibinfo {author} {\bibfnamefont {V.~B.}\ \bibnamefont
  {Geshkenbein}}, \bibinfo {author} {\bibfnamefont {M.~V.}\ \bibnamefont
  {Feigel'man}}, \ and\ \bibinfo {author} {\bibfnamefont {G.}~\bibnamefont
  {Blatter}},\ }\href@noop {} {\bibfield  {journal} {\bibinfo  {journal}
  {Phys.\ Rev.\ Lett.}\ }\textbf {\bibinfo {volume} {71}},\ \bibinfo {pages}
  {1242} (\bibinfo {year} {1993})}\BibitemShut {NoStop}%
\bibitem [{\citenamefont {Blatter}\ \emph {et~al.}(1994)\citenamefont
  {Blatter}, \citenamefont {Feigelman}, \citenamefont {Geshkenbein},
  \citenamefont {Larkin},\ and\ \citenamefont {Vinokur}}]{BLATTER:1994p494}%
  \BibitemOpen
  \bibfield  {author} {\bibinfo {author} {\bibfnamefont {G.}~\bibnamefont
  {Blatter}}, \bibinfo {author} {\bibfnamefont {M.}~\bibnamefont {Feigelman}},
  \bibinfo {author} {\bibfnamefont {V.}~\bibnamefont {Geshkenbein}}, \bibinfo
  {author} {\bibfnamefont {A.}~\bibnamefont {Larkin}}, \ and\ \bibinfo {author}
  {\bibfnamefont {V.}~\bibnamefont {Vinokur}},\ }\href@noop {} {\bibfield
  {journal} {\bibinfo  {journal} {Rev.\ Mod.\ Phys.}\ }\textbf {\bibinfo
  {volume} {66}},\ \bibinfo {pages} {1125} (\bibinfo {year}
  {1994})}\BibitemShut {NoStop}%
\bibitem [{\citenamefont {Parks}\ \emph {et~al.}(1995)\citenamefont {Parks},
  \citenamefont {Spielman}, \citenamefont {Orenstein}, \citenamefont {Nemeth},
  \citenamefont {Ludwig}, \citenamefont {Clarke}, \citenamefont {Merchant},\
  and\ \citenamefont {Lew}}]{Parks:1995p189}%
  \BibitemOpen
  \bibfield  {author} {\bibinfo {author} {\bibfnamefont {B.}~\bibnamefont
  {Parks}}, \bibinfo {author} {\bibfnamefont {S.}~\bibnamefont {Spielman}},
  \bibinfo {author} {\bibfnamefont {J.}~\bibnamefont {Orenstein}}, \bibinfo
  {author} {\bibfnamefont {D.~T.}\ \bibnamefont {Nemeth}}, \bibinfo {author}
  {\bibfnamefont {F.}~\bibnamefont {Ludwig}}, \bibinfo {author} {\bibfnamefont
  {J.}~\bibnamefont {Clarke}}, \bibinfo {author} {\bibfnamefont
  {P.}~\bibnamefont {Merchant}}, \ and\ \bibinfo {author} {\bibfnamefont
  {D.~J.}\ \bibnamefont {Lew}},\ }\href@noop {} {\bibfield  {journal} {\bibinfo
   {journal} {Phys.\ Rev.\ Lett.}\ }\textbf {\bibinfo {volume} {74}},\ \bibinfo
  {pages} {3265} (\bibinfo {year} {1995})}\BibitemShut {NoStop}%
\bibitem [{\citenamefont {Golosovsky}\ \emph {et~al.}(1996)\citenamefont
  {Golosovsky}, \citenamefont {Tsindlekht},\ and\ \citenamefont
  {Davidov}}]{GOLOSOVSKY:1996p1}%
  \BibitemOpen
  \bibfield  {author} {\bibinfo {author} {\bibfnamefont {M.}~\bibnamefont
  {Golosovsky}}, \bibinfo {author} {\bibfnamefont {M.}~\bibnamefont
  {Tsindlekht}}, \ and\ \bibinfo {author} {\bibfnamefont {D.}~\bibnamefont
  {Davidov}},\ }\href@noop {} {\bibfield  {journal} {\bibinfo  {journal} {Sup.\
  Sci.\ Tech}\ }\textbf {\bibinfo {volume} {9}},\ \bibinfo {pages} {1}
  (\bibinfo {year} {1996})}\BibitemShut {NoStop}%
\bibitem [{\citenamefont {Kopnin}\ and\ \citenamefont
  {Kravtsov}(1976{\natexlab{b}})}]{KOPNIN:1976vi}%
  \BibitemOpen
  \bibfield  {author} {\bibinfo {author} {\bibfnamefont {N.~B.}\ \bibnamefont
  {Kopnin}}\ and\ \bibinfo {author} {\bibfnamefont {V.~E.}\ \bibnamefont
  {Kravtsov}},\ }\href@noop {} {\bibfield  {journal} {\bibinfo  {journal}
  {JETP.\ Lett.}\ }\textbf {\bibinfo {volume} {23}},\ \bibinfo {pages} {578}
  (\bibinfo {year} {1976}{\natexlab{b}})}\BibitemShut {NoStop}%
\bibitem [{\citenamefont {Josephson}(1965)}]{Josephson:1965bo}%
  \BibitemOpen
  \bibfield  {author} {\bibinfo {author} {\bibfnamefont {B.~D.}\ \bibnamefont
  {Josephson}},\ }\href@noop {} {\bibfield  {journal} {\bibinfo  {journal}
  {Phys.\ Lett.}\ }\textbf {\bibinfo {volume} {16}},\ \bibinfo {pages} {242}
  (\bibinfo {year} {1965})}\BibitemShut {NoStop}%
\bibitem [{\citenamefont {Larkin}\ and\ \citenamefont
  {Ovchinnikov}(1979)}]{Larkin:1979ta}%
  \BibitemOpen
  \bibfield  {author} {\bibinfo {author} {\bibfnamefont {A.~I.}\ \bibnamefont
  {Larkin}}\ and\ \bibinfo {author} {\bibfnamefont {Y.~N.}\ \bibnamefont
  {Ovchinnikov}},\ }\href@noop {} {\bibfield  {journal} {\bibinfo  {journal}
  {J.\ Low Temp.\ Phys.}\ }\textbf {\bibinfo {volume} {34}},\ \bibinfo {pages}
  {409} (\bibinfo {year} {1979})}\BibitemShut {NoStop}%
\bibitem [{\citenamefont {Kim}\ \emph {et~al.}(1965)\citenamefont {Kim},
  \citenamefont {Hempstead},\ and\ \citenamefont {Strnad}}]{Kim:1965vl}%
  \BibitemOpen
  \bibfield  {author} {\bibinfo {author} {\bibfnamefont {Y.~B.}\ \bibnamefont
  {Kim}}, \bibinfo {author} {\bibfnamefont {C.~F.}\ \bibnamefont {Hempstead}},
  \ and\ \bibinfo {author} {\bibfnamefont {A.~R.}\ \bibnamefont {Strnad}},\
  }\href@noop {} {\bibfield  {journal} {\bibinfo  {journal} {Phys.\ Rev.}\
  }\textbf {\bibinfo {volume} {139}},\ \bibinfo {pages} {A1163} (\bibinfo
  {year} {1965})}\BibitemShut {NoStop}%
\bibitem [{\citenamefont {Kunchur}\ \emph {et~al.}(1993)\citenamefont
  {Kunchur}, \citenamefont {Christen},\ and\ \citenamefont
  {Phillips}}]{Kunchur:1993ie}%
  \BibitemOpen
  \bibfield  {author} {\bibinfo {author} {\bibfnamefont {M.~N.}\ \bibnamefont
  {Kunchur}}, \bibinfo {author} {\bibfnamefont {D.~K.}\ \bibnamefont
  {Christen}}, \ and\ \bibinfo {author} {\bibfnamefont {J.~M.}\ \bibnamefont
  {Phillips}},\ }\href@noop {} {\bibfield  {journal} {\bibinfo  {journal}
  {Phys.\ Rev.\ Lett.}\ }\textbf {\bibinfo {volume} {70}},\ \bibinfo {pages}
  {998} (\bibinfo {year} {1993})}\BibitemShut {NoStop}%
\bibitem [{\citenamefont {Owliaei}\ \emph {et~al.}(1992)\citenamefont
  {Owliaei}, \citenamefont {Sridhar},\ and\ \citenamefont
  {Talvacchio}}]{Owliaei:1992tt}%
  \BibitemOpen
  \bibfield  {author} {\bibinfo {author} {\bibfnamefont {J.}~\bibnamefont
  {Owliaei}}, \bibinfo {author} {\bibfnamefont {S.}~\bibnamefont {Sridhar}}, \
  and\ \bibinfo {author} {\bibfnamefont {J.}~\bibnamefont {Talvacchio}},\
  }\href@noop {} {\bibfield  {journal} {\bibinfo  {journal} {Phys.\ Rev.\
  Lett.}\ }\textbf {\bibinfo {volume} {69}},\ \bibinfo {pages} {3366} (\bibinfo
  {year} {1992})}\BibitemShut {NoStop}%
\bibitem [{\citenamefont {Pambianchi}\ \emph {et~al.}(1993)\citenamefont
  {Pambianchi}, \citenamefont {Wu}, \citenamefont {Ganapathi},\ and\
  \citenamefont {Anlage}}]{Pambianchi:1993jv}%
  \BibitemOpen
  \bibfield  {author} {\bibinfo {author} {\bibfnamefont {M.~S.}\ \bibnamefont
  {Pambianchi}}, \bibinfo {author} {\bibfnamefont {D.~H.}\ \bibnamefont {Wu}},
  \bibinfo {author} {\bibfnamefont {L.}~\bibnamefont {Ganapathi}}, \ and\
  \bibinfo {author} {\bibfnamefont {S.~M.}\ \bibnamefont {Anlage}},\
  }\href@noop {} {\bibfield  {journal} {\bibinfo  {journal} {IEEE Trans.\
  Appl.\ Supercond.}\ }\textbf {\bibinfo {volume} {3}},\ \bibinfo {pages}
  {2774} (\bibinfo {year} {1993})}\BibitemShut {NoStop}%
\bibitem [{\citenamefont {Morgan}(1993)}]{DCMorgan1993}%
  \BibitemOpen
  \bibfield  {author} {\bibinfo {author} {\bibfnamefont {D.~C.}\ \bibnamefont
  {Morgan}},\ }\emph {\bibinfo {title} {Studies of the Flux Flow Resistivity in
  YBa$_2$Cu$_3$O$_{6.95}$ by Microwave Techniques}},\ \href@noop {} {Ph.D.
  thesis},\ \bibinfo  {school} {University of British Columbia} (\bibinfo
  {year} {1993})\BibitemShut {NoStop}%
\bibitem [{\citenamefont {Morgan}\ \emph {et~al.}(1994)\citenamefont {Morgan},
  \citenamefont {Zhang}, \citenamefont {Bonn}, \citenamefont {Liang},
  \citenamefont {Hardy}, \citenamefont {Kallin},\ and\ \citenamefont
  {Berlinsky}}]{Morgan:1994p2404}%
  \BibitemOpen
  \bibfield  {author} {\bibinfo {author} {\bibfnamefont {D.~C.}\ \bibnamefont
  {Morgan}}, \bibinfo {author} {\bibfnamefont {K.}~\bibnamefont {Zhang}},
  \bibinfo {author} {\bibfnamefont {D.~A.}\ \bibnamefont {Bonn}}, \bibinfo
  {author} {\bibfnamefont {R.}~\bibnamefont {Liang}}, \bibinfo {author}
  {\bibfnamefont {W.~N.}\ \bibnamefont {Hardy}}, \bibinfo {author}
  {\bibfnamefont {C.}~\bibnamefont {Kallin}}, \ and\ \bibinfo {author}
  {\bibfnamefont {A.~J.}\ \bibnamefont {Berlinsky}},\ }\href@noop {} {\bibfield
   {journal} {\bibinfo  {journal} {Physica C}\ }\textbf {\bibinfo {volume}
  {235--240}},\ \bibinfo {pages} {2015} (\bibinfo {year} {1994})}\BibitemShut
  {NoStop}%
\bibitem [{\citenamefont {Revenaz}\ \emph {et~al.}(1994)\citenamefont
  {Revenaz}, \citenamefont {Oates}, \citenamefont {Labbe-Lavigne},
  \citenamefont {Dresselhaus},\ and\ \citenamefont
  {Dresselhaus}}]{Revenaz:1994fg}%
  \BibitemOpen
  \bibfield  {author} {\bibinfo {author} {\bibfnamefont {S.}~\bibnamefont
  {Revenaz}}, \bibinfo {author} {\bibfnamefont {D.~E.}\ \bibnamefont {Oates}},
  \bibinfo {author} {\bibfnamefont {D.}~\bibnamefont {Labbe-Lavigne}}, \bibinfo
  {author} {\bibfnamefont {G.}~\bibnamefont {Dresselhaus}}, \ and\ \bibinfo
  {author} {\bibfnamefont {M.~S.}\ \bibnamefont {Dresselhaus}},\ }\href@noop {}
  {\bibfield  {journal} {\bibinfo  {journal} {Phys.\ Rev.\ B}\ }\textbf
  {\bibinfo {volume} {50}},\ \bibinfo {pages} {1178} (\bibinfo {year}
  {1994})}\BibitemShut {NoStop}%
\bibitem [{\citenamefont {Golosovsky}\ \emph {et~al.}(1994)\citenamefont
  {Golosovsky}, \citenamefont {Tsindlekht}, \citenamefont {Chayet},\ and\
  \citenamefont {Davidov}}]{GOLOSOVSKY:1994p169}%
  \BibitemOpen
  \bibfield  {author} {\bibinfo {author} {\bibfnamefont {M.}~\bibnamefont
  {Golosovsky}}, \bibinfo {author} {\bibfnamefont {M.}~\bibnamefont
  {Tsindlekht}}, \bibinfo {author} {\bibfnamefont {H.}~\bibnamefont {Chayet}},
  \ and\ \bibinfo {author} {\bibfnamefont {D.}~\bibnamefont {Davidov}},\
  }\href@noop {} {\bibfield  {journal} {\bibinfo  {journal} {Phys.\ Rev.\ B}\
  }\textbf {\bibinfo {volume} {50}},\ \bibinfo {pages} {470} (\bibinfo {year}
  {1994})}\BibitemShut {NoStop}%
\bibitem [{\citenamefont {Powell}\ \emph {et~al.}(1996)\citenamefont {Powell},
  \citenamefont {Porch}, \citenamefont {Wellh{\"o}fer}, \citenamefont
  {Humphreys},\ and\ \citenamefont {Gough}}]{Powell:1996tb}%
  \BibitemOpen
  \bibfield  {author} {\bibinfo {author} {\bibfnamefont {J.~R.}\ \bibnamefont
  {Powell}}, \bibinfo {author} {\bibfnamefont {A.}~\bibnamefont {Porch}},
  \bibinfo {author} {\bibfnamefont {F.}~\bibnamefont {Wellh{\"o}fer}}, \bibinfo
  {author} {\bibfnamefont {R.~G.}\ \bibnamefont {Humphreys}}, \ and\ \bibinfo
  {author} {\bibfnamefont {C.~E.}\ \bibnamefont {Gough}},\ }\href@noop {}
  {\bibfield  {journal} {\bibinfo  {journal} {Czech J.\ Phys.}\ }\textbf
  {\bibinfo {volume} {46}},\ \bibinfo {pages} {1089} (\bibinfo {year}
  {1996})}\BibitemShut {NoStop}%
\bibitem [{\citenamefont {Belk}\ \emph {et~al.}(1997)\citenamefont {Belk},
  \citenamefont {Oates}, \citenamefont {Feld}, \citenamefont {Dresselhaus},\
  and\ \citenamefont {Dresselhaus}}]{Belk:1997hn}%
  \BibitemOpen
  \bibfield  {author} {\bibinfo {author} {\bibfnamefont {N.}~\bibnamefont
  {Belk}}, \bibinfo {author} {\bibfnamefont {D.~E.}\ \bibnamefont {Oates}},
  \bibinfo {author} {\bibfnamefont {D.~A.}\ \bibnamefont {Feld}}, \bibinfo
  {author} {\bibfnamefont {G.}~\bibnamefont {Dresselhaus}}, \ and\ \bibinfo
  {author} {\bibfnamefont {M.~S.}\ \bibnamefont {Dresselhaus}},\ }\href@noop {}
  {\bibfield  {journal} {\bibinfo  {journal} {Phys.\ Rev.\ B}\ }\textbf
  {\bibinfo {volume} {56}},\ \bibinfo {pages} {11966} (\bibinfo {year}
  {1997})}\BibitemShut {NoStop}%
\bibitem [{\citenamefont {Ghosh}\ \emph {et~al.}(1999)\citenamefont {Ghosh},
  \citenamefont {Cohen},\ and\ \citenamefont {Gallop}}]{Ghosh:1997p170}%
  \BibitemOpen
  \bibfield  {author} {\bibinfo {author} {\bibfnamefont {I.~S.}\ \bibnamefont
  {Ghosh}}, \bibinfo {author} {\bibfnamefont {L.~F.}\ \bibnamefont {Cohen}}, \
  and\ \bibinfo {author} {\bibfnamefont {J.~C.}\ \bibnamefont {Gallop}},\
  }\href@noop {} {\bibfield  {journal} {\bibinfo  {journal} {Supercond.\ Sci.\
  Tech.}\ }\textbf {\bibinfo {volume} {10}},\ \bibinfo {pages} {936} (\bibinfo
  {year} {1999})}\BibitemShut {NoStop}%
\bibitem [{\citenamefont {Hanaguri}\ \emph {et~al.}(1999)\citenamefont
  {Hanaguri}, \citenamefont {Tsuboi}, \citenamefont {Tsuchiya}, \citenamefont
  {Sasaki},\ and\ \citenamefont {Maeda}}]{Hanaguri:1999fn}%
  \BibitemOpen
  \bibfield  {author} {\bibinfo {author} {\bibfnamefont {T.}~\bibnamefont
  {Hanaguri}}, \bibinfo {author} {\bibfnamefont {T.}~\bibnamefont {Tsuboi}},
  \bibinfo {author} {\bibfnamefont {Y.}~\bibnamefont {Tsuchiya}}, \bibinfo
  {author} {\bibfnamefont {K.~I.}\ \bibnamefont {Sasaki}}, \ and\ \bibinfo
  {author} {\bibfnamefont {A.}~\bibnamefont {Maeda}},\ }\href@noop {}
  {\bibfield  {journal} {\bibinfo  {journal} {Phys.\ Rev.\ Lett.}\ }\textbf
  {\bibinfo {volume} {82}},\ \bibinfo {pages} {1273} (\bibinfo {year}
  {1999})}\BibitemShut {NoStop}%
\bibitem [{\citenamefont {Silva}\ \emph {et~al.}(2000)\citenamefont {Silva},
  \citenamefont {Fastampa}, \citenamefont {Giura}, \citenamefont {Marcon},
  \citenamefont {Neri},\ and\ \citenamefont {Sarti}}]{Silva:2000ia}%
  \BibitemOpen
  \bibfield  {author} {\bibinfo {author} {\bibfnamefont {E.}~\bibnamefont
  {Silva}}, \bibinfo {author} {\bibfnamefont {R.}~\bibnamefont {Fastampa}},
  \bibinfo {author} {\bibfnamefont {M.}~\bibnamefont {Giura}}, \bibinfo
  {author} {\bibfnamefont {R.}~\bibnamefont {Marcon}}, \bibinfo {author}
  {\bibfnamefont {D.}~\bibnamefont {Neri}}, \ and\ \bibinfo {author}
  {\bibfnamefont {S.}~\bibnamefont {Sarti}},\ }\href@noop {} {\bibfield
  {journal} {\bibinfo  {journal} {Supercond.\ Sci.\ Tech.}\ }\textbf {\bibinfo
  {volume} {13}},\ \bibinfo {pages} {1186} (\bibinfo {year}
  {2000})}\BibitemShut {NoStop}%
\bibitem [{\citenamefont {Tsuchiya}\ \emph {et~al.}(2001)\citenamefont
  {Tsuchiya}, \citenamefont {Iwaya}, \citenamefont {Kinoshita}, \citenamefont
  {Hanaguri}, \citenamefont {Kitano}, \citenamefont {Maeda}, \citenamefont
  {Shibata}, \citenamefont {Nishizaki},\ and\ \citenamefont
  {Kobayashi}}]{Tsuchiya:2001p200}%
  \BibitemOpen
  \bibfield  {author} {\bibinfo {author} {\bibfnamefont {Y.}~\bibnamefont
  {Tsuchiya}}, \bibinfo {author} {\bibfnamefont {K.}~\bibnamefont {Iwaya}},
  \bibinfo {author} {\bibfnamefont {K.}~\bibnamefont {Kinoshita}}, \bibinfo
  {author} {\bibfnamefont {T.}~\bibnamefont {Hanaguri}}, \bibinfo {author}
  {\bibfnamefont {H.}~\bibnamefont {Kitano}}, \bibinfo {author} {\bibfnamefont
  {A.}~\bibnamefont {Maeda}}, \bibinfo {author} {\bibfnamefont
  {K.}~\bibnamefont {Shibata}}, \bibinfo {author} {\bibfnamefont
  {T.}~\bibnamefont {Nishizaki}}, \ and\ \bibinfo {author} {\bibfnamefont
  {N.}~\bibnamefont {Kobayashi}},\ }\href@noop {} {\bibfield  {journal}
  {\bibinfo  {journal} {Phys.\ Rev.\ B}\ }\textbf {\bibinfo {volume} {63}},\
  \bibinfo {pages} {184517} (\bibinfo {year} {2001})}\BibitemShut {NoStop}%
\bibitem [{\citenamefont {Matsuda}\ \emph {et~al.}(2002)\citenamefont
  {Matsuda}, \citenamefont {Shibata}, \citenamefont {Izawa}, \citenamefont
  {Ikuta}, \citenamefont {Hasegawa},\ and\ \citenamefont
  {Kato}}]{MATSUDA:2002p2718}%
  \BibitemOpen
  \bibfield  {author} {\bibinfo {author} {\bibfnamefont {Y.}~\bibnamefont
  {Matsuda}}, \bibinfo {author} {\bibfnamefont {A.}~\bibnamefont {Shibata}},
  \bibinfo {author} {\bibfnamefont {K.}~\bibnamefont {Izawa}}, \bibinfo
  {author} {\bibfnamefont {H.}~\bibnamefont {Ikuta}}, \bibinfo {author}
  {\bibfnamefont {M.}~\bibnamefont {Hasegawa}}, \ and\ \bibinfo {author}
  {\bibfnamefont {Y.}~\bibnamefont {Kato}},\ }\href@noop {} {\bibfield
  {journal} {\bibinfo  {journal} {Phys.\ Rev.\ B}\ }\textbf {\bibinfo {volume}
  {66}},\ \bibinfo {pages} {014527} (\bibinfo {year} {2002})}\BibitemShut
  {NoStop}%
\bibitem [{\citenamefont {Silva}\ \emph {et~al.}(2004)\citenamefont {Silva},
  \citenamefont {Marcon}, \citenamefont {Muzzi}, \citenamefont {Pompeo},
  \citenamefont {Fastampa}, \citenamefont {Giura}, \citenamefont {Sarti},
  \citenamefont {Boffa}, \citenamefont {Cucolo},\ and\ \citenamefont
  {Cucolo}}]{Silva:2004bh}%
  \BibitemOpen
  \bibfield  {author} {\bibinfo {author} {\bibfnamefont {E.}~\bibnamefont
  {Silva}}, \bibinfo {author} {\bibfnamefont {R.}~\bibnamefont {Marcon}},
  \bibinfo {author} {\bibfnamefont {L.}~\bibnamefont {Muzzi}}, \bibinfo
  {author} {\bibfnamefont {N.}~\bibnamefont {Pompeo}}, \bibinfo {author}
  {\bibfnamefont {R.}~\bibnamefont {Fastampa}}, \bibinfo {author}
  {\bibfnamefont {M.}~\bibnamefont {Giura}}, \bibinfo {author} {\bibfnamefont
  {S.}~\bibnamefont {Sarti}}, \bibinfo {author} {\bibfnamefont
  {M.}~\bibnamefont {Boffa}}, \bibinfo {author} {\bibfnamefont {A.~M.}\
  \bibnamefont {Cucolo}}, \ and\ \bibinfo {author} {\bibfnamefont {M.~C.}\
  \bibnamefont {Cucolo}},\ }\href@noop {} {\bibfield  {journal} {\bibinfo
  {journal} {Physica C}\ }\textbf {\bibinfo {volume} {404}},\ \bibinfo {pages}
  {350} (\bibinfo {year} {2004})}\BibitemShut {NoStop}%
\bibitem [{\citenamefont {Morgan}(2005)}]{BMorgan2005}%
  \BibitemOpen
  \bibfield  {author} {\bibinfo {author} {\bibfnamefont {B.}~\bibnamefont
  {Morgan}},\ }\emph {\bibinfo {title} {Microwave Surface Impedance of
  YBa$_2$Cu$_3$O$_{6.95}$ in the Mixed State}},\ \href@noop {} {Ph.D. thesis},\
  \bibinfo  {school} {University of Cambridge} (\bibinfo {year}
  {2005})\BibitemShut {NoStop}%
\bibitem [{\citenamefont {Pompeo}\ \emph {et~al.}(2008)\citenamefont {Pompeo},
  \citenamefont {Silva}, \citenamefont {Ausloos},\ and\ \citenamefont
  {Cloots}}]{Pompeo:2008cd}%
  \BibitemOpen
  \bibfield  {author} {\bibinfo {author} {\bibfnamefont {N.}~\bibnamefont
  {Pompeo}}, \bibinfo {author} {\bibfnamefont {E.}~\bibnamefont {Silva}},
  \bibinfo {author} {\bibfnamefont {M.}~\bibnamefont {Ausloos}}, \ and\
  \bibinfo {author} {\bibfnamefont {R.}~\bibnamefont {Cloots}},\ }\href@noop {}
  {\bibfield  {journal} {\bibinfo  {journal} {J.\ Appl.\ Phys.}\ }\textbf
  {\bibinfo {volume} {103}},\ \bibinfo {pages} {103912} (\bibinfo {year}
  {2008})}\BibitemShut {NoStop}%
\bibitem [{\citenamefont {Narduzzo}\ \emph {et~al.}(2008)\citenamefont
  {Narduzzo}, \citenamefont {Grbi{\'c}}, \citenamefont {Pozek}, \citenamefont
  {Dul{\v c}i{\'c}}, \citenamefont {Paar}, \citenamefont {Kondrat},
  \citenamefont {Hess}, \citenamefont {Hellmann}, \citenamefont {Klingeler},
  \citenamefont {Werner}, \citenamefont {K{\"o}hler}, \citenamefont {Behr},\
  and\ \citenamefont {B{\"u}chner}}]{Narduzzo:2008io}%
  \BibitemOpen
  \bibfield  {author} {\bibinfo {author} {\bibfnamefont {A.}~\bibnamefont
  {Narduzzo}}, \bibinfo {author} {\bibfnamefont {M.}~\bibnamefont {Grbi{\'c}}},
  \bibinfo {author} {\bibfnamefont {M.}~\bibnamefont {Pozek}}, \bibinfo
  {author} {\bibfnamefont {A.}~\bibnamefont {Dul{\v c}i{\'c}}}, \bibinfo
  {author} {\bibfnamefont {D.}~\bibnamefont {Paar}}, \bibinfo {author}
  {\bibfnamefont {A.}~\bibnamefont {Kondrat}}, \bibinfo {author} {\bibfnamefont
  {C.}~\bibnamefont {Hess}}, \bibinfo {author} {\bibfnamefont {I.}~\bibnamefont
  {Hellmann}}, \bibinfo {author} {\bibfnamefont {R.}~\bibnamefont {Klingeler}},
  \bibinfo {author} {\bibfnamefont {J.}~\bibnamefont {Werner}}, \bibinfo
  {author} {\bibfnamefont {A.}~\bibnamefont {K{\"o}hler}}, \bibinfo {author}
  {\bibfnamefont {G.}~\bibnamefont {Behr}}, \ and\ \bibinfo {author}
  {\bibfnamefont {B.}~\bibnamefont {B{\"u}chner}},\ }\href@noop {} {\bibfield
  {journal} {\bibinfo  {journal} {Phys.\ Rev.\ B}\ }\textbf {\bibinfo {volume}
  {78}},\ \bibinfo {pages} {012507} (\bibinfo {year} {2008})}\BibitemShut
  {NoStop}%
\bibitem [{\citenamefont {Ikebe}\ \emph {et~al.}(2009)\citenamefont {Ikebe},
  \citenamefont {Shimano}, \citenamefont {Ikeda}, \citenamefont {Fukumura},\
  and\ \citenamefont {Kawasaki}}]{Ikebe:2009it}%
  \BibitemOpen
  \bibfield  {author} {\bibinfo {author} {\bibfnamefont {Y.}~\bibnamefont
  {Ikebe}}, \bibinfo {author} {\bibfnamefont {R.}~\bibnamefont {Shimano}},
  \bibinfo {author} {\bibfnamefont {M.}~\bibnamefont {Ikeda}}, \bibinfo
  {author} {\bibfnamefont {T.}~\bibnamefont {Fukumura}}, \ and\ \bibinfo
  {author} {\bibfnamefont {M.}~\bibnamefont {Kawasaki}},\ }\href@noop {}
  {\bibfield  {journal} {\bibinfo  {journal} {Phys.\ Rev.\ B}\ }\textbf
  {\bibinfo {volume} {79}},\ \bibinfo {pages} {174525} (\bibinfo {year}
  {2009})}\BibitemShut {NoStop}%
\bibitem [{\citenamefont {Pompeo}\ and\ \citenamefont
  {Silva}(2008)}]{Pompeo:2008p2717}%
  \BibitemOpen
  \bibfield  {author} {\bibinfo {author} {\bibfnamefont {N.}~\bibnamefont
  {Pompeo}}\ and\ \bibinfo {author} {\bibfnamefont {E.}~\bibnamefont {Silva}},\
  }\href@noop {} {\bibfield  {journal} {\bibinfo  {journal} {Phys.\ Rev.\ B}\
  }\textbf {\bibinfo {volume} {78}},\ \bibinfo {pages} {094503} (\bibinfo
  {year} {2008})}\BibitemShut {NoStop}%
\bibitem [{\citenamefont {Zhou}(2009)}]{Zhou2009}%
  \BibitemOpen
  \bibfield  {author} {\bibinfo {author} {\bibfnamefont {X.~Q.}\ \bibnamefont
  {Zhou}},\ }\emph {\bibinfo {title} {Microwave Flux-Flow Impedance
  Measurements of Type-{II} Superconductors}},\ \href@noop {} {Ph.D. thesis},\
  \bibinfo  {school} {Simon Fraser University} (\bibinfo {year}
  {2009})\BibitemShut {NoStop}%
\bibitem [{\citenamefont {Bean}\ and\ \citenamefont
  {Livingston}(1964)}]{Bean:1964wr}%
  \BibitemOpen
  \bibfield  {author} {\bibinfo {author} {\bibfnamefont {C.~P.}\ \bibnamefont
  {Bean}}\ and\ \bibinfo {author} {\bibfnamefont {J.~D.}\ \bibnamefont
  {Livingston}},\ }\href@noop {} {\bibfield  {journal} {\bibinfo  {journal}
  {Phys.\ Rev.\ Lett.}\ }\textbf {\bibinfo {volume} {12}},\ \bibinfo {pages}
  {14} (\bibinfo {year} {1964})}\BibitemShut {NoStop}%
\bibitem [{\citenamefont {London}\ and\ \citenamefont
  {London}(1935)}]{London:1935uf}%
  \BibitemOpen
  \bibfield  {author} {\bibinfo {author} {\bibfnamefont {F.}~\bibnamefont
  {London}}\ and\ \bibinfo {author} {\bibfnamefont {H.}~\bibnamefont
  {London}},\ }\href@noop {} {\bibfield  {journal} {\bibinfo  {journal} {Proc.\
  R.\ Soc.\ Lon.\ Ser.-A}\ }\textbf {\bibinfo {volume} {149}},\ \bibinfo
  {pages} {71} (\bibinfo {year} {1935})}\BibitemShut {NoStop}%
\bibitem [{\citenamefont {Caroli}\ \emph {et~al.}(1964)\citenamefont {Caroli},
  \citenamefont {de~Gennes},\ and\ \citenamefont {Matricon}}]{Caroli1964}%
  \BibitemOpen
  \bibfield  {author} {\bibinfo {author} {\bibfnamefont {C.}~\bibnamefont
  {Caroli}}, \bibinfo {author} {\bibfnamefont {P.~G.}\ \bibnamefont
  {de~Gennes}}, \ and\ \bibinfo {author} {\bibfnamefont {J.}~\bibnamefont
  {Matricon}},\ }\href@noop {} {\bibfield  {journal} {\bibinfo  {journal}
  {Phys.\ Lett.}\ }\textbf {\bibinfo {volume} {9}},\ \bibinfo {pages} {307}
  (\bibinfo {year} {1964})}\BibitemShut {NoStop}%
\bibitem [{\citenamefont {Ichioka}\ \emph {et~al.}(1999)\citenamefont
  {Ichioka}, \citenamefont {Hasegawa},\ and\ \citenamefont
  {Machida}}]{Ichioka:1999p359}%
  \BibitemOpen
  \bibfield  {author} {\bibinfo {author} {\bibfnamefont {M.}~\bibnamefont
  {Ichioka}}, \bibinfo {author} {\bibfnamefont {A.}~\bibnamefont {Hasegawa}}, \
  and\ \bibinfo {author} {\bibfnamefont {K.}~\bibnamefont {Machida}},\
  }\href@noop {} {\bibfield  {journal} {\bibinfo  {journal} {J Supercond}\
  }\textbf {\bibinfo {volume} {12}},\ \bibinfo {pages} {571} (\bibinfo {year}
  {1999})}\BibitemShut {NoStop}%
\bibitem [{\citenamefont {Coffey}\ and\ \citenamefont
  {Clem}(1991)}]{COFFEY:1991p156}%
  \BibitemOpen
  \bibfield  {author} {\bibinfo {author} {\bibfnamefont {M.~W.}\ \bibnamefont
  {Coffey}}\ and\ \bibinfo {author} {\bibfnamefont {J.~R.}\ \bibnamefont
  {Clem}},\ }\href@noop {} {\bibfield  {journal} {\bibinfo  {journal} {Phys.\
  Rev.\ Lett.}\ }\textbf {\bibinfo {volume} {67}},\ \bibinfo {pages} {386}
  (\bibinfo {year} {1991})}\BibitemShut {NoStop}%
\bibitem [{\citenamefont {Brandt}(1991)}]{BRANDT:1991p149}%
  \BibitemOpen
  \bibfield  {author} {\bibinfo {author} {\bibfnamefont {E.~H.}\ \bibnamefont
  {Brandt}},\ }\href@noop {} {\bibfield  {journal} {\bibinfo  {journal} {Phys.\
  Rev.\ Lett.}\ }\textbf {\bibinfo {volume} {67}},\ \bibinfo {pages} {2219}
  (\bibinfo {year} {1991})}\BibitemShut {NoStop}%
\bibitem [{Note1()}]{Note1}%
  \BibitemOpen
  \bibinfo {note} {The local electrodynamic relation, $\rho = Z_s^2/\protect
  \mathrm {i}\omega \mu _0$, is obtained by solving Maxwell's equations for
  phasor fields (Amp\`ere and Faraday laws) at the interface between vacuum and
  a conductor with local electrodynamics, \unhbox \voidb@x \hbox {$\protect
  \mathbf {E}(\protect \mathbf {r}) = \rho \protect \mathbf {J}(\protect
  \mathbf {r})$}. The surface impedance $Z_s$ is defined as the ratio of the
  tangential components of electric and magnetic field at the
  interface}\BibitemShut {NoStop}%
\bibitem [{\citenamefont {Yip}\ and\ \citenamefont
  {Sauls}(1992)}]{YIP:1992p3020}%
  \BibitemOpen
  \bibfield  {author} {\bibinfo {author} {\bibfnamefont {S.~K.}\ \bibnamefont
  {Yip}}\ and\ \bibinfo {author} {\bibfnamefont {J.~A.}\ \bibnamefont
  {Sauls}},\ }\href@noop {} {\bibfield  {journal} {\bibinfo  {journal} {Phys.\
  Rev.\ Lett.}\ }\textbf {\bibinfo {volume} {69}},\ \bibinfo {pages} {2264}
  (\bibinfo {year} {1992})}\BibitemShut {NoStop}%
\bibitem [{\citenamefont {Bidinosti}\ \emph {et~al.}(1999)\citenamefont
  {Bidinosti}, \citenamefont {Hardy}, \citenamefont {Bonn},\ and\ \citenamefont
  {Liang}}]{Bidinosti:1999p2720}%
  \BibitemOpen
  \bibfield  {author} {\bibinfo {author} {\bibfnamefont {C.~P.}\ \bibnamefont
  {Bidinosti}}, \bibinfo {author} {\bibfnamefont {W.~N.}\ \bibnamefont
  {Hardy}}, \bibinfo {author} {\bibfnamefont {D.~A.}\ \bibnamefont {Bonn}}, \
  and\ \bibinfo {author} {\bibfnamefont {R.}~\bibnamefont {Liang}},\
  }\href@noop {} {\bibfield  {journal} {\bibinfo  {journal} {Phys.\ Rev.
  Lett.}\ }\textbf {\bibinfo {volume} {83}},\ \bibinfo {pages} {3277} (\bibinfo
  {year} {1999})}\BibitemShut {NoStop}%
\bibitem [{\citenamefont {Sonier}\ \emph {et~al.}(2007)\citenamefont {Sonier},
  \citenamefont {Sabok-Sayr}, \citenamefont {Callaghan}, \citenamefont
  {Kaiser}, \citenamefont {Pacradouni}, \citenamefont {Brewer}, \citenamefont
  {Stubbs}, \citenamefont {Hardy}, \citenamefont {Bonn}, \citenamefont
  {Liang},\ and\ \citenamefont {Atkinson}}]{Sonier:2007p1185}%
  \BibitemOpen
  \bibfield  {author} {\bibinfo {author} {\bibfnamefont {J.~E.}\ \bibnamefont
  {Sonier}}, \bibinfo {author} {\bibfnamefont {S.~A.}\ \bibnamefont
  {Sabok-Sayr}}, \bibinfo {author} {\bibfnamefont {F.~D.}\ \bibnamefont
  {Callaghan}}, \bibinfo {author} {\bibfnamefont {C.~V.}\ \bibnamefont
  {Kaiser}}, \bibinfo {author} {\bibfnamefont {V.}~\bibnamefont {Pacradouni}},
  \bibinfo {author} {\bibfnamefont {J.~H.}\ \bibnamefont {Brewer}}, \bibinfo
  {author} {\bibfnamefont {S.~L.}\ \bibnamefont {Stubbs}}, \bibinfo {author}
  {\bibfnamefont {W.~N.}\ \bibnamefont {Hardy}}, \bibinfo {author}
  {\bibfnamefont {D.~A.}\ \bibnamefont {Bonn}}, \bibinfo {author}
  {\bibfnamefont {R.}~\bibnamefont {Liang}}, \ and\ \bibinfo {author}
  {\bibfnamefont {W.~A.}\ \bibnamefont {Atkinson}},\ }\href@noop {} {\bibfield
  {journal} {\bibinfo  {journal} {Phys. Rev. B}\ }\textbf {\bibinfo {volume}
  {76}},\ \bibinfo {pages} {134518} (\bibinfo {year} {2007})}\BibitemShut
  {NoStop}%
\bibitem [{\citenamefont {Liang}\ \emph {et~al.}(2012)\citenamefont {Liang},
  \citenamefont {Bonn},\ and\ \citenamefont {Hardy}}]{Liang:2012va}%
  \BibitemOpen
  \bibfield  {author} {\bibinfo {author} {\bibfnamefont {R.}~\bibnamefont
  {Liang}}, \bibinfo {author} {\bibfnamefont {D.~A.}\ \bibnamefont {Bonn}}, \
  and\ \bibinfo {author} {\bibfnamefont {W.~N.}\ \bibnamefont {Hardy}},\
  }\href@noop {} {\bibfield  {journal} {\bibinfo  {journal} {Philos.\ Mag.}\
  }\textbf {\bibinfo {volume} {92}},\ \bibinfo {pages} {2563} (\bibinfo {year}
  {2012})}\BibitemShut {NoStop}%
\bibitem [{\citenamefont {Huttema}\ \emph {et~al.}(2006)\citenamefont
  {Huttema}, \citenamefont {Morgan}, \citenamefont {Turner}, \citenamefont
  {Hardy}, \citenamefont {Zhou}, \citenamefont {Bonn}, \citenamefont {Liang},\
  and\ \citenamefont {Broun}}]{Huttema:2006p344}%
  \BibitemOpen
  \bibfield  {author} {\bibinfo {author} {\bibfnamefont {W.~A.}\ \bibnamefont
  {Huttema}}, \bibinfo {author} {\bibfnamefont {B.}~\bibnamefont {Morgan}},
  \bibinfo {author} {\bibfnamefont {P.~J.}\ \bibnamefont {Turner}}, \bibinfo
  {author} {\bibfnamefont {W.~N.}\ \bibnamefont {Hardy}}, \bibinfo {author}
  {\bibfnamefont {X.}~\bibnamefont {Zhou}}, \bibinfo {author} {\bibfnamefont
  {D.~A.}\ \bibnamefont {Bonn}}, \bibinfo {author} {\bibfnamefont
  {R.}~\bibnamefont {Liang}}, \ and\ \bibinfo {author} {\bibfnamefont {D.~M.}\
  \bibnamefont {Broun}},\ }\href@noop {} {\bibfield  {journal} {\bibinfo
  {journal} {Rev.\ Sci.\ Instrum.}\ }\textbf {\bibinfo {volume} {77}},\
  \bibinfo {pages} {023901} (\bibinfo {year} {2006})}\BibitemShut {NoStop}%
\bibitem [{\citenamefont {Sridhar}\ and\ \citenamefont
  {Kennedy}(1988)}]{Sridhar:1988p495}%
  \BibitemOpen
  \bibfield  {author} {\bibinfo {author} {\bibfnamefont {S.}~\bibnamefont
  {Sridhar}}\ and\ \bibinfo {author} {\bibfnamefont {W.~L.}\ \bibnamefont
  {Kennedy}},\ }\href@noop {} {\bibfield  {journal} {\bibinfo  {journal} {Rev.\
  Sci.\ Instrum.}\ }\textbf {\bibinfo {volume} {59}},\ \bibinfo {pages} {531}
  (\bibinfo {year} {1988})}\BibitemShut {NoStop}%
\bibitem [{\citenamefont {Altshuler}(1963)}]{altshuler1963}%
  \BibitemOpen
  \bibfield  {author} {\bibinfo {author} {\bibfnamefont {H.~M.}\ \bibnamefont
  {Altshuler}},\ }in\ \href@noop {} {\emph {\bibinfo {booktitle} {Handbook of
  Microwave Measurements}}}\ (\bibinfo  {publisher} {Polytechnic Institute of
  Brooklyn},\ \bibinfo {address} {Brooklyn, New York},\ \bibinfo {year}
  {1963})\ pp.\ \bibinfo {pages} {495--548}\BibitemShut {NoStop}%
\bibitem [{\citenamefont {Pereg-Barnea}\ \emph {et~al.}(2004)\citenamefont
  {Pereg-Barnea}, \citenamefont {Turner}, \citenamefont {Harris}, \citenamefont
  {Mullins}, \citenamefont {Bobowski}, \citenamefont {Raudsepp}, \citenamefont
  {Liang}, \citenamefont {Bonn},\ and\ \citenamefont
  {Hardy}}]{PeregBarnea:2004p761}%
  \BibitemOpen
  \bibfield  {author} {\bibinfo {author} {\bibfnamefont {T.}~\bibnamefont
  {Pereg-Barnea}}, \bibinfo {author} {\bibfnamefont {P.~J.}\ \bibnamefont
  {Turner}}, \bibinfo {author} {\bibfnamefont {R.}~\bibnamefont {Harris}},
  \bibinfo {author} {\bibfnamefont {G.~K.}\ \bibnamefont {Mullins}}, \bibinfo
  {author} {\bibfnamefont {J.}~\bibnamefont {Bobowski}}, \bibinfo {author}
  {\bibfnamefont {M.}~\bibnamefont {Raudsepp}}, \bibinfo {author}
  {\bibfnamefont {R.}~\bibnamefont {Liang}}, \bibinfo {author} {\bibfnamefont
  {D.~A.}\ \bibnamefont {Bonn}}, \ and\ \bibinfo {author} {\bibfnamefont
  {W.~N.}\ \bibnamefont {Hardy}},\ }\href@noop {} {\bibfield  {journal}
  {\bibinfo  {journal} {Phys.\ Rev.\ B}\ }\textbf {\bibinfo {volume} {69}},\
  \bibinfo {pages} {184513} (\bibinfo {year} {2004})}\BibitemShut {NoStop}%
\bibitem [{\citenamefont {Prozorov}\ \emph {et~al.}(2000)\citenamefont
  {Prozorov}, \citenamefont {Giannetta}, \citenamefont {Carrington},\ and\
  \citenamefont {Araujo-Moreira}}]{Prozorov:2000tj}%
  \BibitemOpen
  \bibfield  {author} {\bibinfo {author} {\bibfnamefont {R.}~\bibnamefont
  {Prozorov}}, \bibinfo {author} {\bibfnamefont {R.~W.}\ \bibnamefont
  {Giannetta}}, \bibinfo {author} {\bibfnamefont {A.}~\bibnamefont
  {Carrington}}, \ and\ \bibinfo {author} {\bibfnamefont {F.~M.}\ \bibnamefont
  {Araujo-Moreira}},\ }\href@noop {} {\bibfield  {journal} {\bibinfo  {journal}
  {Phys.\ Rev.\ B}\ }\textbf {\bibinfo {volume} {62}},\ \bibinfo {pages} {115}
  (\bibinfo {year} {2000})}\BibitemShut {NoStop}%
\bibitem [{\citenamefont {Ramshaw}\ \emph {et~al.}(2012)\citenamefont
  {Ramshaw}, \citenamefont {Day}, \citenamefont {Vignolle}, \citenamefont
  {LeBoeuf}, \citenamefont {Dosanjh}, \citenamefont {Proust}, \citenamefont
  {Taillefer}, \citenamefont {Liang}, \citenamefont {Hardy},\ and\
  \citenamefont {Bonn}}]{Ramshaw2012}%
  \BibitemOpen
  \bibfield  {author} {\bibinfo {author} {\bibfnamefont {B.~J.}\ \bibnamefont
  {Ramshaw}}, \bibinfo {author} {\bibfnamefont {J.}~\bibnamefont {Day}},
  \bibinfo {author} {\bibfnamefont {B.}~\bibnamefont {Vignolle}}, \bibinfo
  {author} {\bibfnamefont {D.}~\bibnamefont {LeBoeuf}}, \bibinfo {author}
  {\bibfnamefont {P.}~\bibnamefont {Dosanjh}}, \bibinfo {author} {\bibfnamefont
  {C.}~\bibnamefont {Proust}}, \bibinfo {author} {\bibfnamefont
  {L.}~\bibnamefont {Taillefer}}, \bibinfo {author} {\bibfnamefont
  {R.}~\bibnamefont {Liang}}, \bibinfo {author} {\bibfnamefont {W.~N.}\
  \bibnamefont {Hardy}}, \ and\ \bibinfo {author} {\bibfnamefont {D.~A.}\
  \bibnamefont {Bonn}},\ }\href@noop {} {\bibfield  {journal} {\bibinfo
  {journal} {Phys.\ Rev.\ B}\ }\textbf {\bibinfo {volume} {86}},\ \bibinfo
  {pages} {174501} (\bibinfo {year} {2012})}\BibitemShut {NoStop}%
\bibitem [{Note2()}]{Note2}%
  \BibitemOpen
  \bibinfo {note} {Note that vortex dissipation is due to the coupling of
  induced vortex electric fields to charge excitations in the vicinity of the
  vortex cores: $\eta ^\prime (\omega )$ therefore reflects the dynamics of
  these excitations, rather than the dynamics of the vortices
  themselves.}\BibitemShut {Stop}%
\bibitem [{\citenamefont {Senoussi}\ \emph {et~al.}(1988)\citenamefont
  {Senoussi}, \citenamefont {Ouss{\'e}na}, \citenamefont {Collin},\ and\
  \citenamefont {Campbell}}]{Senoussi:1988jf}%
  \BibitemOpen
  \bibfield  {author} {\bibinfo {author} {\bibfnamefont {S.}~\bibnamefont
  {Senoussi}}, \bibinfo {author} {\bibfnamefont {M.}~\bibnamefont
  {Ouss{\'e}na}}, \bibinfo {author} {\bibfnamefont {G.}~\bibnamefont {Collin}},
  \ and\ \bibinfo {author} {\bibfnamefont {I.~A.}\ \bibnamefont {Campbell}},\
  }\href@noop {} {\bibfield  {journal} {\bibinfo  {journal} {Phys.\ Rev.\ B}\
  }\textbf {\bibinfo {volume} {37}},\ \bibinfo {pages} {9792} (\bibinfo {year}
  {1988})}\BibitemShut {NoStop}%
\bibitem [{\citenamefont {Shi}\ \emph {et~al.}(1994)\citenamefont {Shi},
  \citenamefont {Chaikin}, \citenamefont {Ong},\ and\ \citenamefont
  {Wang}}]{Shi:1994cn}%
  \BibitemOpen
  \bibfield  {author} {\bibinfo {author} {\bibfnamefont {X.~D.}\ \bibnamefont
  {Shi}}, \bibinfo {author} {\bibfnamefont {P.~M.}\ \bibnamefont {Chaikin}},
  \bibinfo {author} {\bibfnamefont {N.~P.}\ \bibnamefont {Ong}}, \ and\
  \bibinfo {author} {\bibfnamefont {Z.~Z.}\ \bibnamefont {Wang}},\ }\href@noop
  {} {\bibfield  {journal} {\bibinfo  {journal} {Phys.\ Rev.\ B}\ }\textbf
  {\bibinfo {volume} {50}},\ \bibinfo {pages} {13845} (\bibinfo {year}
  {1994})}\BibitemShut {NoStop}%
\bibitem [{\citenamefont {Feigel'man}\ and\ \citenamefont
  {Vinokur}(1990)}]{Feigelman:1990gp}%
  \BibitemOpen
  \bibfield  {author} {\bibinfo {author} {\bibfnamefont {M.~V.}\ \bibnamefont
  {Feigel'man}}\ and\ \bibinfo {author} {\bibfnamefont {V.~M.}\ \bibnamefont
  {Vinokur}},\ }\href@noop {} {\bibfield  {journal} {\bibinfo  {journal}
  {Phys.\ Rev.\ B}\ }\textbf {\bibinfo {volume} {41}},\ \bibinfo {pages} {8986}
  (\bibinfo {year} {1990})}\BibitemShut {NoStop}%
\bibitem [{\citenamefont {Fisher}\ \emph {et~al.}(1991)\citenamefont {Fisher},
  \citenamefont {Fisher},\ and\ \citenamefont {Huse}}]{FISHER:1991p681}%
  \BibitemOpen
  \bibfield  {author} {\bibinfo {author} {\bibfnamefont {D.~S.}\ \bibnamefont
  {Fisher}}, \bibinfo {author} {\bibfnamefont {M.~P.~A.}\ \bibnamefont
  {Fisher}}, \ and\ \bibinfo {author} {\bibfnamefont {D.~A.}\ \bibnamefont
  {Huse}},\ }\href@noop {} {\bibfield  {journal} {\bibinfo  {journal} {Phys.\
  Rev.\ B}\ }\textbf {\bibinfo {volume} {43}},\ \bibinfo {pages} {130}
  (\bibinfo {year} {1991})}\BibitemShut {NoStop}%
\bibitem [{\citenamefont {Charalambous}\ \emph {et~al.}(1992)\citenamefont
  {Charalambous}, \citenamefont {Chaussy},\ and\ \citenamefont
  {Lejay}}]{Charalambous:1992gf}%
  \BibitemOpen
  \bibfield  {author} {\bibinfo {author} {\bibfnamefont {M.}~\bibnamefont
  {Charalambous}}, \bibinfo {author} {\bibfnamefont {J.}~\bibnamefont
  {Chaussy}}, \ and\ \bibinfo {author} {\bibfnamefont {P.}~\bibnamefont
  {Lejay}},\ }\href@noop {} {\bibfield  {journal} {\bibinfo  {journal} {Phys.\
  Rev.\ B}\ }\textbf {\bibinfo {volume} {45}},\ \bibinfo {pages} {5091}
  (\bibinfo {year} {1992})}\BibitemShut {NoStop}%
\bibitem [{\citenamefont {Kwok}\ \emph {et~al.}(1992)\citenamefont {Kwok},
  \citenamefont {Fleshler}, \citenamefont {Welp}, \citenamefont {Vinokur},
  \citenamefont {Downey}, \citenamefont {Crabtree},\ and\ \citenamefont
  {Miller}}]{Kwok:1992ug}%
  \BibitemOpen
  \bibfield  {author} {\bibinfo {author} {\bibfnamefont {W.~K.}\ \bibnamefont
  {Kwok}}, \bibinfo {author} {\bibfnamefont {S.}~\bibnamefont {Fleshler}},
  \bibinfo {author} {\bibfnamefont {U.}~\bibnamefont {Welp}}, \bibinfo {author}
  {\bibfnamefont {V.~M.}\ \bibnamefont {Vinokur}}, \bibinfo {author}
  {\bibfnamefont {J.}~\bibnamefont {Downey}}, \bibinfo {author} {\bibfnamefont
  {G.~W.}\ \bibnamefont {Crabtree}}, \ and\ \bibinfo {author} {\bibfnamefont
  {M.~M.}\ \bibnamefont {Miller}},\ }\href@noop {} {\bibfield  {journal}
  {\bibinfo  {journal} {Phys.\ Rev.\ Lett.}\ }\textbf {\bibinfo {volume}
  {69}},\ \bibinfo {pages} {3370} (\bibinfo {year} {1992})}\BibitemShut
  {NoStop}%
\bibitem [{\citenamefont {Safar}\ \emph {et~al.}(1992)\citenamefont {Safar},
  \citenamefont {Gammel}, \citenamefont {Huse}, \citenamefont {Bishop},
  \citenamefont {Rice},\ and\ \citenamefont {Ginsberg}}]{Safar:1992ep}%
  \BibitemOpen
  \bibfield  {author} {\bibinfo {author} {\bibfnamefont {H.}~\bibnamefont
  {Safar}}, \bibinfo {author} {\bibfnamefont {P.~L.}\ \bibnamefont {Gammel}},
  \bibinfo {author} {\bibfnamefont {D.~A.}\ \bibnamefont {Huse}}, \bibinfo
  {author} {\bibfnamefont {D.~J.}\ \bibnamefont {Bishop}}, \bibinfo {author}
  {\bibfnamefont {J.~P.}\ \bibnamefont {Rice}}, \ and\ \bibinfo {author}
  {\bibfnamefont {D.~M.}\ \bibnamefont {Ginsberg}},\ }\href@noop {} {\bibfield
  {journal} {\bibinfo  {journal} {Phys.\ Rev.\ Lett.}\ }\textbf {\bibinfo
  {volume} {69}},\ \bibinfo {pages} {824} (\bibinfo {year} {1992})}\BibitemShut
  {NoStop}%
\bibitem [{\citenamefont {Mackenzie}\ \emph {et~al.}(1993)\citenamefont
  {Mackenzie}, \citenamefont {Julian}, \citenamefont {Lonzarich}, \citenamefont
  {Carrington}, \citenamefont {Hughes}, \citenamefont {Liu},\ and\
  \citenamefont {Sinclair}}]{MACKENZIE:1993p197}%
  \BibitemOpen
  \bibfield  {author} {\bibinfo {author} {\bibfnamefont {A.~P.}\ \bibnamefont
  {Mackenzie}}, \bibinfo {author} {\bibfnamefont {S.~R.}\ \bibnamefont
  {Julian}}, \bibinfo {author} {\bibfnamefont {G.~G.}\ \bibnamefont
  {Lonzarich}}, \bibinfo {author} {\bibfnamefont {A.}~\bibnamefont
  {Carrington}}, \bibinfo {author} {\bibfnamefont {S.~D.}\ \bibnamefont
  {Hughes}}, \bibinfo {author} {\bibfnamefont {R.~S.}\ \bibnamefont {Liu}}, \
  and\ \bibinfo {author} {\bibfnamefont {D.~C.}\ \bibnamefont {Sinclair}},\
  }\href@noop {} {\bibfield  {journal} {\bibinfo  {journal} {Phys.\ Rev.\
  Lett.}\ }\textbf {\bibinfo {volume} {71}},\ \bibinfo {pages} {1238} (\bibinfo
  {year} {1993})}\BibitemShut {NoStop}%
\bibitem [{\citenamefont {Durst}\ and\ \citenamefont
  {Lee}(2000)}]{Durst:2000p963}%
  \BibitemOpen
  \bibfield  {author} {\bibinfo {author} {\bibfnamefont {A.~C.}\ \bibnamefont
  {Durst}}\ and\ \bibinfo {author} {\bibfnamefont {P.~A.}\ \bibnamefont
  {Lee}},\ }\href@noop {} {\bibfield  {journal} {\bibinfo  {journal} {Phys.\
  Rev.\ B}\ }\textbf {\bibinfo {volume} {62}},\ \bibinfo {pages} {1270}
  (\bibinfo {year} {2000})}\BibitemShut {NoStop}%
\bibitem [{\citenamefont {Volovik}(1993)}]{VOLOVIK:1993p201}%
  \BibitemOpen
  \bibfield  {author} {\bibinfo {author} {\bibfnamefont {G.~E.}\ \bibnamefont
  {Volovik}},\ }\href@noop {} {\bibfield  {journal} {\bibinfo  {journal} {JETP
  Lett.}\ }\textbf {\bibinfo {volume} {58}},\ \bibinfo {pages} {469} (\bibinfo
  {year} {1993})}\BibitemShut {NoStop}%
\end{thebibliography}

%

\end{document}